\documentclass[12pt,letterpaper]{article}
\pdfoutput=1

\usepackage[compress]{cite}
\usepackage{geometry}
\usepackage{tikz}
\usepackage{float}
\usepackage{color}
\usepackage{slashed}
\definecolor{refkey}{rgb}{1,0,0}
\definecolor{labelkey}{rgb}{0,0,1}
\usepackage{amsfonts,amssymb,amsmath}
\usepackage{bbold}
\usepackage{tcolorbox}
\usepackage{esint}
\usepackage{rotating}
\usepackage[utf8]{inputenc}

\usepackage{graphicx}
\usepackage{caption}
\usepackage{subcaption}
\usepackage{booktabs}

\usepackage{listings}

\usepackage{hyperref}

\newcommand{\be}{\begin{equation}}
\newcommand{\ee}{\end{equation}}
\newcommand{\ben}{\begin{displaymath}}
\newcommand{\een}{\end{displaymath}}
\newcommand{\bea}{\begin{equation}\begin{aligned}}
\newcommand{\eea}{\end{aligned}\end{equation}}
\newcommand{\bean}{\begin{eqnarray*}}
\newcommand{\eean}{\end{eqnarray*}}

\def\a {\alpha}

\newcommand{\bra}[1]{\mbox{$\langle #1 |$}}
\newcommand{\ket}[1]{\mbox{$| #1 \rangle$}}

\newcommand{\eg}{{\it e.g.}}
\newcommand{\ie}{{\it i.e.}}

\newcommand{\tr}{\mbox{Tr}}

\newcommand{\commentout}[1]{}

\newcommand{\beq}{\begin{equation}}
\newcommand{\eeq}{\end{equation}}
\newcommand{\beqr}{\begin{displaymath}}
\newcommand{\eeqr}{\end{displaymath}}
\newcommand{\beqa}{\begin{eqnarray}}
\newcommand{\eeqa}{\end{eqnarray}}
\newcommand{\beqar}{\begin{eqnarray*}}
\newcommand{\eeqar}{\end{eqnarray*}}

\newcommand{\cO}{{\cal O}}
\newcommand{\cF}{{\cal F}}
\newcommand{\cL}{{\cal L}}

\newcommand{\non}{\nonumber}

\newcommand{\half}{\ensuremath{\frac{1}{2}}}

\renewcommand{\Re}{\ensuremath{\mathrm{Re}}}
\renewcommand{\Im}{\ensuremath{\mathrm{Im}}}

\newcommand{\intxy}{\ensuremath{\int_4^\infty\!\!\!\!\!\! dx \int_4^\infty\!\!\!\!\!\! dy\,}}
\newcommand{\intx}{\ensuremath{\int_4^\infty\!\!\!\!\!\! dx\,}}

\newcommand{\MeV}{\ensuremath{\mbox{MeV}}}
\newcommand{\GeV}{\ensuremath{\mbox{GeV}}}

\definecolor{gblue}{RGB}{15,10,220}
\definecolor{gred}{RGB}{220,10,15}
\newcommand{\gtbblue}[1]{\color{gblue}#1}
\newcommand{\gtbred}[1]{\color{gred}#1}

\newcommand{\fpi}{\ensuremath{f_{\pi}}}
\newcommand{\mpi}{\ensuremath{m_{\pi}}}

\begin{document}

\title{\Large \bf Gauge Theory Bootstrap: \\ Pion amplitudes and low energy parameters}

\author{
	Yifei He$^\text{1}$,
	Martin Kruczenski$^\text{2}$ \thanks{E-mail: \texttt{yifei.he@ens.fr, markru@purdue.edu.}} \\
[1.0mm]
$^1$ \small Laboratoire de Physique de l'\'Ecole Normale Supérieure, ENS, Université PSL,\\
\small CNRS, Sorbonne Université, Université Paris Cité, F-75005 Paris, France \\
$^2$ \small Department of Physics and Astronomy and PQSEI\thanks{Purdue Quantum Science and Engineering Institute}, \\
\small Purdue University, West Lafayette, IN 47907, USA.}

\date{\today}

\maketitle

\begin{abstract}
	Following the Gauge Theory Bootstrap method proposed in our previous work \cite{GTBPRL,GTBPRD}, we compute pion scattering phase shifts for all partial waves with angular momentum $\ell\le 3$ up to 2 GeV and calculate the low energy $\chi$PT coefficients $\bar{\ell}_{1,2,4,6}$. The method looks for the most general S-matrix that matches at low energy the tree level amplitudes of the non-linear sigma model and at high energy, QCD sum rules and form factors. This is a theoretical/numerical calculation that uses as only data the pion mass $m_\pi$, pion decay constant $f_\pi$ and the QCD parameters $N_c=3$, $N_f=2$, $m_q$ and $\alpha_s$. All results are in reasonable agreement with experiment. In particular, we find the $\rho(770)$, $f_2(1270)$ and $\rho(1450)$ resonances. The interplay between the UV gauge theory and low energy pion physics is an example of a general situation where we know the microscopic theory as well as the effective theory of long wavelength fluctuations but we want to solve the strongly coupled dynamics at intermediate energies. The bootstrap builds a bridge  between the low and high energy by determining the consistent S-matrix that matches both and provides, in this case, a new direction to understand the strongly coupled physics of gauge theories.	
\end{abstract}

\clearpage

\tableofcontents

\newpage
\newpage

\numberwithin{equation}{section}

\section{Introduction} 

 We recently proposed the Gauge Theory Bootstrap \cite{GTBPRL,GTBPRD} a method to address the important problem of computing pion scattering amplitudes in the strongly coupled regime of asymptotically free gauge theories under the assumption that they undergo confinement and chiral symmetry breaking. In QCD language, ignoring all other interactions, the pion and the nucleons are the only stable particles. Pions, in particular, can be thought as the pseudo-Goldstone bosons of chiral symmetry breaking. Their interactions are severely restricted by symmetry considerations and their low energy effective field theory is a non-linear sigma model with coupling $f_\pi$ (pion decay constant) and pion mass $m_\pi$. In pion scattering, nucleons can be ignored below the nucleon anti-nucleon threshold $\sim 2\,\GeV$. Thus, we have a UV description of the theory in terms of weakly coupled quarks and gluons and a low energy effective field theory (EFT) description in terms of weakly coupled pions. However, this is \emph{not sufficient} to compute pion scattering since the effective pion coupling increases with the energy and the EFT description breaks down before we reach the asymptotically free regime. 
 \begin{figure}[H]
 	\centering
 	\includegraphics[height=0.5\textwidth]{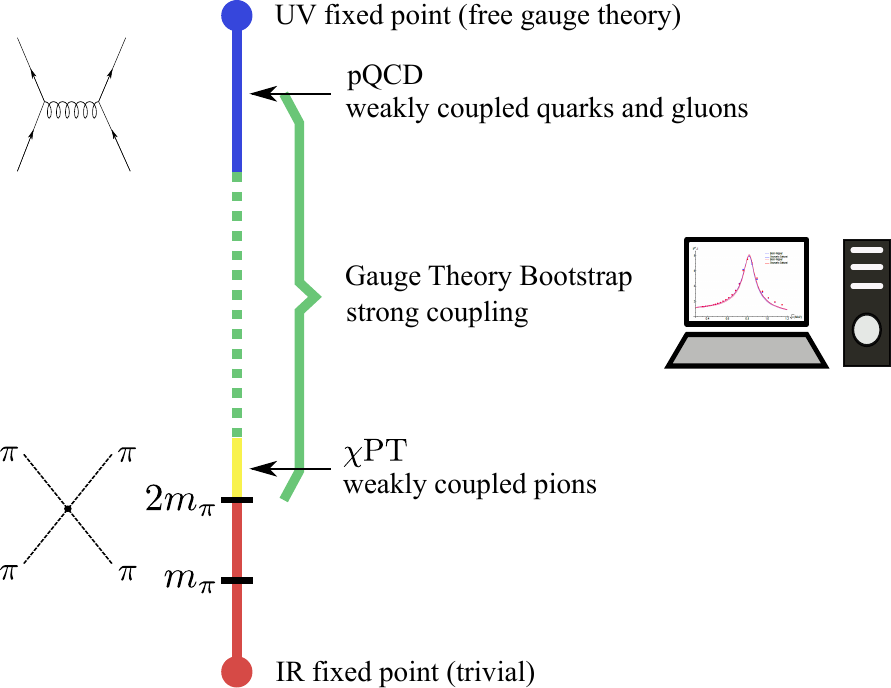}
 	\caption{The Gauge Theory Bootstrap builds a (strongly coupled) bridge between the weakly coupled QCD at high energies and a weakly coupled EFT of pions at low energies.}
 	\label{UVIR}
 \end{figure}
 In this intermediate regime, both the gauge theory and the theory of pions are strongly coupled, making it an ideal target for bootstrap methods. The situation is sketched in fig.\ref{UVIR}. The Gauge Theory Bootstrap \cite{GTBPRL,GTBPRD} addresses this problem by looking for the most general S-matrix that matches the low and high energy theories. This is done in three steps. First we parameterize the space of S-matrices and form factors that satisfy the general conditions of analyticity, crossing, unitarity and global symmetries. In order to do that we use bootstrap ideas \cite{chew1966analytic,Eden:1966dnq} as recently reformulated for the S-matrix by Paulos, Penedodes, Toledo, van Rees and Vieira \cite{Paulos:2016fap,Paulos:2016but,Paulos:2017fhb,Kruczenski:2022lot} 
 and augmented by Karateev, Kuhn and Penedones \cite{Karateev:2019ymz} to incorporate form factors and current spectral densities (or vacuum polarizations).
 
 Second, we consider that pions are the pseudo-Goldstone of chiral symmetry breaking and are described at low energy by a non-linear sigma model (Weinberg model \cite{PhysRevLett.17.616}). Thus, we require that the partial waves match (within a tolerance) the tree level predictions of the non-linear sigma model. This drastically reduces the space of S-matrices and produces partial waves that, in some channels, display properties similar to experiment. A notable exception is the absence of a $\rho$ meson resonance in the $P1$ channel, showing that pion scattering is not just a result of chiral dynamics but requires a deeper  understanding of its relation to the high energy gauge theory.  
 
 Therefore, the third and final step is to introduce information about the high energy gauge theory by using the SVZ sum rules (Shifman, Vainshtein, Zakharov) \cite{SHIFMAN1979385,SHIFMAN1979448,Shifman:1978bw,Novikov:1977dq,Reinders:1981bq,Reinders:1984sr,Gubler:2018ctz} together with the high energy behavior of the form factors (from Brodsky and Lepage \cite{osti_1447331,Pire:1996bc}).  
 
\begin{figure}[h]
 	\centering
 	\includegraphics[width=0.9\textwidth]{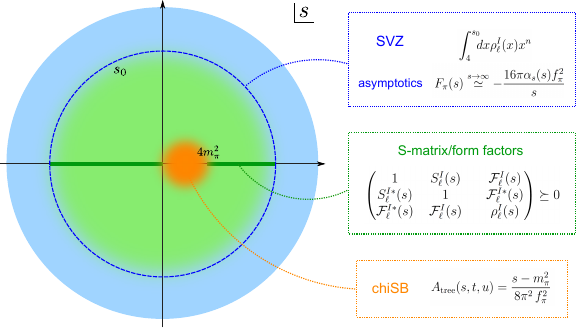}
 	\caption{Schematic of the Gauge Theory Bootstrap.}
 	\label{GTB}
 \end{figure}

 All these ingredients together allow to set up a numerical procedure to search for the S-matrix and compute the phase shifts of pion scattering. A Schematic of the Gauge Theory Bootstrap is depicted in fig. \ref{GTB}.
 
 It is clear that, if enough conditions are imposed, we should \emph{inevitably} find, from all S-matrices allowed by the basic constraints, the unique S-matrix that corresponds to the given UV theory. Perhaps surprisingly in the previous paper \cite{GTBPRL,GTBPRD} we found that a small set of IR and UV constraints drastically reduce the space of allowed S-matrices leading to partial waves that qualitatively and quantitatively match the experimental results.  Notice that other than symmetries and gauge theory parameters, the only input are the values of $m_\pi$ and $f_\pi$. Since no experimental input on pion scattering is used, the bootstrap numerical calculation can be considered a \emph{first principles calculation} from the pion physics point of view. 
 
 It is useful to recall how an effective field theory calculation would address the same problem. Initially we can attempt a standard perturbative calculation and start by computing tree level amplitudes using the lowest order chiral Lagrangian (sigma model). Since tree level amplitudes do not satisfy unitarity we have to incorporate loops. However the theory is non-renormalizable and we also have to add more and more counter-terms with new couplings that we do not know. In other words the UV completion is not unique. However the ambiguity is removed if we require that the theory matches QCD at high energy since the gauge theory provides the UV completion. The bootstrap shortcuts this procedure by focusing on the physically meaningful S-matrix, and allows us to pick out the one that corresponds to the given UV gauge theory. Moreover, the numerical implementation of this idea does not require expensive computations, in fact, computing the partial waves requires only a few minutes on an average PC. 

\smallskip 
  
In understanding why this method works and how it connects to other approaches, it is important to keep in mind that general considerations of analyticity, crossing and unitarity have played a significant role in studying pion scattering since the early days \cite{chew1966analytic} and have motivated various bootstrap approaches to the problem. Indeed, phenomenological studies by Pelaez and Yndurain \cite{Pelaez:2004vs,Yndurain:2002ud} using conformal mapping and by Colangelo, Gasser and Leutwyler \cite{Colangelo:2001df} using Roy equations have shown that the experimental phase shifts can be described in terms of analytic functions satisfying crossing and unitarity. More recently, the low energy effective field theory was extensively studied with the S-matrix bootstrap in order to put generic bounds on its parameters. In the case of massive and massless pions in \cite{Guerrieri:2020bto,Guerrieri:2018uew} the allowed parameter space was determined using the full constraints of unitarity, crossing and analyticity. There is also a number of interesting related work on the large $N_c$ limit of the effective theory using bootstrap \cite{Albert:2022oes,Fernandez:2022kzi,Albert:2023jtd,Ma:2023vgc,Albert:2023seb}. For example, using the large $N_c$ QCD positivity bootstrap,  \cite{Albert:2023seb} found a point in the space of S-matrices that displays a full Regge trajectory. Another related topic is that the hadronic vacuum polarization or spectral densities that we consider are related to the $g-2$ experiment (see \eg\ the review \cite{KESHAVARZI2022115675}), although we cannot yet compute them with the required precision to compete for example with lattice calculations \cite{Davier:2023cyp}.  

 What seems particularly interesting is that, by using UV theory information, our method allows to identify the low energy S-matrix corresponding to a specific gauge theory, in some sense going back to the roots of the bootstrap program  \cite{chew1966analytic} which was to \emph{compute} the S-matrix of the strong interactions.  
An important consequence of having such a method is that it allows a numerical exploration of different gauge theories (number of colors, quark masses, etc.) and also of different phenomena in the same theory. By including and removing constraints we can determine what physics is given by chiral symmetry alone (seemingly the $S0$ and $S2$ partial waves) and which is a consequence of QCD dynamics (like the $\rho$ meson). For example removing the $P1$ sum rules eliminates the $\rho$ meson but removing the $D0$ sum rule does not completely eliminate the $f_2$ resonance in that channel. These naive explorations that the method allow may become an important tool in understanding the interplay between the non-linear sigma model and the gauge theory.

\bigskip

\noindent {\bf Organization of the paper:}\\
In section \ref{sec:gtbsummary} we review the main idea of the method and give a short summary of its implementation.  The main results are described in section \ref{sec:phaseshifts}, namely, the scattering phase shifts for partial waves with angular momentum $\ell\le 3$ up to a center of mass energy $\sqrt{s}=2\,\GeV$. Analyzing those results in more detail  allow the computation of the scattering lengths, effective range parameter, charge radii of the pion and finally the chiral Lagrangian coefficients usually denoted as $\bar{l}_j$ following Gasser and Leutwyler \cite{GASSER1984142,GASSER198765}. The results are presented in section \ref{sec:parameters}. In the following section \ref{sec:resonances}, we study the resonances that are evident in the phase shifts, \ie\ $\rho(770)$, $f_2(1270)$ and $\rho(1450)$, find their masses and widths as well as couplings to pions and form factors. This completes the description of the results. For the reader interested in the perturbative QCD calculations, we include section \ref{sec:FFl} where we review the computation of the asymptotic form factor pioneered by Brodsky and Lepage \cite{osti_1447331,Pire:1996bc} and section \ref{sec:jj} where we review the calculation of the two point functions that we use in the SVZ sum rules \cite{SHIFMAN1979385}.  For the readers interested in the numerical procedure and reproducing the results, we provide a detailed description in section \ref{sec:nummeth} and accompanying Mathematica and Matlab notebooks with this submission.\footnote{The Mathematica program computes the coefficients needed to setup the bootstrap maximization problem. Having those, the matlab program takes $\sim 20$ min on an average laptop to find the partial waves.} Finally we give our conclusions in section \ref{sec:conclusions}. In the appendices we include a recalculation of the scattering lengths based on the $\chi{\mathrm{PT}}$ parameters $\bar{l}_j$, a set of tests to understand how the solutions depend on the parameters associated with the numerical method (\eg\ discretization), a more detailed example of a Feynman diagram that contributes to the asymptotic form factors and some analytical integrals that are useful in the main text. 

\section{Summary of the gauge theory bootstrap}\label{sec:gtbsummary}

 The basic idea of the gauge theory bootstrap is to find the S-matrix (here we concentrate on the elastic $2\rightarrow 2$ S-matrix) of pion scattering that agrees with lowest order chiral perturbation theory at low energy and perturbative QCD at high energy. To implement this idea we follow the procedure already described in \cite{GTBPRL,GTBPRD} but extend it to a larger range of energy and several more partial waves. In this section, we describe the main ideas of this approach and give a brief summary of its implementation. A more detailed description of the numerical method is given in section \ref{sec:nummeth}.
 
\subsection{S-matrix/form factor bootstrap}\label{Sformb}

We start with the basic setup where we parameterize the scattering amplitudes and form factors so that all linear constraints of analyticity, crossing and global symmetries are satisfied. Unitarity, a non-linear constraint is imposed later numerically. Thus, we consider a scattering amplitude analytic in the Mandelstam variables $(s,t,u)$ and assume maximal (or Mandelstam) analyticity which means that we only consider singularities required by physical reasons. Those can be poles due to stable bound states (that we do not have here) and two-particle (and other multi-particle) cuts.  Under this conditions we write the $2\rightarrow 2$ pion scattering amplitude as
\beqa\label{Adef}
A(s,t,u) &=& T_0 + \frac{1}{\pi} \intx  \frac{\sigma_1(x)}{x-s}+\frac{1}{\pi} \intx \sigma_2(x) \left[\frac{1}{x-t}+\frac{1}{x-u}\right]\nonumber \\ 
&& +  \frac{1}{\pi^2}  \intxy  \frac{\rho_1(x,y)}{x-s}\left[\frac{1}{y-t}+\frac{1}{y-u}\right] \nonumber\\
&& +  \frac{1}{\pi^2}  \intxy   \frac{\rho_2(x,y)}{(x-t)(y-u)}
\label{a11}
\eeqa
where $s+t+u=4$ (setting $m_{\pi}=1$). The amplitude is parametrized by a constant $T_0$, two single spectral densities $\sigma_{1,2}$ and two double spectral densities $\rho_{1,2}$. It satisfies the crossing condition $A(s,t,u)=A(s,u,t)$ by requiring $\rho_2(x,y)=\rho_2(y,x)$. 
In each isospin channel $I=0,1,2$ the amplitude reads:
\begin{subequations}\label{a10}
	\beqa	
	T^{I=0}(s,t,u) &=& 3 A(s,t,u)+A(t,s,u)+A(u,t,s) \\ 
	T^{I=1}(s,t,u) &=& A(t,s,u)-A(u,t,s) \\ 
	T^{I=2}(s,t,u) &=& A(t,s,u)+A(u,t,s) 
	\eeqa
\end{subequations}
Projecting onto fixed angular momentum we find the partial waves
\beq\label{fdef}
f^I_\ell(s) = \frac{1}{4} \int_{-1}^1\!\!\!\! d\cos\theta P_{\ell}(\cos\theta) T^I(s,t),\;\; \ell=\begin{cases}
	\text{even},& I=0,2\\
	\text{odd}& I=1
\end{cases}
\eeq
with
\begin{equation}
	t=-\frac{(s-4)(1-\cos\theta)}{2}
\end{equation}
The partial waves \eqref{fdef} are analytic functions of $s$ with cuts for $s\in(-\infty,0)\cup(4,\infty)$. Finally, unitarity is imposed on the partial waves
\begin{equation}\label{h4}
	S_\ell^I(s^+)=1+i\pi\sqrt{\frac{s-4}{s}}f^I_\ell(s)=\eta^I_\ell(s)\, e^{2i\delta^I_{\ell}(s)},\;\;s\in \mathbb{R}_{\ge 4}
\end{equation}
as
\begin{equation}\label{uni}
	\eta^I_\ell(s)=|S_\ell^I(s^+)|\le 1,\;\;\ s\in \mathbb{R}_{>4},\;\;\forall I,\ell
\end{equation}

In addition to the scattering amplitudes, we consider the 2-particle form factors associated with gauge theory local operators $\cO^I_\ell$ with same quantum numbers $I,\ell$ that we are considering for the partial wave (see for e.g. \eqref{S0op}, \eqref{P1op}, \eqref{D0op} with quantum numbers $S0, P1, D0$). The form factors $F_\ell^I=\bra{\pi\pi} \cO_\ell^I \ket{0}$ (see more details in section \ref{FFdef}) are analytic functions of one complex variable with a cut on the real axis $(4,\infty)$.
Assuming convergence with one subtraction, the form factors can be parametrized by their imaginary part on the real axis
\begin{equation}\label{Fdisp}
	F^I_{\ell}(s)=1+\frac{1}{\pi}\int_4^{\infty}\!\!\! dx\bigg(\frac{1}{x-s}-\frac{1}{x}\bigg)\text{Im}F^I_{\ell}(x)
\end{equation}
where we use the condition $F^I_\ell(0)=1$ (otherwise we change the constant). The two point function, i.e., vacuum polarization $\Pi^I_{\ell}(s)$ of the operator $\cO^I_\ell$ itself has the same analytic structure with a cut $s\ge 4$ and the discontinuity across the cut defines the spectral density $\rho^I_{\ell}(x),x\in(4,\infty)$. See section \ref{2ptdef} for more details.

As initially discussed by Karateev, Kuhn and Penedones \cite{Karateev:2019ymz} (see also \cite{GTBPRL,GTBPRD} in this context), the amplitudes $S^I_{\ell}(s)$, the rescaled form factors $\mathcal{F}^I_{\ell}(s)$ (see section \ref{FFdef}) and the spectral densityies $\rho^I_{\ell}(s)$ obey the positivity condition
\begin{equation}\label{pos}
\begin{pmatrix}
	1 &\ S^I_\ell(s)\ &\ \ \cF^I_\ell(s) \\
	S_\ell^{I*}(s) &\ 1\ & \ \ \cF^{I*}_\ell(s) \\
	\cF^{I*}_\ell(s) &\ \cF^I_\ell(s)\ &\ \ \rho^I_\ell(s)
\end{pmatrix}\succeq 0,\;\;\; s>4,\; \; \forall \ell,I
\end{equation}
Thus, we define the variables of the gauge theory bootstrap procedure:
\begin{equation}\label{ffApara}
\{T_0,\;\; \sigma_{\a=1,2}(x),\;\; \rho_{\a=1,2}(x,y),\;\; \text{Im}F_{\ell}(x), \;\; \rho^I_{\ell}(x)\},\;\; x,y\in(4,\infty),\forall \ell,I
\end{equation} 

\bigskip

For numerical implementation, we discretize the bootstrap variables \eqref{ffApara} and use a set of $\mathbf{M}$ real variables
 \beq\label{discretevars}
 [T_0,\;\; \sigma_{\a=1,2;j},\;\; \rho_{\a=1,2;j},\;\; \text{Im}F^I_{\ell;j}, \;\; \rho^I_{\ell;i}]\in \mathbb{R}^{\mathbf{M}}
 \eeq
for some $\mathbf{M}$ usually large ($\sim 5,000$ in our numerics). The set of constraints defines a convex space in $\mathbb{R}^{\mathbf{M}}$ of scattering amplitudes, form factors and spectral densities that satisfy the most generic constraints; a space that we want to explore. We give the detailed numerical method for this implementation in section \ref{sec:nummeth}.

\subsection{Chiral symmetry breaking (low energy input)}\label{chiSB}

The low energy description of QCD is in terms of a theory of pions, the pseudo Goldstone bosons of chiral symmetry breaking.  To understand the results it is useful to set up the calculation in the context of the full chiral perturbation theory framework initiated by Gasser and Leutwyler \cite{GASSER1984142} even if in our calculation, the only input we use is the non-linear sigma model \cite{PhysRevLett.17.616} that has only terms with up to two derivatives:
\beq\label{Weinberg}
 \cL_2 =  \frac{f_\pi^2}{4}\tr[D_\mu U D^\mu U^\dagger] + \frac{f_\pi^2}{4} \tr(\chi U^\dagger+U \chi^\dagger)
\eeq
where $U(x)=e^{\frac{i\pi_a\tau_a}{f_\pi}}\in SU(2)$ parameterizes a 3-sphere, the coset space $SU(2)_L\times SU(2)_R/SU(2)_V$. Therefore, the only low energy parameters we require are the pion decay constant $f_\pi$ and the pion mass $m_\pi$ that is introduced by taking $\chi=m_\pi \mathbb{1}_{2\times 2}$. The pion decay constant $f_\pi$ is defined in terms of the matrix element of the axial current 
\beq\label{axialc}
\bra{0} j^{I,\mu}_5(x) \ket{\pi^{I'}(p)} = -i   p^\mu f_\pi\,\delta^{II'}\ e^{-ipx}
\eeq 
and can therefore be determined independently from weak pion decay. The Lagrangian \eqref{Weinberg} includes external gauge fields to identify the scalar, vector and axial currents. At the next order in the expansion in derivatives and pion mass new terms appear:   
\beqa\label{L4}
 \cL_4 &=& \frac{l_1}{4}\left\{\tr[D_\mu U (D^\mu U)^\dagger]\right\}^2 + \frac{l_2}{4}\tr[D_\mu U(D_\nu U)^\dagger]\tr[D^\mu U(D^\nu U)^\dagger] \nonumber\\
 && +\frac{l_3}{16}[\tr(\chi U^\dagger+U\chi^\dagger)]^2 +\frac{l_4}{4}\tr[D_\mu U(D^\mu\chi)^\dagger+D_\mu\chi(D^\mu U)^\dagger] \nonumber\\
 && +l_5\left[\tr(f_{\mu\nu}^R U f_L^{\mu\nu} U^\dagger) -\half \tr(f_{\mu\nu}^L f^{\mu\nu}_L +f^R_{\mu\nu}f^{\mu\nu}_R) \right] \nonumber\\
 && +i\frac{l_6}{2}\tr[f_{\mu\nu}^R D^\mu U (D^\nu U)^\dagger + f_{\mu\nu}^L (D^\mu U)^\dagger D^\nu U ] \nonumber\\
 && -\frac{l_7}{16} [\tr(\chi U^\dagger - U \chi^\dagger)]^2 
\eeqa
as described in \cite{GASSER1984142,chiPTintro} that the reader can consult for a more detailed description. Let us just note that we omitted contact terms that do not depend on $U$ and that $f_{\mu\nu}^{R,L}$ are the field strengths associated with the external spectator gauge fields (sources). 
In the usual approach the couplings $l_{1\ldots 7}$ are determined from experiment, for example $l_{1,2}$ are related to the scattering lengths of the $D$-waves and $l_{4,6}$ to the scalar and vector charge radius of the pion. More recently, positivity methods have been used to put generic bounds on those parameters (see \eg \cite{Albert:2022oes}). On the other hand, in our gauge theory bootstrap approach, these parameters are \emph{calculable}. In this paper we determine $\bar{l}_{1,2,4,6}$, the renormalized values of the couplings ($l_j$ in $\cL_4$ are divergent bare couplings).

 Being able to compute (approximately) the low energy couplings is already quite interesting progress. However, the EFT becomes strongly coupled as we increase the energy and predictions become unreliable long before we reach the asymptotically free regime. For example the $\rho$ resonance in the $P1$ channel will not appear from the EFT. In fact, the main purpose of this and the previous paper \cite{GTBPRL,GTBPRD} is to find the phase shifts in that region by using gauge theory input.
 
 We want to emphasize that, from the point of view of pion physics, the bootstrap calculation is a \emph{first principles} calculation of pion scattering. Usually what that means is that we compute pion scattering starting from a given Lagrangian or Hamiltonian. In this case the Lagrangian is $\cL_2$ that contains the pion mass $m_\pi$ and coupling $f_\pi$. Using this Lagrangian, however, is not sufficient since it has to be supplemented by an infinite number of higher derivative terms with unknown couplings that have to be measured. Instead of doing that, we use the information that pions describe the low-energy dynamics of a gauge theory (QCD). In that way we do not require any extra experimental input for the computation.  
 
Now let's see how we incorporate the Lagrangian $\cL_2$ in the bootstrap. At tree level $\cL_2$ determines the function $A(s,t,u)$ in \eqref{Adef} to be    
\beq\label{h5}
A_{\text{tree}}(s,t,u) =  \frac{s-m_{\pi}^2}{8\pi^2\,f_\pi^2}
\eeq
 This was already derived by Weinberg \cite{PhysRevLett.17.616} and later confirmed by experiment even before the advent of QCD \cite{TheBook}. In our case we want to consider amplitudes that satisfy the generic constraints of the previous subsection and that match this low energy behavior. It turns out to be more convenient to put the constraints in terms of the partial waves that we want to compute. We also expect this tree level computation (linearized approximation) to be better at small values of $s$, in particular in the unphysical region $0<s<4$ where the partial waves are real. The only non-zero tree level partial waves are  (see \eg\ \cite{donoghue_golowich_holstein_2014}):
\begin{equation}\label{chiralratio}
f^0_{0,\text{tree}}(s)=\frac{2}{\pi}\frac{2s-m_{\pi}^2}{32\pi f_{\pi}^2},\;\; f^1_{1,\text{tree}}(s)=\frac{2}{\pi}\frac{s-4m_{\pi}^2}{96\pi f_{\pi}^2},\;\; f^2_{0,\text{tree}}(s)=\frac{2}{\pi}\frac{2m_{\pi}^2-s}{32\pi f_{\pi}^2}
\end{equation}

\bigskip

To implement this low energy match, we choose a small set of points $s_j$ in the very low energy unphysical region $0<s\le 2$, and require that
\begin{equation}\label{chiralconstraints} 
\begin{aligned}
||f^0_0(s_j)-R^{\text{tree}}_{01}(s_j)f^1_1(s_j)||&\le \epsilon^{\chi},\\
||f^2_0(s_j)-R^{\text{tree}}_{21}(s_j)f^1_1(s_j)||&\le \epsilon^{\chi},\\
R^{\text{tree}}_{01}=\frac{f^0_{0,\text{tree}}}{f^1_{1,\text{tree}}},\;\;R^{\text{tree}}_{21}=&\frac{f^2_{0,\text{tree}}}{f^1_{1,\text{tree}}}
\end{aligned}
\end{equation}
with some norm and a tolerance $\epsilon^\chi$. Note that the expressions \ref{chiralconstraints} use only the ratios of the tree level partial waves which eliminate the $f_{\pi}$ so the matching applies to arbitrary $f_{\pi}$. For each given tolerance this drastically reduces the convex space of amplitudes inside $\mathbb{R}^{\mathbf{M}}$ defined by the generic constraints in the previous step. Finally, we use the physical value of $f_{\pi}=92$ MeV to choose an appropriate tolerance $\epsilon^{\chi}$ for our computation. The precise description of how to implement these constraints is given in section \ref{sec:nummeth}.

\bigskip
An important part of the method is to include form factors and spectral densities in the bootstrap in order to include information from high energy perturbative QCD computations. For that reason we also have to understand their low energy behavior from the pion lagrangian. The main results that we need can be found in \cite{GASSER1984142} where it is important to notice that they are kinematical, namely they follow from a free theory of pions. The calculation starts by identifying the pion operators corresponding to the currents we use:
\beq
\begin{aligned}
	S          &= m_q\bar{\psi}_A \psi_A =\ \frac{m^2_{\pi}}{2}\pi^a \pi^a + \cO(\pi^4) \\
V_\mu^a    &=\bar{\psi}_A \gamma_\mu \frac{\tau_{AB}^a}{2} \psi_B =\ \epsilon^{abc} \pi^b \partial_\mu \pi^c + \cO(\pi^4) \\
T_{\mu\nu}&= T^g_{\mu\nu} + T^f_{\mu\nu} \\
&= \partial_\mu \pi^a \partial_\nu \pi^a-\half(\partial_{\alpha} \pi^a \partial^{\alpha} \pi^a - \mpi^2 \pi^a\pi^a)\eta_{\mu\nu}  + \cO(\pi^4) 
\end{aligned}
\eeq
with the gluonic and fermionic parts of the energy momentum begin
\beqa
 T^g _{\mu\nu} &=& -F^a_{\mu\alpha}F^{a\,\alpha}{}_{\nu} +F^a_{\alpha\beta}F^{a\,\alpha\beta}\, \eta_{\mu\nu} \\
 T^f_{\mu\nu} &=& \frac{i}{2}\bar{\psi}_A(\gamma_\mu D_\nu+\gamma_\nu D_\mu)\psi_A - \bar{\psi}_A(i\slashed{D}-m_q)  \psi_A\, \eta_{\mu\nu} 
\eeqa
The two point functions of such operators give for $\rho^I_{\ell}$: 
\begin{subequations}\label{rholowenergy}
\beqa
\rho^0_0(s) &\simeq& \frac{m^4_{\pi}}{(2\pi)^4}\, \frac{3}{16\pi}\,\left(1-\frac{4}{s}\right)^{\frac{1}{2}} \\ 
\rho^1_1(s) &\simeq& \frac{1}{(2\pi)^4}\,\frac{s}{24\pi}\, \left(1-\frac{4}{s}\right)^{\frac{3}{2}} \\ 
\rho^0_2(s) &\simeq&  \frac{1}{(2\pi)^4}\,\frac{s^2}{160\pi}\left(1-\frac{4}{s}\right)^{\frac{5}{2}}
\eeqa
\end{subequations}
that we use to constrain the spectral densities numerically.\footnote{The spectral densities can be parameterized in a way that incorporates these low energy behavior. This results in a small space of solutions as shown in \cite{He:2025gws}.} It is also important to notice that below the four pion threshold and for operators that create an even number of pions, only two pion states can be created leading to the relation $\rho_\ell(s) = |\cF_\ell(s)|^2$, $4\le s\le 16$.

\subsection{Sum rules and form factor asymptotics (QCD input)}\label{SVZasympt}

Even after significantly reducing the allowed space using low energy information, there is still a large number of allowed S-matrices, each of which may (or may not) correspond to (different) consistent UV completions. For our purpose of determining the strongly coupled dynamics of the gauge theory, we need to incorporate high energy information, and this is done by constraining spectral densities and form factors variables in \eqref{discretevars}. In order to do that we choose a scale $s_0$ where we assume the theory is properly described at least approximately\footnote{Since going to very large energies is challenging numerically due to the larger number of interpolation points required, we make a compromise in choosing $s_0$.} by perturbative QCD. In our previous paper we took $\sqrt{s_0}=1.2\,\GeV$ and here we take $\sqrt{s_0}=2\, \GeV$. An important test of the method is that, below $1.2\,\GeV$ the phase shifts agree with the previously computed ones. If $\sqrt{s_0}$ were setting the energy scale of all features (\eg\ resonances), a factor of almost 2 in $\sqrt{s_0}$ should be quite noticeable. 

Having chosen $s_0$, we impose constraints on the spectral density given by a finite energy version of the Shifman, Vainshtein, Zakharov (SVZ) sum rules \cite{SHIFMAN1979385} and constraints on the asymptotic behavior of the form factors as given by Brodsky and Lepage \cite{osti_1447331}. Notice that this asymptotic behavior of the form factors should be reached at very high energies and therefore we use it for initial guidance as described in the numerical section \ref{sec:nummeth}. 
Below, we give a brief summary of the form factors and spectral densities with quantum numbers $S0$ (scalar current), $P1$ (vector current) and $D0$ (energy momentum tensor) we use in this paper. The relevant quantities are properly defined in more detail later in sections \ref{FFdef} and \ref{2ptdef}. Here we simply present the definitions of the current $j^I_{\ell}(x)$, their two-particle form factor $F^I_{\ell}(s)$, the form factor $\cF^I_{\ell}$ for states with definite angular momentum and isospin, and the low energy kinematic behavior of $\rho^I_{\ell}$.  We also present the large $s$ (short distance) expansion of the two-point function $\Pi^I_{\ell}(s)$ leading to the SVZ sum rules on $\rho^I_{\ell}$ as well as the asymptotic form of the form factors, all for the case of $N_c=3$, $N_f=2$. The detailed computations are given in sections \ref{sec:FFl} and \ref{sec:jj}.\footnote{We were not able to find in the literature the computation of the scalar form factor so we include a crude estimate (up to numerical factors).}
{\allowdisplaybreaks
\begin{enumerate}
	\item[\tcbox{$S0$}]
	\begin{subequations}
	\beqa
	j^0_0 &=&  m_q(\bar{u}u+\bar{d}d)\label{S0op}\\
	\bra{\pi^+(p_1) \pi^{-}(p_2)} j^0_0\ket{0} &=& F_0^0(s) \label{scalarF}\\
	F_0^0(0) &=& 1 \\
	\bra{I=0,P\ell\sigma} j^0_0\ket{0} &=& \delta_{\ell 0}\delta_{\sigma 0} \cF_0^0(s) \\
	\cF_0^0(s) &=& \frac{\sqrt{6\pi}}{16\pi^3}\frac{1}{s^{\frac{1}{4}}}\left(\frac{s-4}{4}\right)^{\frac{1}{4}} F_0^0(s) \label{FFS0scale}\\
	\rho^0_0(s) &\simeq& \frac{1}{(2\pi)^4}\, \frac{3}{16\pi}\,\left(1-\frac{4}{s}\right)^{\frac{1}{2}}\ \ \ \mbox{[low energy]}\label{S0lr}\\
	\Pi^0_0(s) &\simeq& - \frac{N_f m_q^2}{(2\pi)^4}  \frac{3}{8\pi^2}\left(1+\frac{13}{3}\frac{\alpha_s}{\pi}\right)s\ln\left(-\frac{s}{\mu^2}\right)  \label{S02pt}\\
	\frac{1}{s_0^{n+2}}\int_4^{s_0} \rho^0_0(x) x^n dx &\simeq&  \frac{N_f m_q^2}{(2\pi)^4} \frac{3}{4\pi(n+2)}\left(1+\frac{13}{3}\frac{\alpha_s}{\pi}\right),\ n\ge 0 \label{S0SR} \\
	|F^0_0(s)| &\sim &   4\pi\alpha_s  f_\pi^2\frac{m_q^2}{s}\ln\left(\frac{s}{m_\pi^2}\right), \ \ (s\rightarrow\infty) \label{S0asymp}
	\eeqa
	\end{subequations}	
	\rule{\textwidth}{1.pt}
	\item[\tcbox{$P1$}]
	\begin{subequations}
	\beqa
	j^1_{1} &=&  \half(\bar{u}\gamma_+ u -\bar{d} \gamma_+ d)\label{P1op}\\
	\bra{\pi^+(p_1) \pi^{-}(p_2)} j^1_{1,\mu}\ket{0} &=& (p_{2,\mu}-p_{1,\mu}) F_1^1(s) \\
	F_1^1(0) &=& 1 \\
	\bra{I=1,P\ell\sigma} j^1_{1}\ket{0} &=& \delta_{\ell 1}\delta_{\sigma 1} \cF_1^1(s) \\
	\cF_1^1(s) &=& \sqrt{\frac{4\pi}{3}} \frac{1}{8\pi^3} \frac{1}{s^{\frac{1}{4}}}\left(\frac{s-4}{4}\right)^{\frac{3}{4}} F_1^1(s) \label{FFP1scale}\\
	\rho^1_1(s) &\simeq& \frac{1}{(2\pi)^4}\,\frac{s}{24\pi}\, \left(1-\frac{4}{s}\right)^{\frac{3}{2}}\ \ \ \mbox{[low energy]}\label{P1lr}\\
	\Pi^1_1(s) &\simeq& -\frac{1}{(2\pi)^4}  \frac{1}{8\pi^2}\left(1+\frac{\alpha_s}{\pi}\right)s\ln\left(-\frac{s}{\mu^2}\right) \label{P12pt}\\ 
	\frac{1}{s_0^{n+2}}\int_4^{s_0} \rho^1_1(x) x^n dx &\simeq&  \frac{1}{(2\pi)^4} \frac{1}{4\pi(n+2)}\left(1+\frac{\alpha_s}{\pi}\right),\ \ n\ge -1 \label{P1SR}\\
	|F^1_1(s)| &\simeq &   16\pi\frac{\alpha_s  f_\pi^2}{s}, \ \ (s\rightarrow\infty) \label{P1asymp}
	\eeqa
		\end{subequations}	
	\rule{\textwidth}{1.pt}
	\item[\tcbox{$D0$}]
	\begin{subequations}
	\beqa
	j^0_2 &=& T^{++}(0) \label{D0op}\\
	\bra{\pi^+(p_1) \pi^{-}(p_2)} j^0_2\ket{0} &=&  4 |\vec{p}_1|^2  \sqrt{\frac{2\pi}{15}}\, Y_{22}(\hat{p}_1)\ F^0_2(s) \\
	F^0_2(s) &=& 1 \\
	\bra{I=0,P\ell\sigma} j^0_2\ket{0} &=& \delta_{\ell 2}\delta_{\sigma 2} \cF^0_2(s) \\
	\cF^0_2(s) &=& \frac{1}{4\pi^3}\sqrt{\frac{\pi}{5}}\frac{1}{s^{\frac{1}{4}}}\left(\frac{s-4}{4}\right)^{\frac{5}{4}}\ F^0_2(s) \label{FFD0scale}\\
	\rho^0_2(s) &\simeq& \frac{1}{(2\pi)^4}\,\frac{s^2}{160\pi}\left(1-\frac{4}{s}\right)^{\frac{5}{2}}\ \ \ \mbox{[low energy]}\label{D0lr}\\
	\Pi^0_2(s) &\simeq&  - \frac{1}{(2\pi)^4} \frac{1}{8\pi^2}\left(\frac{11}{10}-\frac{17}{18}\frac{\alpha_s}{\pi}\right)\,s^2\, \ln\left(-\frac{s}{\mu^2}\right) \label{D02pt}\\
	\frac{1}{s_0^{n+3}} \int_4^{s_0}\!\rho^0_2(x) x^n dx&\simeq&\frac{1}{(2\pi)^4}\frac{1}{4\pi}\frac{1}{n+3}\left(\frac{11}{10}-\frac{17}{18}\frac{\alpha_s}{\pi}\right), \ n\ge -2 \label{D0SR}\\
	| F^0_2(s) | &\simeq& \frac{48\pi\alpha_s  f_\pi^2}{s},\; (s\rightarrow\infty)  \label{D0asymp}
	\eeqa
	\end{subequations}	
	\rule{\textwidth}{1.pt}
\end{enumerate}
}
In these formulas, $+$ indicates the component of a vector $v_+ = \frac{1}{\sqrt{2}}(v_1-i v_2)$, also $P=p_1+p_2$ and $s=P^2$. Equivalently, as we do later, we can contract all currents with a vector $\Delta=\frac{1}{\sqrt{2}}(0,1,i,0)$.
\bigskip

We implement  the SVZ sum rules \eqref{S0SR}, \eqref{P1SR}, \eqref{D0SR} and the low energy kinematic behavior \eqref{S0lr}, \eqref{P1lr}, \eqref{D0lr} as constraints on the spectral densities. It is worth noting that the SVZ expansion of the two point functions contains expectation values (condensates) of higher order operators that we do not include in \eqref{S02pt}, \eqref{P12pt}, \eqref{D02pt} because they are suppressed by the large energy scale $s_0$. We comment on this further in section \ref{condensate}. The precise implementation of the sum rules is described in section \ref{sec:nummeth}.

 For the numerical computation to work we have to take into account that the form factors vanish at infinity as predicted by QCD. In practice, we require that the form factors above the energy scale $s_0$ are small, with an initial estimate of their value following from the asymptotic forms \eqref{S0asymp}, \eqref{P1asymp} and \eqref{D0asymp}. More details of this are given in section \ref{sec:nummeth}.

\subsection{Allowed space and extremal amplitudes}\label{shapesection}

\begin{figure}[h]
	\centering
	\includegraphics[width=0.9\textwidth]{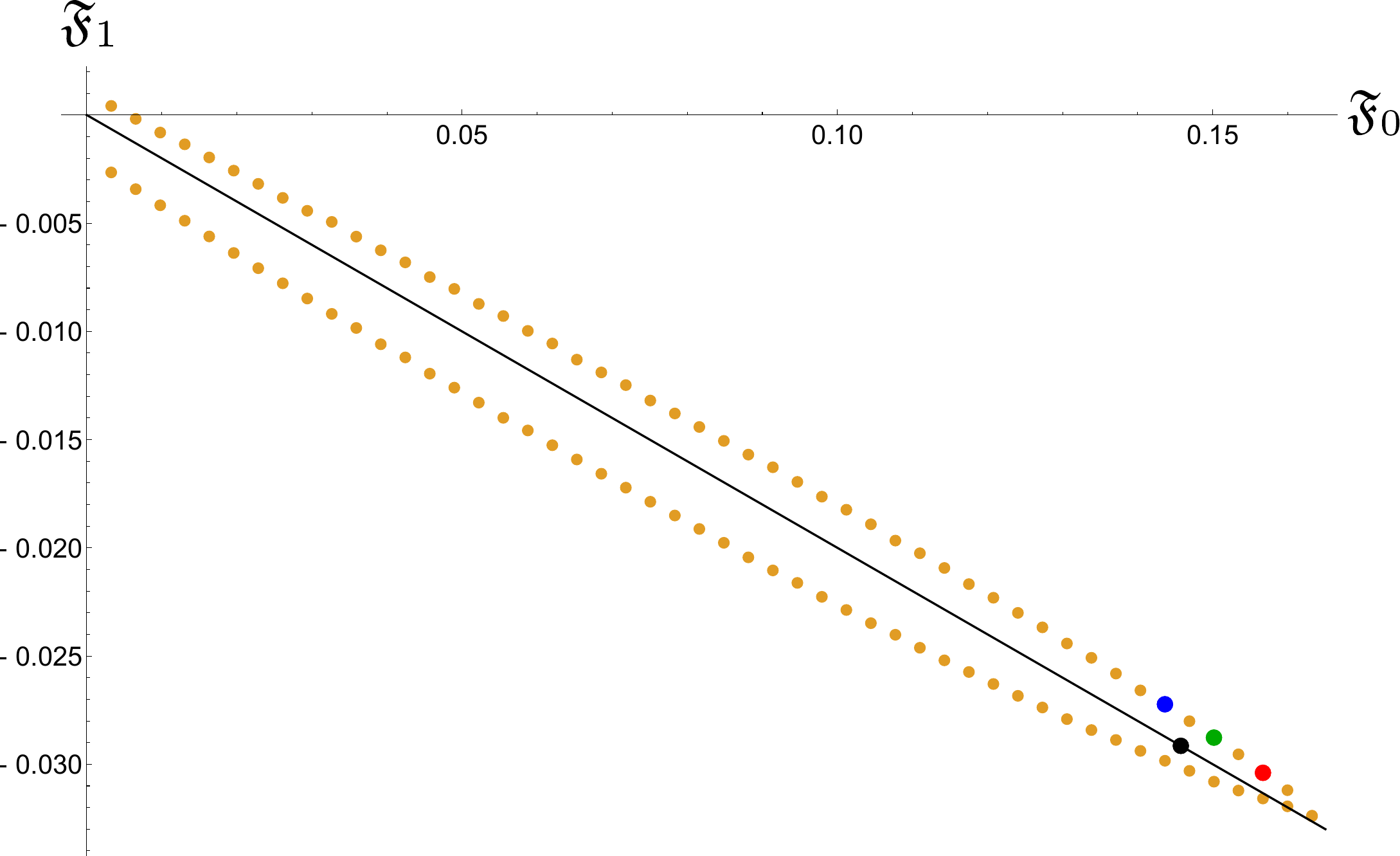}
	\caption{The projected space of amplitudes characterized by two linear functionals \eqref{fnals} in the gauge theory bootstrap. The black line indicates the two functionals evaluated using tree level amplitudes whereas the black dot is for the physical value $f_{\pi}=92\,\MeV$, the initial approximation we start with. We take three points (blue, green and red) on the boundary of the allowed space closest to the black point.}
	\label{shape}
\end{figure}

We started by discretizing the variables and parameterizing the amplitudes/form factors by a point in $\mathbb{R}^{\mathbf{M}}$ for some large $\mathbf{M}\sim 5,000$. After imposing many constraints, the dimension and volume of such space is considerably reduced. At this point we hope that the set of remaining points describe similar S-matrices but how do we focus on one? A common practice of the bootstrap is to project the space onto a lower dimensional subspace \cite{Cordova:2019lot}, for example a plane. In that subspace we plot the ``shadow" or projection of the higher dimensional space. Intuitively, the edges or boundaries of the shadow will lift to unique points on the boundary of the higher dimensional space. Such a point corresponds to a set of bootstrap variables that describe a particular extremal amplitudes/form factors living on the boundary of the space. Finding the boundary of the region can be done by maximizing linear functionals of the bootstrap variables. While this idea was used extensively in the pure S-matrix bootstrap to put generic bounds on certain physical quantities, here we do not consider these linear functionals to carry specific physical meanings, but we use them simply as a way to access scattering amplitudes/form factors on the boundary of the allowed space under constraints. The naive belief is that given enough physical conditions from low energy and high energy, there is eventually a unique (within errors) physical amplitude that matches these physical requirement as well as bootstrap consistency conditions. Although, ideally, the specific functional used to find this amplitude should be irrelevant, in practice we have to choose functionals that include relevant physical information and lead to a well defined numerical problem.  
In this work, we consider two linear functionals (related to the forward amplitudes): \footnote{This is a slight modification of the functionals $f^0_0(s=3),f^1_1(s=3)$ we used in previous work \cite{GTBPRL,GTBPRD} to include some contributions from the $D$ and $F$ partial waves.}
\begin{equation}\label{fnals}
	\mathfrak{F}_0=2\big(f^0_0(s=3)+5f^0_2(s=3)\big),\;\;\; \mathfrak{F}_1=2\big(3f^1_1(s=3)+7f^1_3(s=3)\big)
\end{equation}
Preliminary tests on other similar functionals give similar results and we hope to report on a full exploration of this aspect in the future. The space of amplitudes projected out on the two dimensional space characterized by \eqref{fnals} is given in figure \ref{shape}. The region inside the yellow shape is allowed by the constraints but no specific amplitude is associated with the interior points (there are many for each point).    
On the other hand, each yellow dot at the boundary \emph{is} associated with a particular extremal amplitude/form factor and therefore we choose among them the one that we expect to represent the correct amplitude. In order to do that we use the physical value $f_\pi=92\,\MeV$ as follows: The black line indicates the values of $\mathfrak{F}_0,\mathfrak{F}_1$ for the Weinberg model or $\chi$PT tree level amplitudes \eqref{chiralratio} evaluated for different values of $f_{\pi}$. The black dot corresponds to $f_{\pi}=92\,\MeV$, namely the initial approximation we start with. Using this input it is then natural to focus on a range of points closest to the black dot (the blue, green and red points) and take the corresponding amplitudes as the result of our computation where the span of the three points give a rough estimate of our computational errors. It is worth pointing out that the points at the lower edge could be alternative candidates but, following the previous paper \cite{GTBPRL,GTBPRD} we concentrate on the upper edge. The rest of the results are for the three points, or just the red and the blue. Since at this stage we cannot identify which exact point corresponds to QCD we use this spread as an estimate of the error until a better identification procedure is found. 


\subsection{Watsonian unitarization}\label{Wunit}

As discussed in \cite{GTBPRL,GTBPRD}, an important constraint in the Gauge Theory Bootstrap is the positivity of the matrix \eqref{pos}. When maximizing a linear functional, the maximum is obtained at the boundary of the allowed space where one expects the constraints to be saturated. In this case it means that the matrices \eqref{pos} have two zero eigenvalues implying the conditions:
\begin{equation}\label{W}
	|S_\ell^I(s)|=1, \ \ S^I_{\ell}(s)=\frac{\cF^I_{\ell}(s)}{\cF^{*I}_{\ell}(s)}, \ \ \rho_\ell^I(s) = |\cF^I_\ell(s)|^2 \ \ \ \ \mathrm{(saturation)}
\end{equation}
This implies Watson's theorem, namely that the phase-shift and the phase of the form factor are equal: $\delta_\ell^I(s)=\alpha_\ell^I(s)$. Here $S_\ell^I(s) = e^{2i\delta_\ell^I(s)}$ and $\cF_\ell^I(s) = |\cF_\ell^I(s)| e^{i\alpha_\ell^I(s)}$. Pictorially one can think that the SVZ sum rules and asymptotic values of the form factors obtained from the UV theory (QCD) constrain the modulus of $\cF_\ell^I(s)$ and, through analyticity its phase. Watson's theorem conveys this information to the phase shift thus constraining the partial waves. Physically these saturation conditions are exact below the four pion threshold for $4\le s\le 16$. At larger energies our numerical procedure tries to get as close as possible to saturation with the underlying assumption that particle creation is not important until at least $1\GeV$.     

Numerically, after the maximum is found, it is convenient to do a further improving step by choosing a functional constructed schematically as follows: If the variables of the problem are $x_j$ and the constraints we want to improve are  $\Phi_A(x_j)\le 0$, we maximize 
\beq 
 \cF_P(x_j) = \sum_{A,j} \left. \frac{\partial \Phi_A(x_j)}{\partial x_j}\right|_{\mathrm{old}} x_j
\eeq
That is, we find the maximum in the direction of an average of the gradient of the constraints evaluated on the old solution.
Such maximization increases the value of $\Phi_A(x)$ and gets closer to the boundary. This step can be repeated several times as was shown in \cite{He:2018uxa} in the cased of 2D theories where very good results were obtained. The main constraints we want to improve are $|S^I_\ell(s)|-1\le 0$. Thus, after the first maximization step, we construct a new functional
\beqa
 \cF_P &=& \sum_{\ell,I}\int_4^\infty \!\! ds\ \Re \left[(S_\ell^I(s))^* \frac{\delta S_\ell^I(s)}{\delta x_j} \right]_\mathrm{old} \ x_j \\
   &=&  \sum_{\ell,I}\int_4^\infty \!\! ds\ \Re \left[(\left. S_\ell^I(s))^*\right|_\mathrm{old} (S_\ell^I(s)-1)\right]
\eeqa
where we used that the parameterization of the partial waves is linear in the variables $x_j$. Alternatively, in this paper we introduce a similar but new type of unitarization where we replace the ``old" S-matrix by the phase computed from the form factor:   
\beq
  \cF_W = \sum_{\ell,I}\int_4^\infty \!\! ds\ \Re \left[\left. e^{-2i\alpha_\ell^I(s)}\right|_\mathrm{old} (S_\ell^I(s)-1)\right]
\eeq
 Since we do not compute the form factor for all partial waves, whenever $\alpha_\ell^I(s)$ is not available from the previous (old) solution, then we just use the phase-shift. This unitarization we call {\it Watsonian} unitarization because it leads not only to saturation of unitarity but also it is maximized when the phase shift agrees with the phase of the form factor. If needed, this Watsonian unitarization can be repeated until convergence.\footnote{When iterated, the Watsonian unitarization converges to a certain GTB solution, see \cite{He:2025gws}. The converged solution is largely independent of the functional chosen to find the initial solution.} Since the high energy information is introduced through the form factors and spectral densities the Watsonian unitarization is, at least qualitatively, a way to transfer that information into the partial waves.  

Note that, experimentally, elastic pion scattering saturates unitarity up to energies of order $1.2\,\GeV$. Although it is not completely clear why this is so, it agrees very nicely with the discussion in the last paragraph justifying the last unitarization step.  

\subsection{Comment on the condensates}\label{condensate}

One important point of the original SVZ proposal was the existence of the quark condensate $m_q\bra{0} \bar{q} q\ket{0}$ and gluon condensate or equivalently expectation value of the trace of the energy momentum tensor $\bra{0} T^\mu_\mu \ket{0}$ that contribute to the sum rules. This was taken into account by using the OPE of the currents:
\beq
T\{j(x) j(0)\} = C_{\mathbb{1}}(x)\ \mathbb{1}  + \sum_\cO C_{\cO}(x)\ \cO(0)   \label{h28}
\eeq   
and realizing that, when taking expectation value in the symmetry broken vacuum, other operators contribute. Thus, the SVZ expansion is
\beq\label{svzexp}
\bra{0} T\{j(x) j(0)\} \ket{0} = C_{\mathbb{1}}(x)  +  C_{\bar{q} q}(x)\ \bra{0}j_S(0)\ket{0} + C_{G^2}(x)\ \bra{0} \frac{\alpha_s}{\pi} G_{\mu\nu}^a G^{a\,\mu\nu} \ket{0} + \ldots
\eeq   
where $j_S=m_q(\bar{u}u+\bar{d}d)$. The expectation value $\bra{0} \frac{\alpha_s}{\pi} G_{\mu\nu}^a G^{a\,\mu\nu} \ket{0}$ can be written in terms of the quark condensate and the trace of $T_{\mu\nu}$. These condensates play an important role in the SVZ approach similar to the low energy EFT coefficients $l_j$ do in chiral perturbation theory, they have to be determined from experiment (or lattice). However, in the finite energy sum rules that we used in the bootstrap their contribution is suppressed by inverse powers of the QCD matching scale $s_0$ and we ignore them in the computational setup. The idea is the same as for $l_j$, namely that fitting the high energy to the bootstrap we should find that corrections are required and thus the condensates should be \emph{calculable} in the bootstrap. At the moment our precision is not enough for that so we leave this interesting topic for future work. Notice that this is an improvement on paper \cite{GTBPRL,GTBPRD} where we used condensate values as parameters taken from the literature (although they did not affect the numerics).

\section{Phase shifts for $I=0,1,2$ and $\ell\le 3$ up to $2\ \GeV$}\label{sec:phaseshifts}

\begin{figure}[H]
 	\centering
 	\includegraphics[width=\textwidth]{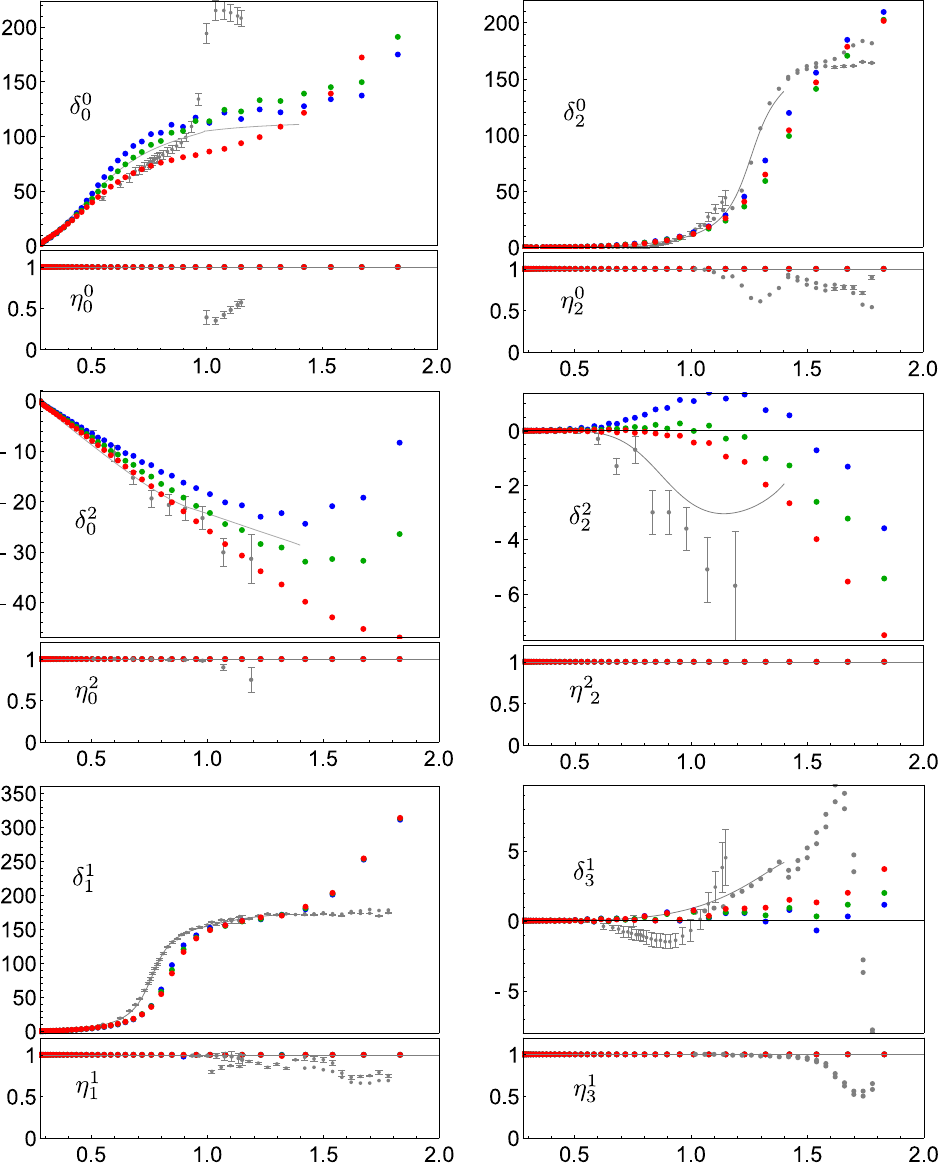}
 	\caption{Phase shifts $\delta^I_{\ell}$ and inelasticity $\eta^I_{\ell}$ from GTB comparing with experiments\cite{Protopopescu:1973sh,LOSTY1974185,Hyams:1975mc} and phenomenological fit \cite{Pelaez:2004vs}. Horizontal axis is $\sqrt{s}$ in $\GeV$. See text.}
 	\label{6pw}
 \end{figure}

Now we apply the procedure to QCD. In this case we consider and $SU(3)$ gauge theory with two quarks of the same mass much lower than the QCD scale. We extend the energy range from $1.2\,\GeV$ in our previous paper \cite{GTBPRL,GTBPRD} to $2\,\GeV$ in this paper. Keep in mind that at $2\,\GeV$, our input from perturbative QCD for the form factors and spectral densities should be more precise. As described in section \ref{shapesection}, we take three points (blue, green and red) in figure \ref{shape} spanning around the black dot that indicates the Weinberg model (tree level $\chi$PT) and examine the corresponding six partial waves $S0$, $D0$, $S2$, $D2$, $P1$, and $F1$. The results for the phase shifts $\delta^I_{\ell}$ and inelasticity $\eta^I_{\ell}$ are plotted in fig.\ref{6pw}, where the blue, green and red points correspond to the points in fig. \ref{shape} of the same color. We also plot together the experimental data (gray dots) from \cite{Protopopescu:1973sh,LOSTY1974185,Hyams:1975mc} and phenomenological fits (gray lines) from \cite{Pelaez:2004vs} for comparison. In this figure, the horizontal axis indicate the center of mass energy $\sqrt{s}$ in GeV.

In addition to the reasonable overall agreement with experiments, it is worth noting several points. First, the experimental data in the $S0$ channel includes Kaon production leading to a rise of the phase shift around $1\,\GeV$ which we do not expect to reproduce as we are not considering the strange quark. Therefore, the phenomenological fit plot for $S0$ is done using the parametrization \cite{Pelaez:2004vs} with the Kaon production removed. Second, note that although the energy range was extended from $1.2\,\GeV$ to $2\,\GeV$, the SVZ sum rules and the form factor asymptotics still determine the $\rho(770)$ meson to be near its physical value. This is further confirmation that our framework does not rely on the $s_0$ scale we choose. With this extension to $2\,\GeV$, we now see a clear $f_2(1270)$ meson resonance in the $D0$ channel. Moreover, the $P1$ channel contains another phase rise at $\simeq 1.6\,\GeV$ (the phase shift values are plotted modulo $\pi$) corresponding to another resonance which should be identified with $\rho(1450)$. See section \ref{1450} below for more details.
 
\subsection{Forward amplitudes}

\begin{figure}[H]
	\centering
	\begin{subfigure}[b]{0.49\textwidth}
		\raggedright
		\includegraphics[width=\textwidth]{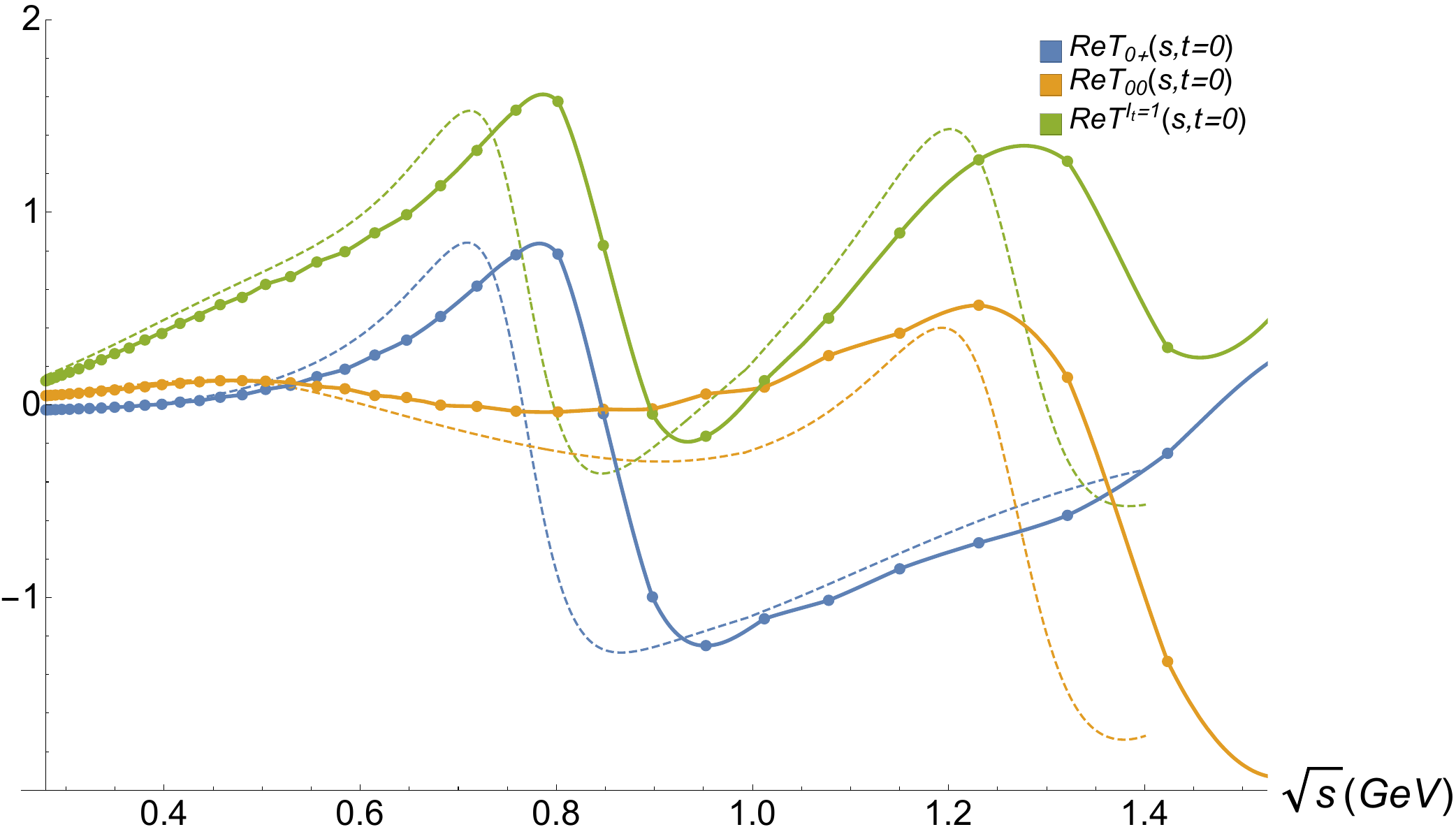}
		\caption{Real part}
	\end{subfigure}
	\begin{subfigure}[b]{0.49\textwidth}
		\centering
		\includegraphics[width=\textwidth]{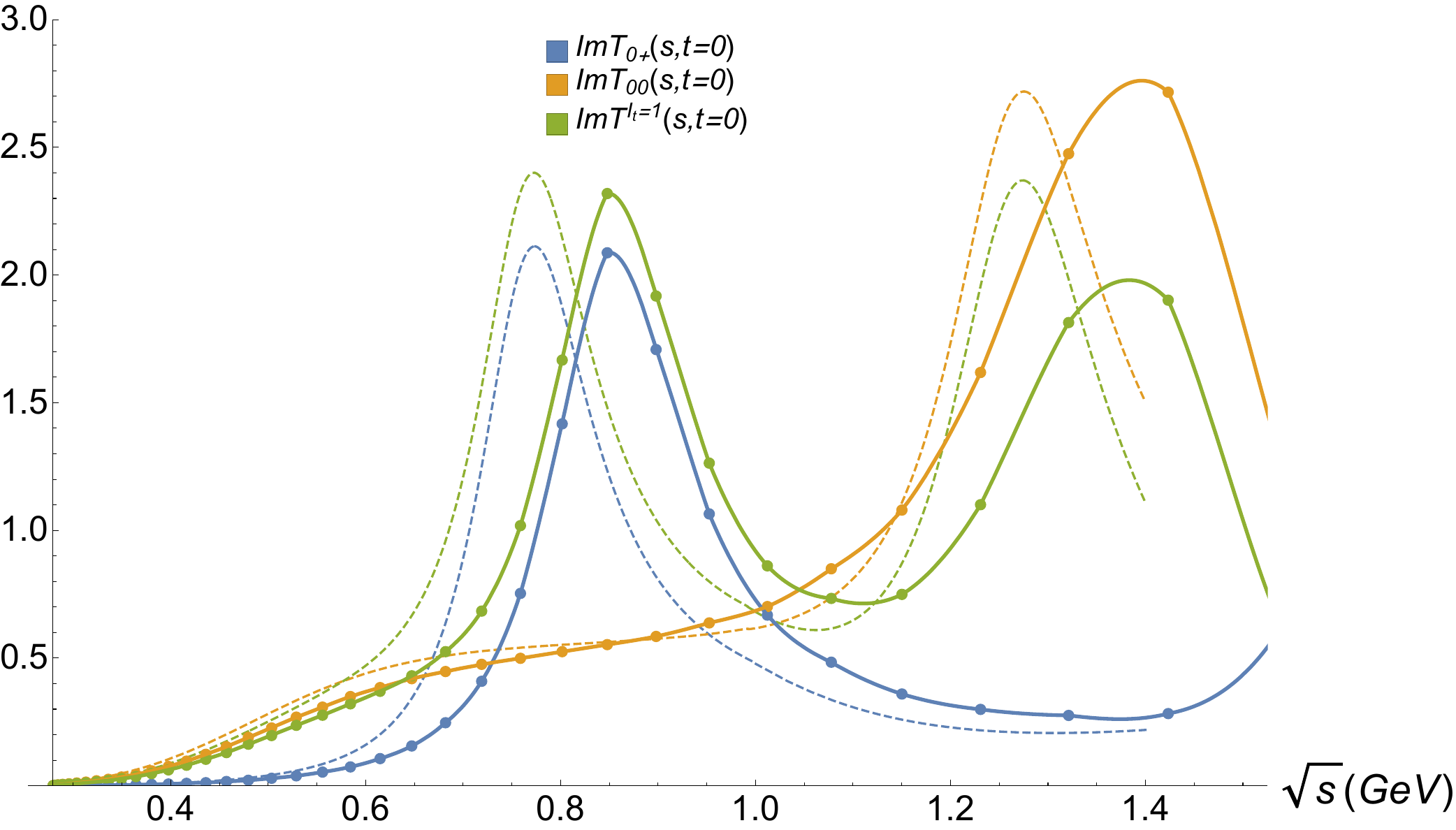}
		\caption{Imaginary part}
	\end{subfigure}
	\caption{Real and Imaginary parts of the forward amplitudes corresponding to the red point in fig.\ref{shape} (points and solid interpolating lines) and the phenomenological analysis of Pelaez and Yndurain (dashed)\cite{Pelaez:2004vs} with Kaon production removed.}
\label{forward}
\end{figure}

The forward amplitudes (zero scattering angle or $t=0$) combine all partial waves and presumably display Regge behavior at larger energies than the ones considered here. In fig. \ref{forward} we plot the amplitudes $T_{0+}(s,0)$, $T_{00}(s,0)$, $T^{I_t=1}(s,0)$ that are $s\leftrightarrow u$ symmetric (or anti--symmetric). In terms of the fixed isospin amplitudes they read 
\begin{subequations}
	\begin{eqnarray}
	T_{0+}\equiv T(\pi^0\pi^+\to\pi^0\pi^+) &=& \frac{1}{2}T^{I_s=1}+\frac{1}{2}T^{I_s=2}\\
	T_{00}\equiv T(\pi^0\pi^0\to\pi^0\pi^0) &=& \frac{1}{3}T^{I_s=0}+\frac{2}{3}T^{I_s=2}\\ 
	T^{I_t=1} &=&\frac{1}{3}T^{I_s=0}+\frac{1}{2}T^{I_s=1}-\frac{5}{6}T^{I_s=2}
	\end{eqnarray}
\end{subequations}


\section{Low energy parameters}\label{sec:parameters}

In the previous section we showed that we find good agreement between the computed phase shifts and the experimental ones. Thus we might expect that all quantities agree. Although this way of thinking is generically correct, low energy parameters like scattering lengths $a^I_{\ell}$, effective range parameters $b^I_{\ell}$ and higher order chiral Lagrangian coefficients $\bar{l}_j$ have sensitive dependence on the detail structure of the amplitudes near threshold and might not necessarily agree with experiment. In this section we analyze these quantities in detail. As previously mentioned, we use the red and blue points as an estimate of the spread of the results and therefore an estimate of the error.  

\subsection{Scattering lengths and effective range parameters}

Near threshold, the scattering is usually characterized by the scattering length and the effective range parameters defined through an expansion of the real part of the partial waves:
\begin{equation}
	\text{Re}f^I_{\ell}(s)\overset{k\to 0}{\simeq}k^{2\ell}\big(a^I_{\ell}+b^I_{\ell}k^2+\ldots\big),\;\; k=\frac{\sqrt{s-4m_{\pi}^2}}{2}
\end{equation}
One success of the Weinberg model \eqref{h5} was that it gave reasonable values for the $S0$, $S2$ and $P1$ scattering lengths. However, as explained in section \ref{chiSB} the gauge theory bootstrap does not impose matching with the Weinberg model very close to threshold and therefore, the scattering lengths for those and the other partial waves resulting from the computation may or may not agree with experimantal values. We displayed the results in table \ref{table:scatlength} that includes the Weinberg model (W) \eqref{h5}, the Gauge Theory Bootstrap (GTB), and the phenomenological results of Colangelo, Gasser and Leutwyler (CGL) \cite{Colangelo:2001df}, and Pelaez and Yndurain (PY) \cite{Pelaez:2004vs}. The blue and red colors correspond to results from the blue and red points depicted in fig.\ref{shape}.

\begin{table}[!h]
	\begin{center}
		\renewcommand{\arraystretch}{1.4}
		\begin{tabular}{|c|c|c|c|c|}
			\hline
			&W&{\bf {\bf GTB}}&CGL&PY\\
			\hline
			$a^{(0)}_0$&0.16  &\gtbblue{0.174},   \gtbred{0.191}    &$0.220\pm 0.005$  &$0.230\pm 0.010$               \\ 
			$a^{(2)}_0$&-0.046&\gtbblue{-0.0357}, \gtbred{-0.0403}  &$-0.0444\pm0.0010$&$-0.0422\pm 0.0022$            \\  
			$b^{(0)}_0$&0.18  &\gtbblue{0.285},   \gtbred{0.282}    &$0.280\pm0.001$   &$0.268\pm 0.010$               \\
			$b^{(2)}_0$&-0.092&\gtbblue{-0.061},  \gtbred{-0.071}   &$-0.080\pm0.001$  &$-0.071\pm0.004$               \\
			$a^{(1)}_1$&   31 &\gtbblue{28.7},    \gtbred{30.7}     &$37.0\pm 0.13$    &$38.1\pm 1.4 \;(\times 10^{-3})$    \\
			$b^{(1)}_1$&0     &\gtbblue{12.4},    \gtbred{7.28}     &$5.67\pm 0.13$    &$ 4.75\pm 0.16 \;(\times 10^{-3})$ \\
			$a^{(0)}_2$&0     &\gtbblue{14.6},    \gtbred{14.9}     &$17.5\pm0.3$      &$18.0\pm 0.2 \;(\times 10^{-4})$    \\
			$a^{(2)}_2$&0     &\gtbblue{3.41},    \gtbred{2.53}     &$1.70\pm0.13$     &$2.2\pm 0.2 \;(\times 10^{-4})$     \\
			\hline
		\end{tabular}
	\end{center}
	\caption{Scattering lengths and effective range parameters in units of $m_\pi$. Red and blue correspond to the amplitudes at points in fig.\ref{shape} with the same color and the others are phenomenological results as described in the text.} \label{table:scatlength}
\end{table}

\subsection{Low energy expansion of form factors}

Since the gauge theory bootstrap also computes the form factors, it is interesting to obtain their low energy expansion given by
\beqa
F^0_0(s) &=& F^0_0(0) \left[ 1 + \frac{1}{6} s \langle r^2 \rangle^\pi_S +\ldots \right] \\
F^1_1(s) &=& 1+ \frac{1}{6} s \langle r^2 \rangle^\pi_V + \ldots\\
F^0_2(s) &=& 1+ \frac{1}{6} s \langle r^2 \rangle^\pi_T + \ldots
\eeqa
where $\sqrt{\langle r^2 \rangle^\pi_{S,V,T}}$ are the scalar, vector and spin 2 radius of the pion. The spin two radius is sometimes called the {\it mechanical} radius. Such radii probe the size of the pion as seen by the various currents. Also recall that $F^0_0(0)=1$ in our units where $m_\pi=1$. The results for the (squared) radii are displayed in table \ref{table:rSV} together with the values derived from experiment \cite{Colangelo:2001df,DONOGHUE1990341,AMENDOLIA1986168}. 
\begin{table}[!h]
	\begin{center}
		\renewcommand{\arraystretch}{1.4}
		\begin{tabular}{|c|c|c|}
			\hline
			&{\bf {\bf GTB}}&Exp. fits\\
			\hline
			$\langle r^2 \rangle^\pi_{S} $&\gtbblue{0.64},  \gtbred{0.55}   &$0.61\pm 0.04 \, \mathrm{fm}^2$ \\ 
			$\langle r^2 \rangle^\pi_{V} $&\gtbblue{0.394}, \gtbred{0.391}  &$0.439\pm 0.008 \,\mathrm{fm}^2$            \\
			$\langle r^2 \rangle^\pi_{T} $&\gtbblue{0.153}, \gtbred{0.140}  &           \\
			\hline
		\end{tabular}
	\end{center}
	\caption{Pion scalar $\langle r^2 \rangle^\pi_{S} $ and vector $\langle r^2 \rangle^\pi_{V} $ (or electromagnetic) radii.} \label{table:rSV}
\end{table}
Although (to our knowledge) there is no direct determination of the spin 2 radius, recent phenomenological calculations based on experimental parton distribution functions give $\langle r^2 \rangle^\pi_{T}= 0.67-0.77\,\mathrm{fm}^2 $ \cite{PhysRevD.97.014020} and $\langle r^2 \rangle^\pi_{T}= 2.1-2.4\,\mathrm{fm}^2 $ \cite{song2018generalizeddistributionamplitudesgravitational}.

\subsection{Chiral Lagrangian coefficients}

As mentioned in section \ref{chiSB}, the EFT method provides a compact way to organize the results in terms of the coefficients of the most general Lagrangian compatible with the symmetries  \eqref{L4}. Usually, these coefficients are determined by comparison with experiment. More recently there has been a large effort in using positivity methods to put generic bounds on them. With the gauge theory bootstrap we have a scheme to actually compute such coefficients based on self-consistency conditions (bootstrap) and gauge theory input. The results are given in table \ref{table:lj}.    
As described in the previous subsections, we have obtained the $D$-waves scattering lengths and the pion scalar and vector radii. Therefore, we can obtain the chiral Lagrangian coefficients $\bar{l}_{1,2,4,6}$ from the formulas \cite{GASSER1984142,chiPTintro}
\beqa
a_{D0} &=& \frac{1}{1440\pi^3f_\pi^4}\left\{\bar{l}_1+4\bar{l}_2-\frac{53}{8} \right\}   +\ldots \\
a_{D2} &=& \frac{1}{1440\pi^3f_\pi^4}\left\{\bar{l}_1+\bar{l}_2-\frac{103}{40} \right\}  +\ldots \\
F^0_0(s) &=& 1 + \frac{s}{16\pi^2f_\pi^2}\left(\bar{l}_4-\frac{13}{12}\right) +\ldots \\
F^1_1(s) &=& 1+ \frac{s}{96\pi^2f_\pi^2}(\bar{l}_6-1) + \ldots
\eeqa
With the values shown in table \ref{table:scatlength} and table \ref{table:rSV}, we find 
\begin{table}[H]
	\begin{center}
		\renewcommand{\arraystretch}{1.4}
		\begin{tabular}{|c|c|c|c|c|}
			\hline
			&{\bf {\bf GTB}}&GL&Bij&CGL\\
			\hline 
			$\bar{l}_1$&\gtbblue{0.96,}\ \gtbred{-0.10}  &$-2.3\pm3.7   $&$-1.7 \pm 1.0     $&$ -0.4\pm0.6 $  \\ 
			$\bar{l}_2$&\gtbblue{4.5,}\ \gtbred{  4.8}  &$ 6.0\pm 1.3  $&$6.1\pm 0.5       $&$ 4.3\pm0.1  $  \\  
			$\bar{l}_4$&\gtbblue{4.8, }\ \gtbred{  4.2}  &$ 4.3\pm0.9   $&$4.4\pm 0.3       $&$ 4.4\pm0.2  $  \\
			$\bar{l}_6$&\gtbblue{14.6,}\ \gtbred{14.5 }  &$ 16.5\pm 1.1 $&$16.0\pm0.5\pm0.7 $&$            $  \\
			\hline
		\end{tabular}
	\end{center}
	\caption{Low energy parameters computed using gauge theory bootstrap (GTB) and phenomenological values (see text). The coefficient $\bar{l}_1$ is challenging to evaluate and very sensitive to the $D$-wave scattering lengths.  } \label{table:lj}
\end{table}
These values may be used to recompute the scattering lengths and effective range parameters to compare with the ones directly extracted in table \ref{table:scatlength} as a consistency check. See appendix \ref{sec:param2}. In table \ref{table:lj}, we also displayed the results compiled in \cite{chiPTintro} from Gasser and Leutwyler (GL) \cite{GASSER1984142}, Bijnens et al. (Bij) \cite{Bijnens:1998fm,Bijnens:1997vq,Bijnens:1994ie} and Colangelo et al.(CGL) \cite{Colangelo:2001df}. 
As we can see, the phenomenological determination of these parameters is also not too precise but there is reasonable agreement except for $\bar{l}_1$ that is challenging to determine. At this stage our procedure is not precise enough to give very accurate figures but it is satisfying that it is within expected values. The main point is, as we mentioned, that we have a scheme to \emph{calculate} the EFT parameters.

\section{Resonances}\label{sec:resonances}

As we have seen in section \ref{sec:phaseshifts} fig.\ref{6pw}, the $P1$, $D0$ and potentially $S0$ channels display resonances. In this section we study their properties more closely.    

\subsection{Scalar form factor}\label{S0FF}

\begin{figure}[H]
	\centering
	\includegraphics[width=0.75\textwidth]{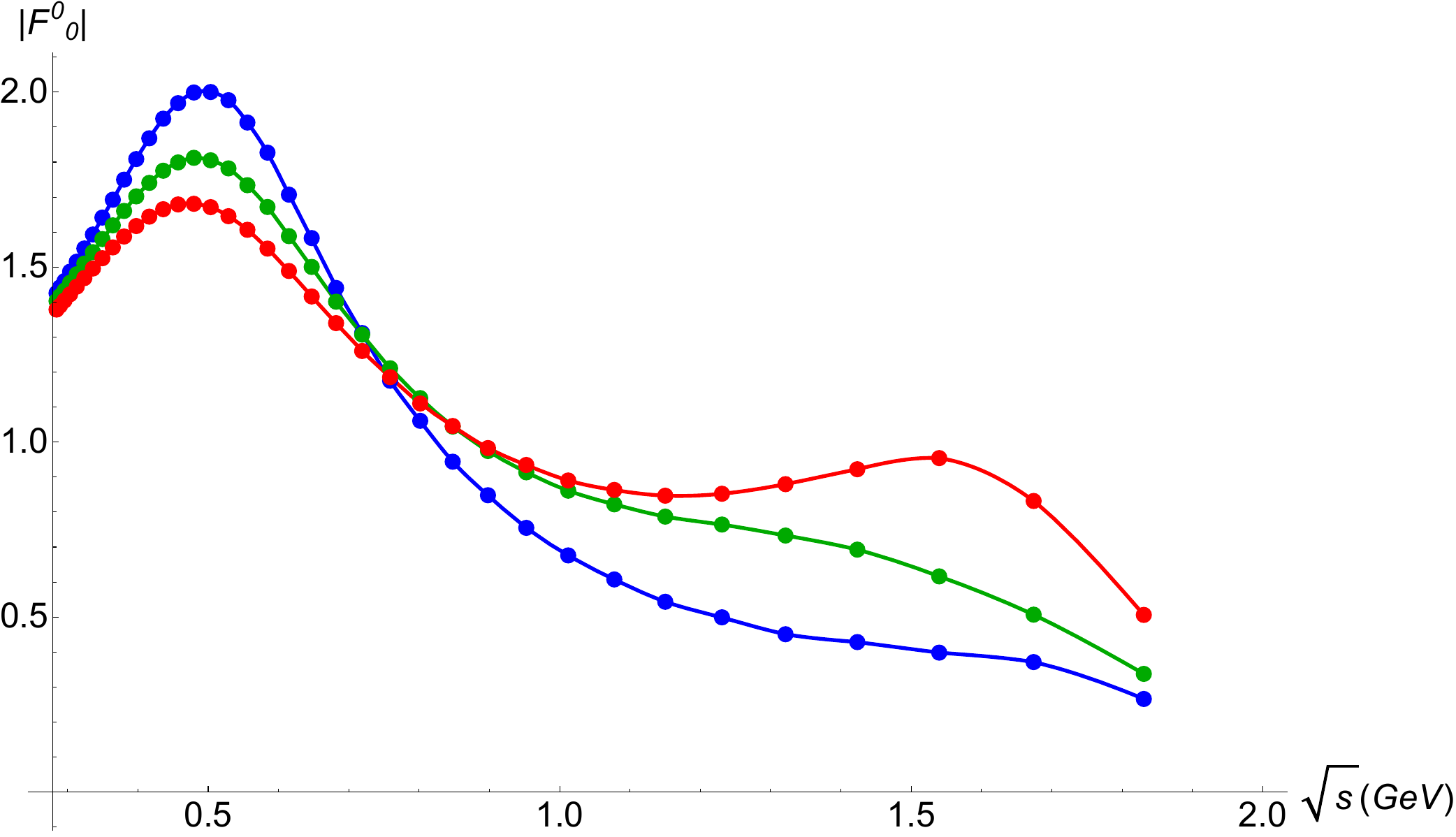}
	\caption{Pion scalar form factor.}
	\label{FFS0}
\end{figure}

In fig. \ref{6pw}, the $S0$ channel phase shift exhibits a sharp rise starting from threshold as also seen in the phenomenological fits. Correspondingly we see, at $\sim 500\,\MeV$ a peak in the scalar form factor plotted in fig.\ref{FFS0} corresponding to the blue and red points in fig. \ref{shape}.
This broad peak can be interpreted as the $\sigma$ resonance but our current numerics is not precise enough to see a pole far from the real axis. This resonance seems to be due to chiral dynamics and therefore sensitive to the low energy input rather than high energy information.

\subsection{The $\rho$ meson}

In our previous work \cite{GTBPRL,GTBPRD}, we found the $\rho$ meson as a resonance in the $P1$ channel. In this paper we see that its properties remain largely unchanged after increasing the QCD matching energy $s_0$ from $\sqrt{s_0}=1.2\,\GeV$ in the previous paper to $\sqrt{s_0}=2\,\GeV$ in this one. This is a good test of the consistency of the procedure in particular because the $\rho$ meson is a signature of QCD and is determined, in this approach, by the gauge theory dynamics imposed at $s_0$. Let us now study its properties in more detail. 

\subsubsection{Pole in the $P1$ wave}

\begin{figure}[h] 
	\centering
	\includegraphics[width=\textwidth,trim={0cm 0 0cm 0},clip]{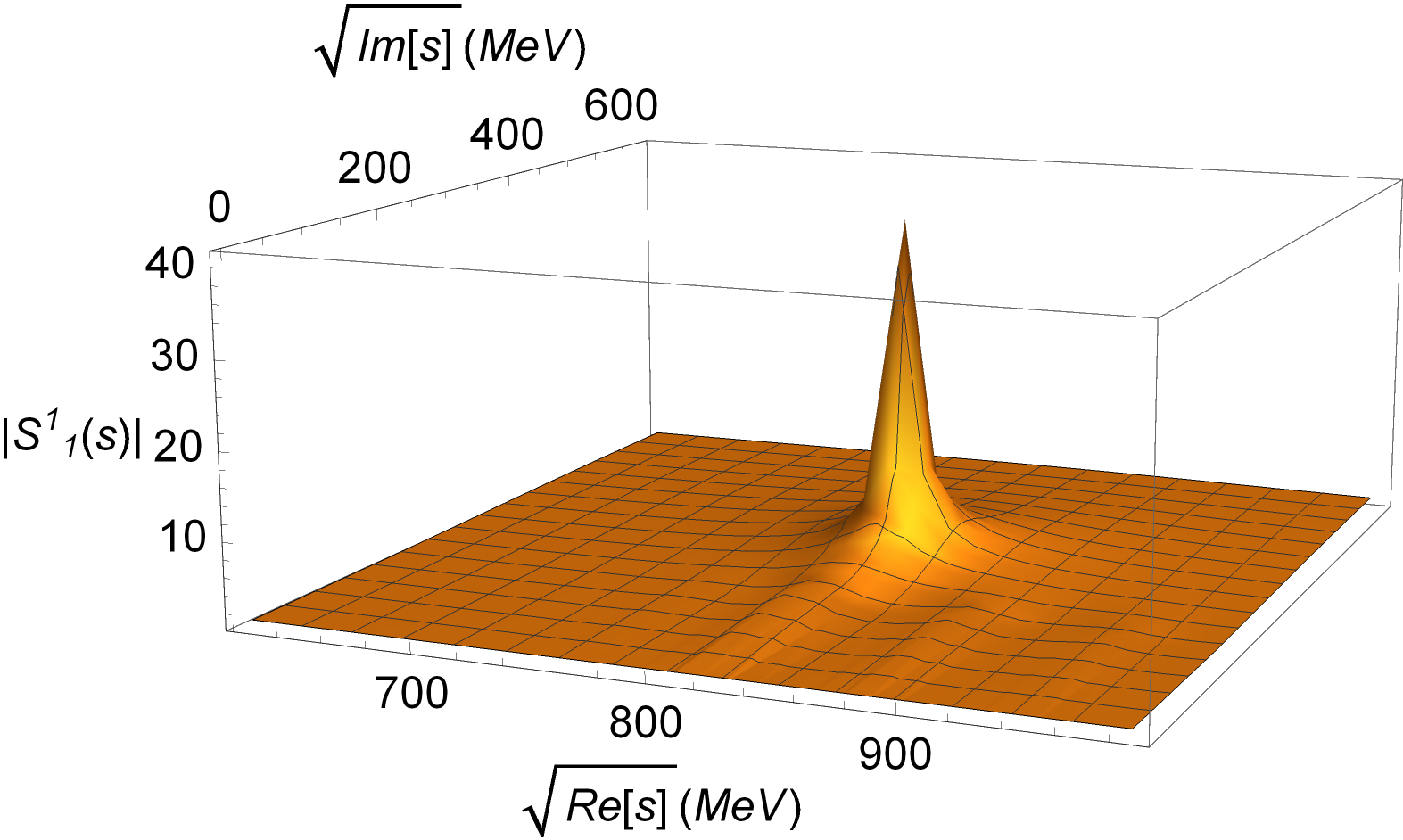}
	\caption{S-matrix pole on the second sheet corresponding to the $\rho$ meson.}
	\label{rhopole}
\end{figure}

The unitarity saturation condition on the $4<s<16$ region of the real axis (or any larger region) for a given partial wave can be written as
\beq\label{SSone}
S_\ell^I(s) (S_\ell^I(\bar{s}))^* = 1
\eeq
Written in this way, the left hand side is an analytic function of $s$ and therefore, if it is equal to 1 in a segment, it is equal to 1 for any value of $s$. Notice however that this means that if $s$ is in the lower half plane, below the cut, $\bar{s}$ has to be above but crossing under the cut, namely on the second sheet. This also explains why $|S(s)|^2=1$ is not valid on the real axis for $s<4$ since $s$ and $\bar{s}$ are on different sheets and neither has to be for $s>16$ since a new cut appears there. In any case, what \eqref{SSone} implies is that a pole in the second sheet appears as a zero in the first sheet. In fig.\ref{rhopole} we plot the modulus of the scattering amplitude $S^1_1(s)$ analytically continued to the second sheet using \eqref{SSone}. A pole is clearly visible in the plot. We extract the position of this pole and the comparison with experiments is given in table \eqref{table:rhopole}.
\begin{table}[!h]
	\begin{center}
		\renewcommand{\arraystretch}{1.4}
		\begin{tabular}{|c|c|c|}
			\hline
			&{\bf {\bf GTB}}& PDG\\
			\hline
			$\Re(\sqrt{s_\rho})    $&\gtbblue{829}, \gtbred{844}  &$761-765\pm 0.23\ \MeV $ \\ 
			$\Im(\sqrt{s_\rho})  $&\gtbblue{71}, \gtbred{82}  &$71-74\pm 0.8  \ \MeV $ \\
			\hline
		\end{tabular}
	\end{center}
	\caption{Position of the $\rho$ meson T-matrix pole $s_\rho$ in $\MeV$.} \label{table:rhopole}
\end{table}
\subsubsection{Vector form factor}

\begin{figure}[h]
	\centering
	\includegraphics[width=0.75\textwidth]{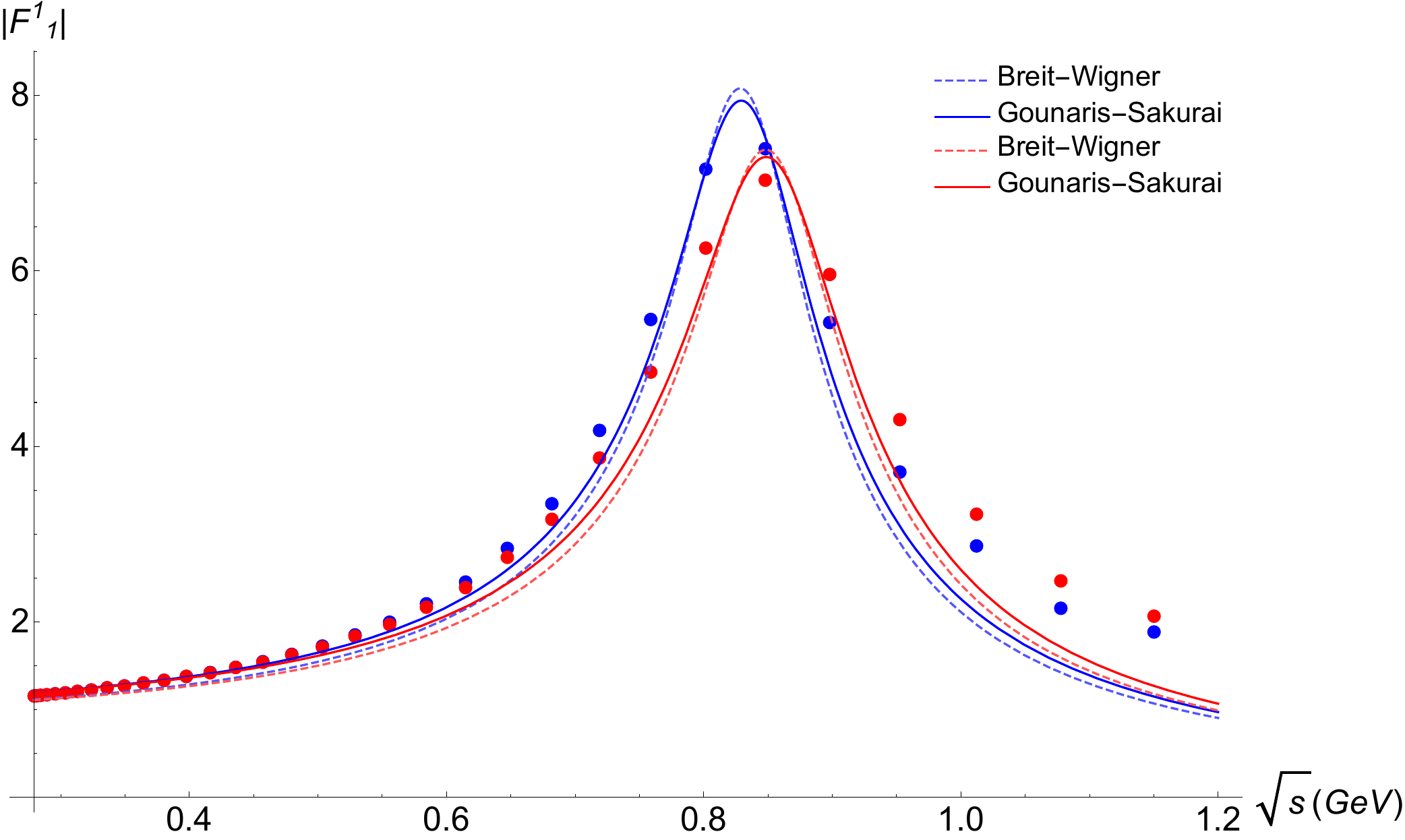}
	\caption{$\rho$ resonance in the vector form factor. The peak is fit with a Breit-Wigner resonance (dashed line) and a Gounrais-Sakurai shape (solid line) that gives a better fit on the low energy side. From here we obtain the mass and width of the $\rho$ meson.}
	\label{FFP1fit}
\end{figure}

\begin{figure}[h]
	\centering
	\includegraphics[width=0.85\textwidth]{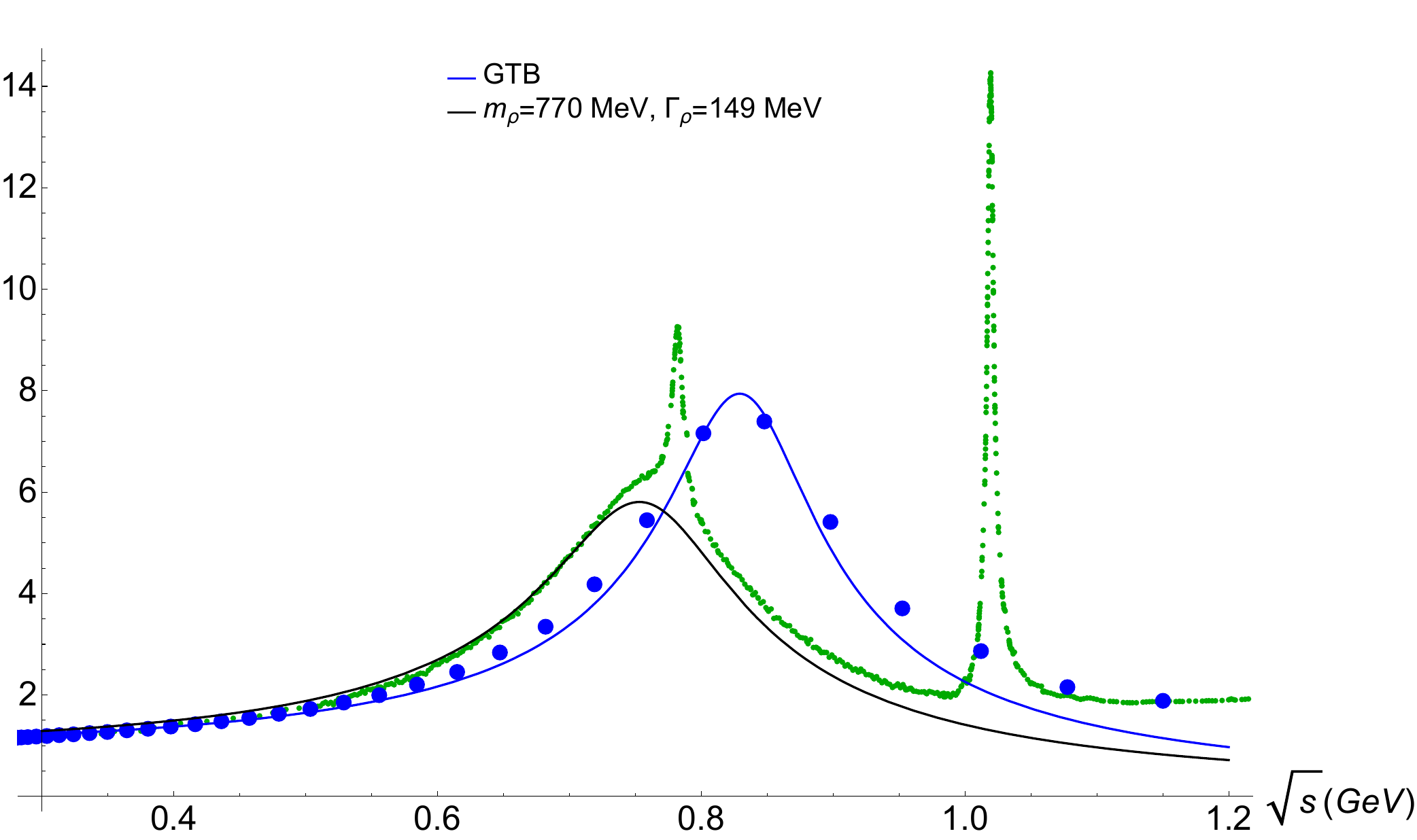}
	\caption{We compare one of the fits (blue) in fig.\ref{FFP1fit} with the Gounaris-Sakurai form using the PDG data (black). The green curve represents the cross section of $e^++e^-\rightarrow $ hadrons (from PDG) properly rescaled to be compared to the form factor of $j_V$. This last one displays two extra narrow peaks, the $\omega$ and $\phi$ resonances that do not appear in the $P1$ channel for $SU(2)$ flavor.}
	\label{FFP1}
\end{figure}

Next, we fit the $P1$ form factor that we obtained from the bootstrap by assuming it is dominated by a $\rho$-meson pole, namely it is of the Breit-Wigner form \cite{DONOGHUE1990341}:
\beq\label{BWform}
F^1_1(s)=-\frac{m_\rho^2}{s-m_{\rho}^2+i m_{\rho}\, \Gamma_\rho\, \theta(s-4m_\pi^2)}
\eeq 
where the function $\theta(x)=1$ if $x>0$ and zero otherwise, ensuring that it is real for $s<4$ and $F^1_1(0)=1$. 
This is a very convenient form since it is given by only two parameters ($m_\rho$, $\Gamma_\rho$) that we can fit. For each of the blue and red points considered (in fig. \ref{shape}) we fit the form factor as shown in figure \ref{FFP1fit} using a Breit-Wigner shape (dashed line). The agreement is reasonable but can be improved. In fact it was noted long ago  that the $\rho$ is not of the Breit-Wigner type. Using general considerations of elastic unitarity and analyticity, a better formula  was developed by Gounaris and Sakurai \cite{Gounaris:1968mw} resulting in
\beq
F^1_1(s) = \frac{m_\rho^2[1+d\Gamma_\rho/m_\rho]}{(m_\rho^2-s)-i m_\rho\Gamma_\rho(q/q_\rho)^3(m_\rho/\sqrt{s})}
\eeq
where $q=\sqrt{\frac{s-4}{4}}$ and 
\beq
d = \frac{3}{\pi}\frac{1}{q_\rho^2} \ln\left(\frac{m_\rho+2q_\rho}{2}\right)+\frac{m_\rho}{2\pi q_\rho} - \frac{m_\rho}{\pi q_\rho^3}. 
\eeq 
Notice that we still have only two constants $(m_\rho,\Gamma_\rho)$ to fit. With this formula we fit the form factor again and find a better agreement on the low energy side of the peak (the Gounaris-Sakurai form takes into account the correct threshold behavior). 
As a result we obtain for the $\rho$ mass and width for each of the two curves compared to the experimental value in table \ref{table:mrho}. It is interesting to note that the associated spectral density is related \cite{Peskin:1995ev,quarksandleptons} to the so called hadronic ratio $R=\frac{\sigma(e^+e^-\rightarrow \mathrm{hadrons})}{\sigma(e^+e^-\rightarrow \mu^+\mu^-)}$  that can be measured very precisely (and for example plays an important role in $g-2$ computations). For completeness we include a plot in fig. \ref{FFP1}. 
\begin{table}[!h]
	\begin{center}
		\renewcommand{\arraystretch}{1.4}
		\begin{tabular}{|c|c|c|}
			\hline
			&{\bf {\bf GTB}}& PDG\\
			\hline
			$m_\rho     $&\gtbblue{838}, \gtbred{859}  &$775\pm 0.23 \ \MeV$ \\ 
			$\Gamma_\rho$&\gtbblue{114}, \gtbred{128}  &$149.1\pm 0.8\ \MeV$ \\
			\hline
		\end{tabular}
	\end{center}
	\caption{$\rho$ meson mass and width in $\MeV$ compared to the data \cite{Workman:2022ynf}.} \label{table:mrho}
\end{table}

Having determined the $\rho$ mass $m_\rho$ and width $\Gamma_\rho$ and given that $\rho$ decays into two pions, we can use these values to determine the $g_{\rho\pi\pi}$ coupling through the formula \cite{TheBook}:
\beq
\Gamma_\rho = g_{\rho\pi\pi}^2 \frac{m_\rho}{48\pi} \left[1-\frac{4m_\pi^2}{m_\rho^2}\right]^{\frac{3}{2}}
\eeq
We find
\beq
g_{\rho\pi\pi} = \text{\gtbblue{4.9}}, \ \ \text{\gtbred{5.2}}
\eeq
compared to the value $g_{\rho\pi\pi}=6$ computed similarly from the PDG values $m_\rho=775\MeV$ and $\Gamma_\rho=149\MeV$. This simply reflects that we have a somewhat narrower resonance.   

Another important constant associated with the $\rho$ meson is $g_\rho$ defined through
\beq
\bra{0} j_V^\mu \ket{\rho} = \frac{m_\rho^2}{g_\rho} \ \epsilon^\mu
\eeq
where $\epsilon$ is the polarization of the $\rho$ meson. It determines the electromagnetic decay width for $\rho\rightarrow e^++e^-$. We can compute $g_\rho$ using (the $(2\pi)^{-4}$ is because of our normalization):
\beq
\int_{\mathrm{\rho\ peak}}\!\!\!\!\!\!\!\!\!\!\! dx\ \frac{|\cF_1^1(x)|^2}{x} \simeq \frac{1}{(2\pi)^4}\,\frac{m_\rho^2}{2}\ \frac{4\pi}{g_\rho^2} 
\eeq
giving $\frac{4\pi}{g_\rho^2}\simeq \gtbblue{0.61},\ \gtbred{0.57}$ to be compared with $\frac{4\pi}{g_\rho^2}=0.507\pm0.011$ \cite{EWERZ201431}. Equivalently $g_\rho=4.8 \simeq g_{\rho\pi\pi}$. Historically \cite{Sakurai}, an early model of the $\rho$ meson consider it to be a gauge boson in which case $g_{\rho}=g_{\rho\pi\pi}$ was the universal gauge coupling. Alternatively, the same equality can be understood from the Vector Meson Dominance (VMD) hypothesis stating that the photon couples to hadrons mainly through the $\rho$ meson and therefore the $\rho$ meson pole dominates the electromagnetic form factor of the pion even far from the resonance. Our results are in agreement with that hypothesis since the fit in fig.\ref{FFP1fit} is good all the way down to threshold.  

\subsubsection{The $\rho(1450)$}\label{1450}

\begin{figure}[H]
	\centering
	\includegraphics[width=0.75\textwidth]{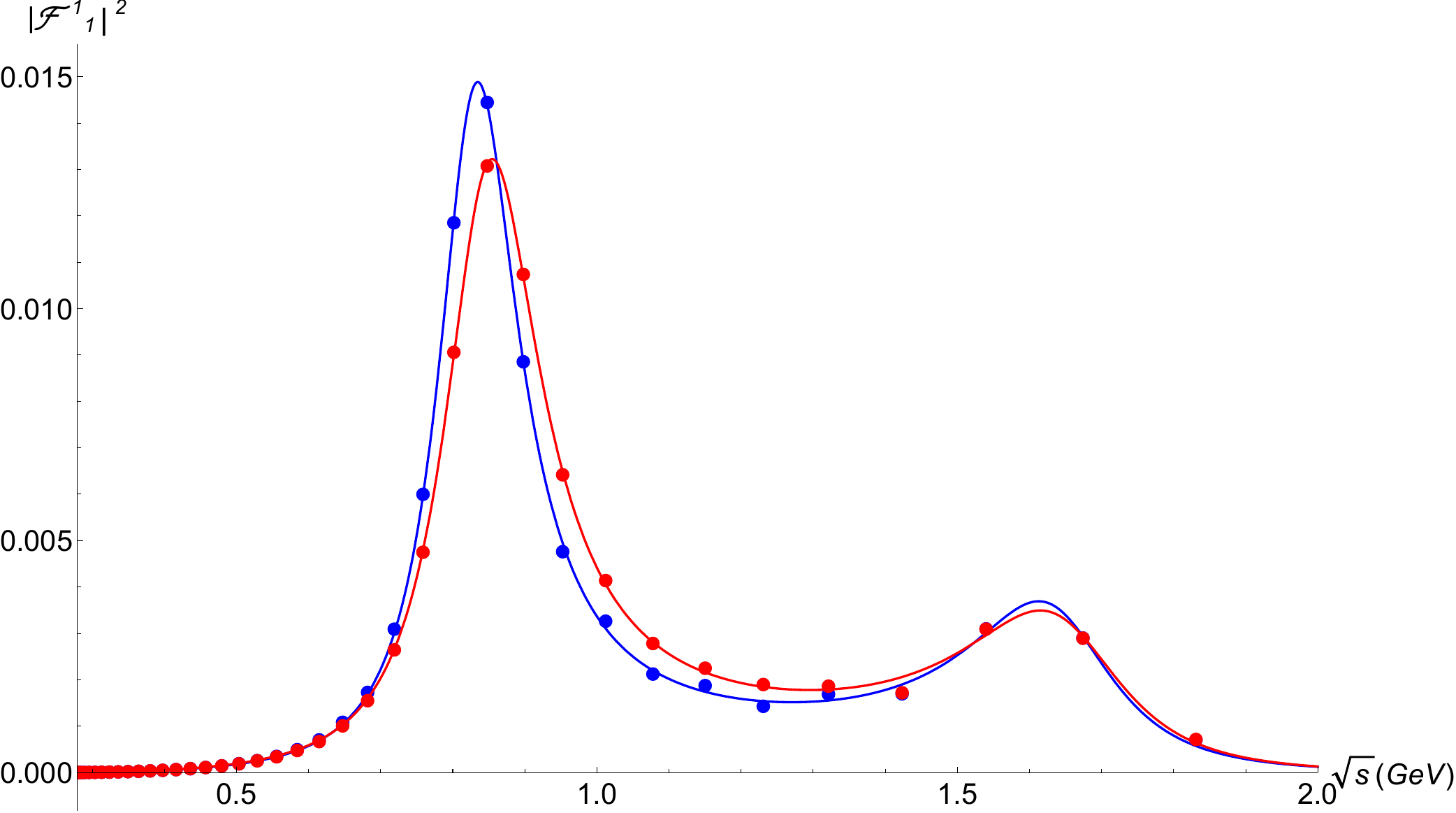}
	\caption{The fit of the bootstrap results of the rescaled form factor $|\mathcal{F}^1_1(s)|^2$ up to 2 GeV with two Breit-Wigner resonances \eqref{twopole}. In addition to the sharper peak we already saw in fig. \ref{FFP1fit} identified with $\rho(770)$, we find a peak at higher energy which we preliminarily identify with $\rho(1450)$.}
	\label{rho1450twopole}
\end{figure}

Although the vector form factor $\mathcal{F}^1_1$ that we obtained (see Fig.\ref{FFP1fit}) is dominated by the $\rho(770)$ meson resonance, at larger energy $\simeq 1.6\,\GeV$ another, smaller, peak becomes visible. It is also apparent in fig. \ref{6pw} as a second rise in the $P1$ phase shift around the same energy. From the PDG meson table \cite{Workman:2022ynf}, it is natural to identify it with the $\rho(1450)$ resonance that has the correct quantum numbers. 
We take this identification as preliminary and leave more precise calculations for future work. In fig. \ref{rho1450twopole}
we fit the form factor we obtained with a a sum of two Breit-Wigner resonances:
\begin{equation}\label{twopole}
	F^1_1(s)=\frac{-A_1 m_1^2}{s-m_1^2+im_1\Gamma_1\theta(s-4m_{\pi}^2)}+\frac{A_2 m_2^2}{s-m_2^2+im_2\Gamma_2\theta(s-4m_{\pi}^2)}
\end{equation}
where $\theta(x)$ is the step function that ensures the form factor is real for $s<4\mpi^2$. The fit parameters are given in table \ref{table:1450}, together with the estimate given by PDG. Note that the fit parameters give
\begin{equation}
	A_1-A_2={\gtbblue{1.03}},{\gtbred{1.11}}
\end{equation}
agreeing nicely with the normalization for the form factor $F^1_1(0)=1$. Assuming that mesons fall into Regge trajectories, we have found the starting point of two distinct Regge trajectories one beginning with the $\rho(770)$ and the second with the $\rho(1450)$. Asymptotically ($s\rightarrow\infty$) the form factor (\ref{twopole}) behaves as
\beq
F^1_1(s)\simeq\frac{-A_1 m_1^2+A_2 m_2^2}{s}
\eeq 
showing that the asymptotic behavior (\ref{P1asymp}) should be the result of cancellations among several resonances. 

\begin{table}[!h]
	\begin{center}
		\renewcommand{\arraystretch}{1.4}
		\begin{tabular}{|c|c|c|}
			\hline
			&{\bf {\bf GTB}}& PDG\\
			\hline
			$m_1     $&\gtbblue{824}, \gtbred{841}  &$775\pm 0.23\ \MeV$ \\ 
			$\Gamma_1$&\gtbblue{137}, \gtbred{164}  &$149.1\pm 0.8\ \MeV$ \\
			$m_2     $&\gtbblue{1635}, \gtbred{1645}  &$1465\pm25\ \MeV$ \\ 
			$\Gamma_2$&\gtbblue{248}, \gtbred{272}  &$400\pm 60\ \MeV$ \\
			$A_1     $&\gtbblue{1.29}, \gtbred{1.39}  &\\ 
			$A_2$&\gtbblue{0.27}, \gtbred{0.28}  &\\
			\hline
		\end{tabular}
	\end{center}
	\caption{the mass and width in MeV of the $\rho(770)$ meson ($m_1,\Gamma_1$) and $\rho(1450)$ meson ($m_2,\Gamma_2$) compared to the data \cite{Workman:2022ynf}.} \label{table:1450}
\end{table}

\subsection{The gravitational form factor and the $f_2$ meson}\label{sec:f2}

The $D0$ wave has isospin $0$ and angular momentum $\ell=2$, the same quantum numbers as the energy momentum tensor and the graviton. Hypothetically gravity could couple to hadrons through the $f_2$ meson resonance in the $D0$ channel as the photon couples through the $\rho$ meson in the $P1$ channel. In this section we study the form factor of the energy momentum tensor a.k.a the gravitational form factor. 

\begin{figure}[h]
	\centering
	\includegraphics[width=0.75\textwidth]{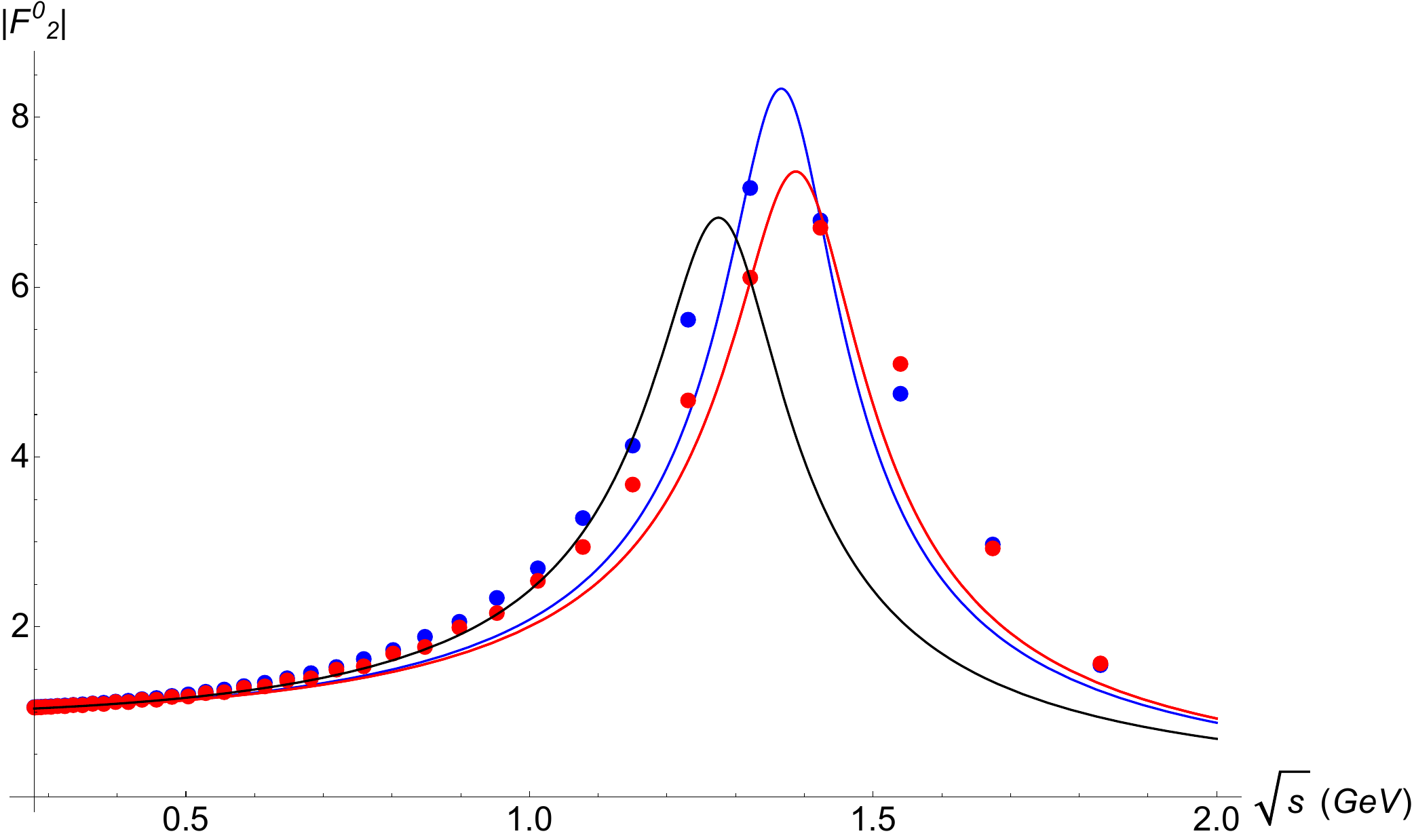}
	\caption{Gravitational form factor, fit with a Breit Wigner form (blue and red), compared to the Breit-Wigner form using PDG data on $f_2$ (black).} 
	\label{D0FFfit}
\end{figure}

The phase shift and form factor of the $D0$ wave clearly show a strong resonance near $1.2\,\GeV$ that we naturally identify with the $f_2(1270)$ meson. Fitting the form factor with a Breit-Wigner form (fig. \ref{D0FFfit}) we determine the mass and width as in table \ref{table:mf2}.

\begin{table}[!h]
	\begin{center}
		\renewcommand{\arraystretch}{1.4}
		\begin{tabular}{|c|c|c|}
			\hline
			&{\bf {\bf GTB}}& PDG\\
			\hline
			$m_{f_2}     $&\gtbblue{1366}, \gtbred{1388}  &$1275.4\pm0.6\ \MeV$ \\ 
			$\Gamma_{f_2}$&\gtbblue{164}, \gtbred{188}  &$186.6\pm 2.3\ \MeV$ \\
			\hline
		\end{tabular}
	\end{center}
	\caption{$f_2$ meson mass and width in $\MeV$ compared to the data \cite{Workman:2022ynf}.} \label{table:mf2}
\end{table}

It should be noted that the $f_2(1270)$ meson decay fraction to $\pi\pi$ is $(84.3^{+2.9}_{-0.9})\%$ with the rest of the time decaying to $\pi\pi\pi\pi$ (10.5\%) and $K\bar{K}$ (4.6\%). This is in constrast to the $\rho$ that decays $\sim 100\%$ into $\pi\pi$. Therefore one has particle production through the resonance ($\pi\pi\rightarrow f_2 \rightarrow \pi\pi\pi\pi$) but it is not clear how that can be incorporated in the bootstrap.   

In any case, using the formula \cite{EWERZ201431}\footnote{The $f_2$ pion coupling is a dimension five operator of the schematic form $\cL \sim \frac{g_{f_2\pi\pi}}{M_0} \phi \partial \pi\partial \pi$. In \cite {EWERZ201431} a reference mass $M_0=1\; \GeV$ is introduced to define a dimensionless coupling.}
\beq
\Gamma(f_2\rightarrow\pi^0\pi^0) = \half\Gamma(f_2\rightarrow \pi^+\pi^-) = \frac{m_{f_2}}{960\pi} |g_{f_2\pi\pi}|^2 \left(\frac{m_{f_2}}{M_0}\right)^2 \left(1-\frac{4m_\pi^2}{m_{f_2}^2}\right)^\frac{5}{2}
\eeq
we find
\beq
g_{f_2\pi\pi} = \text{\gtbblue{8.5}}, \ \ \text{\gtbred{8.9}}
\eeq 
close to the value $g_{f_2\pi\pi}=9.26$ from \cite{EWERZ201431}.

\section{Form factors}\label{sec:FFl}

The form factors are an important part of the numerical method. Given that the results needed for the numerical computation are available in the literature, this section mostly provides some useful definitions and normalizations. It also provides some computations that can be useful in further implementations of the gauge theory bootstrap with more partial waves. The reader may omit sections \ref{sec:FFl} and \ref{sec:jj} initially. 

To each partial wave we associate a gauge theory operator with the same quantum numbers $I,\ell$ (isospin and angular momentum). The operators we choose for $I=0,1$ are
\beq
\begin{aligned}
	I=0, &\ \ell=0                   &j_S            =& m_q \bar{\psi}\psi                                                            \\
	I=1, &\ \ell=1                   &j_V^{a} =& \half\tau^{a}_{AB} \, \bar{\psi}_{A} \slashed\Delta \psi_B                         \\
	I=0, &\ \ell=2                   &j_T =& T^{\mu\nu} \Delta_\mu \Delta_\nu                                              \\
	I=0, &\ \ell> 2, \mathrm{even} &j_\ell      =& \bar{\psi}_{A} \slashed\Delta (D_\Delta)^{\ell-1} \psi_A                        \\
	I=1, &\ \ell> 2, \mathrm{odd}  &j^{a}_\ell     =&\half\tau^{a}_{AB}\,  \bar{\psi}_{A} \slashed\Delta  (  D_\Delta)^{\ell-1}  \psi_B\    
\end{aligned}
\eeq
where $\Delta=\frac{1}{\sqrt{2}}(0,1,i,0)$ is used to consider the operator with largest projection in the direction $x^+=\frac{1}{\sqrt{2}}(x+iy)$ and ensure the angular momentum is $\ell$. The Pauli matrices $\tau^{a}_{AB}$ are used to project to isospin $1$. 
The covariant derivative is $D_\Delta=\Delta^\mu(\partial_\mu -ig A_\mu)$. In the present paper we only use the first three in the list: scalar and vector currents and the energy momentum tensor. They have been extensively studied in the literature \cite{osti_1447331,Pire:1996bc,DONOGHUE1990341,Ananthanarayan_2004,Tong_2021,Raya_2022,Davier:2023cyp,PhysRevD.97.014020,Tong_2022,Hoodbhoy_2004,Tanaka_2018} and we only review their properties. For recent bootstrap applications of form factors specially in 2d systems see \cite{Karateev:2019ymz,Karateev:2020axc,Chen:2021pgx,Correia:2022dyp,Cordova:2023wjp}. For the $\ell>2$ case we give computations that we think will be useful in future implementations of the gauge theory bootstrap. Finally, operators with $I=2$ require at least four quarks, we discuss an example in the next section. 

\subsection{Definitions}\label{FFdef}

\subsubsection{Angular momentum $\ell$}

In the center of mass frame where $p=(\epsilon,\vec{p})$ and $p'=(\epsilon,-\vec{p})$, and then $(p+p')\cdot\Delta=0$, we define the form factor $F_{\ell}(s)$ for an operator of angular momentum $\ell\ge 1$ as follows \footnote{We can use the same definition for $\ell=0$ but the normalization differs from the one used for the scalar current in \eqref{scalarF}.}.

\noindent {\bf Isospin 1:}\\
\beq
\bra{\pi^{a}_{p} \pi^{b}_{p'}} j^{c}_{\ell} \ket{0} = 2 (-ip\Delta)^{\ell}\, \epsilon^{abc} F_\ell^1(s=(p+p')^2)
\eeq 
In the basis of fixed total angular momentum and isospin we find: 
\beqa
\bra{I=1,I_3=0,P\ell \sigma} j_\ell^{c=3} \ket{0} &=&  \delta_{\sigma\ell}\,\cF^{I=1}_\ell(s) \\
\cF^{I=1}_\ell(s) &=& \frac{i(-i)^\ell 2^{\ell/2} \ell! }{8\pi^3}\sqrt{\frac{4\pi}{(2\ell+1)!}} \left(\frac{s-4}{4}\right)^{\frac{2\ell+1}{4}} \frac{1}{s^{\frac{1}{4}}}\ F^{I=1}_\ell(s) \nonumber
\eeqa
To derive this result we used 
\beq
 (-ip\cdot \Delta)^\ell = (-i)^\ell |\vec{p}|^\ell\,2^{\frac{\ell}{2}} \ell!\sqrt{\frac{4\pi}{(2\ell+1)!}} Y_{\ell\ell}(\hat{p})
\eeq
where $Y_{\ell\ell}(\hat{p})$ is a standard spherical harmonic. 

\noindent {\bf Isospin 0:}\\
\beq\label{F0elldefA}
\bra{\pi^{a}_{p}\pi^{b}_{p'}} j_{\ell} \ket{0} = -2 (-ip\Delta)^{\ell}\, \delta^{ab} F_\ell^0(s=(p+p')^2)
\eeq 
and then 
\beqa
\bra{I=0, I_3=0,P\ell \sigma} j_\ell \ket{0} &=&  \delta_{\sigma\ell}\,\cF^{I=0}_\ell(s) \\
\cF^{I=0}_\ell(s) &=& -\frac{(-i)^\ell 2^{\ell/2} \ell!}{8\pi^3}\sqrt{\frac{6\pi}{(2\ell+1)!}}\  \left(\frac{s-4}{4}\right)^{\frac{2\ell+1}{4}} \frac{1}{s^{\frac{1}{4}}}F^{I=0}_\ell(s) \nonumber
\eeqa

\subsubsection{The gravitational form factor}

In the case of the $D0$ wave we need an operator with spin 2 and isospin 0. We can use $\cO=\bar{\psi}_a \gamma^+ D^+ \psi_a$. However that operator mixes with a gluonic one. A particular linear combination is the energy momentum tensor that we are going to use in this case.
From energy momentum conservation we can write (the upper index in the form factor is the isospin)
\beqa\label{Tff}
\bra{\pi^+_{p_1}}T^{\mu\nu}(x)\ket{\pi^+_{p_2}} &=& e^{i(p_1-p_2)x} \left\{ \half F^0_2(t) (p_1+p_2)^\mu (p_1+p_2)^\nu + \right. \\
&& \left. \tilde{F}(t) (t\,\eta^{\mu\nu}-(p_1-p_2)^\mu (p_1-p_2)^\nu)\right\} 
\eeqa
with $t=(p_1-p_2)^2$. Also given that the Hamiltonian is $H=\int d^3 x T^{00}(x)$ and that $H\ket{\pi^+_{p_2}}=\epsilon_{p_2}\ket{\pi^+_{p_2}}$ we find
\beq
\bra{\pi^+_{p_1}}\int\!\! d^3 x\, T^{00}(x)\ket{\pi^+_{p_2}} = \epsilon_{p_2} \bra{\pi^+_{p_1}} \pi^+_{p_2}\rangle = 2\epsilon_{p_1}^2 (2\pi)^3 \delta^{(3)}(\vec{p}_1-\vec{p}_2)
\eeq
Comparing with \eqref{Tff}  we find
\beq
F^0_2(0)=1 
\eeq
From \eqref{Tff} and using crossing symmetry we obtain  
\beq
{}_{\mbox{out}}\bra{\pi^+_{p_1}\pi^-_{p_2}}T^{\mu\nu}(0)\ket{0} = \half F^0_2(s) (p_1-p_2)^\mu (p_1-p_2)^\nu + \tilde{F}(s) (s\,\eta^{\mu\nu}-(p_1+p_2)^\mu (p_1+p_2)^\nu)
\eeq
and in particular
\beq
{}_{\mbox{out}}\bra{\pi^+_{p_1}\pi^-_{p_2}}T^{++}(0)\ket{0} = 4 |\vec{p}_1|^2 F^0_2(s) \sqrt{\frac{2\pi}{15}}\, Y_{22}(\hat{p_1})
\eeq
where $Y_{\ell\sigma}(\hat{p}_1)$ are the usual spherical harmonics evaluated in the direction of $\vec{p}_1$. Now it follows that
\beq
\bra{P\ell\sigma} T^{++}(0)\ket{0} = \cF^0_2(s) \delta_{\ell 2}\delta_{\sigma 2}
\eeq
\beq
\cF^0_2(s) = \frac{1}{4\pi^3}\sqrt{\frac{\pi}{5}}\frac{1}{s^{\frac{1}{4}}}\left(\frac{s-4}{4}\right)^{\frac{5}{4}}\ F^0_2(s) 
\eeq

\subsection{Asymptotic form}\label{FFasymform}

\begin{figure}[H]
	\centering
	\begin{subfigure}[b]{0.2\textwidth}
		\raggedright
		\includegraphics[width=\textwidth]{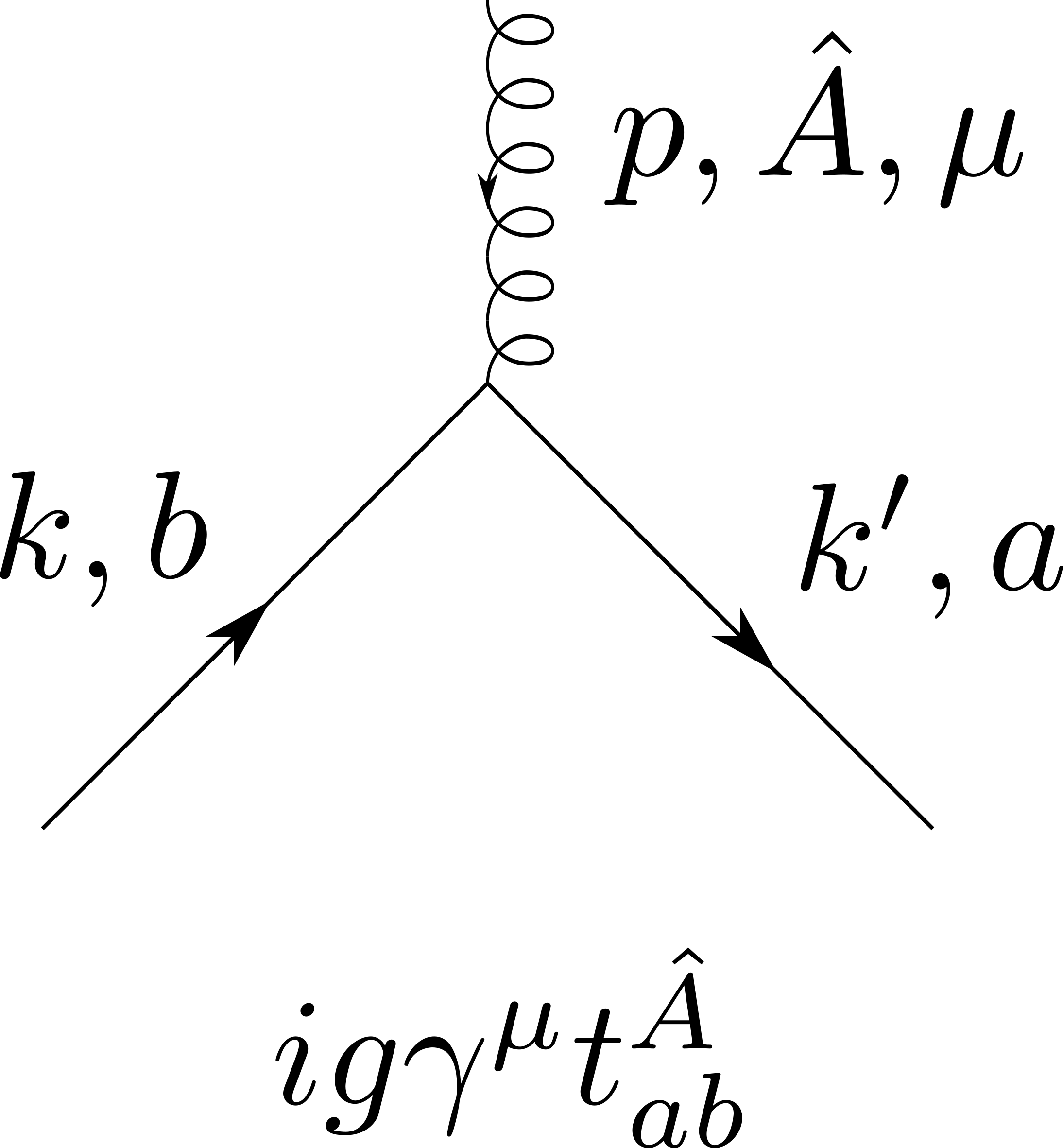}
		\caption{}
\end{subfigure}
\hspace{1cm}
\begin{subfigure}[b]{0.2\textwidth}
	\centering
	\includegraphics[width=\textwidth]{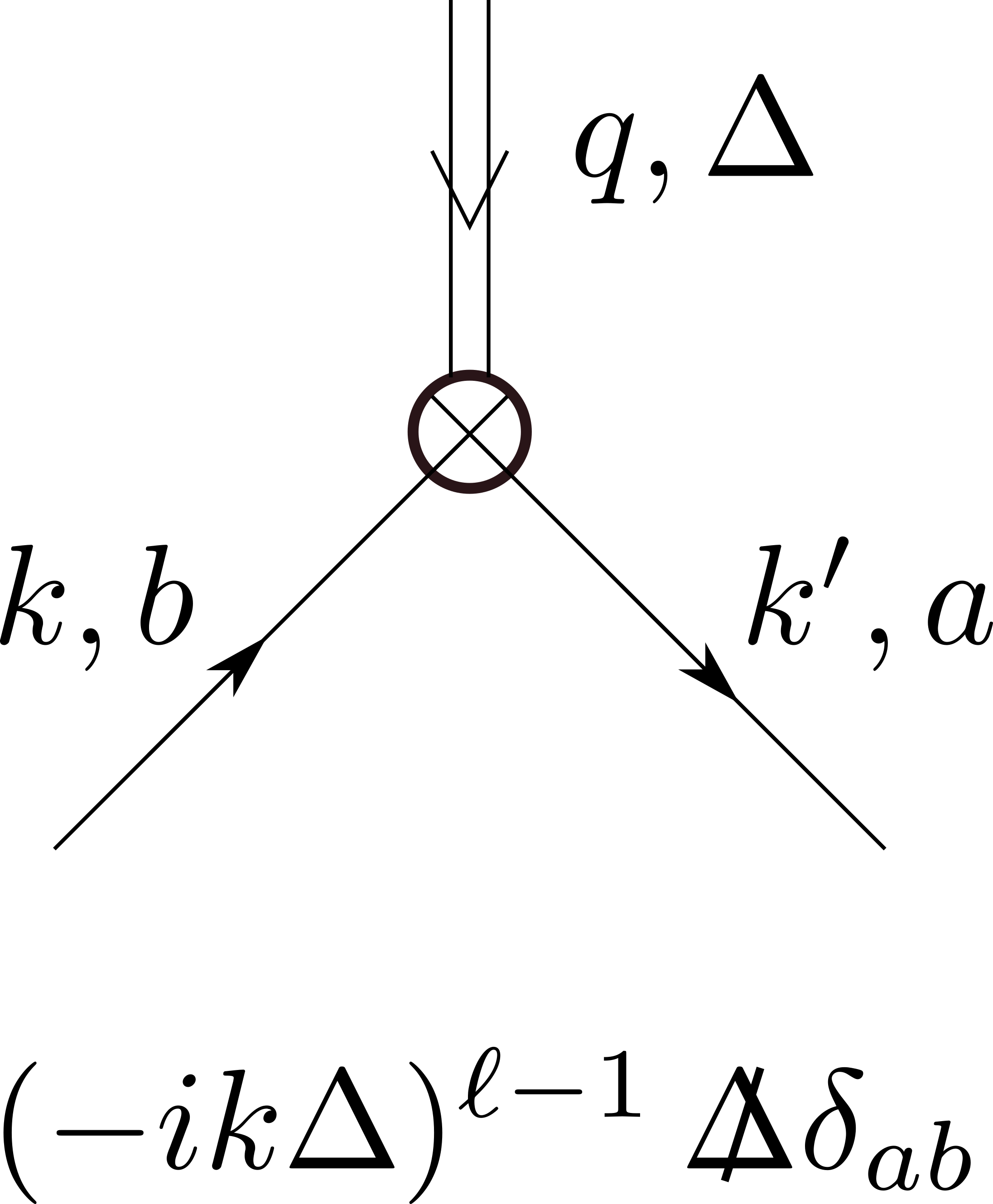}
	\caption{}
	\end{subfigure}
	\hspace{1cm}
	\begin{subfigure}[b]{0.4\textwidth}
\centering
\includegraphics[width=\textwidth]{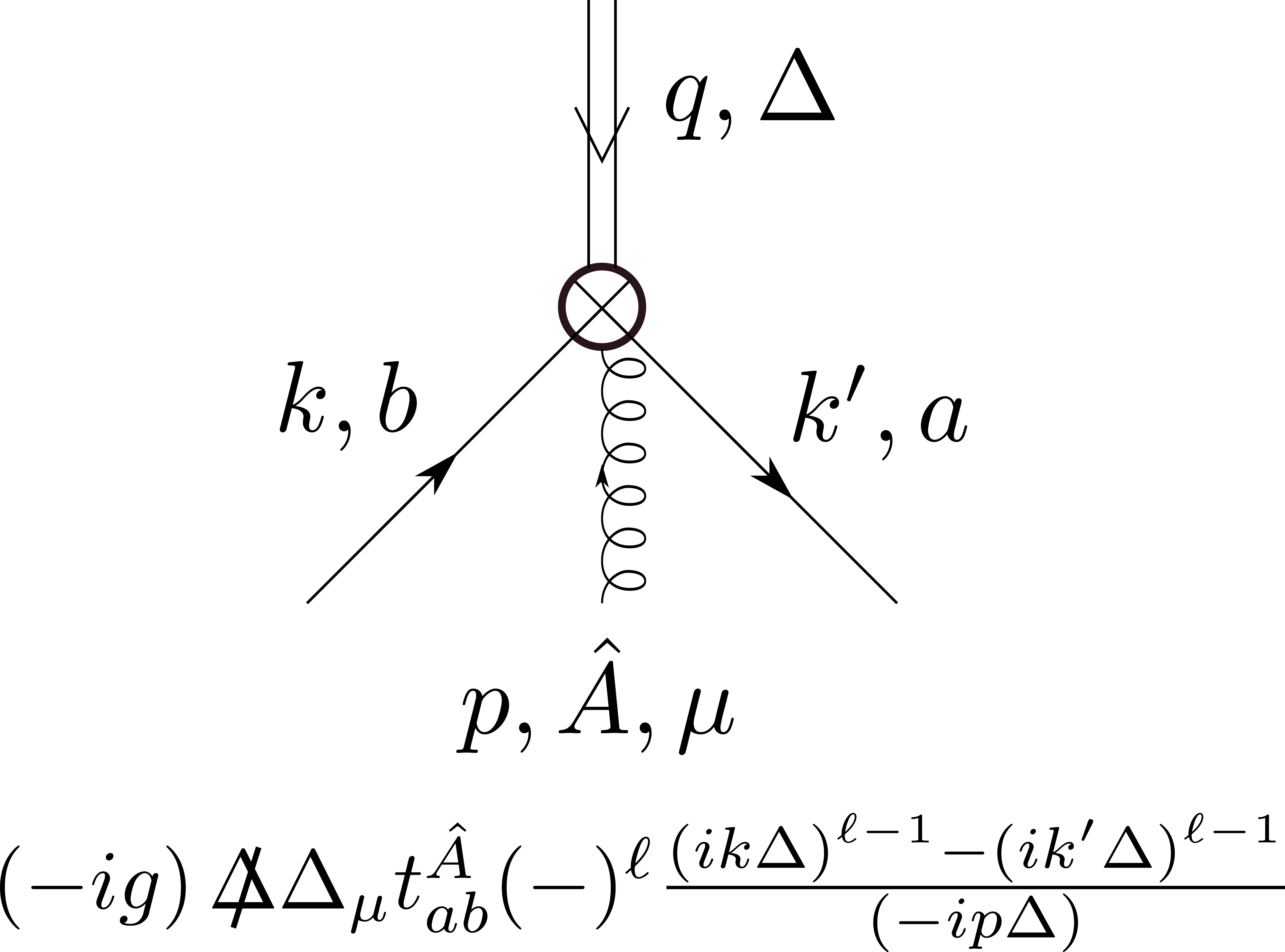}
\caption{}
\end{subfigure}
\caption{Vertices to compute the current propagators and the asymptotic form factors. (a) standard gauge theory vertex (b) Current insertion.  (c) Current insertion with one gluon from the covariant derivatives.}
\label{Vertices}
\end{figure}

\begin{figure}[H]
\centering
\begin{subfigure}{\textwidth}
\centering
\includegraphics[width=0.7\textwidth]{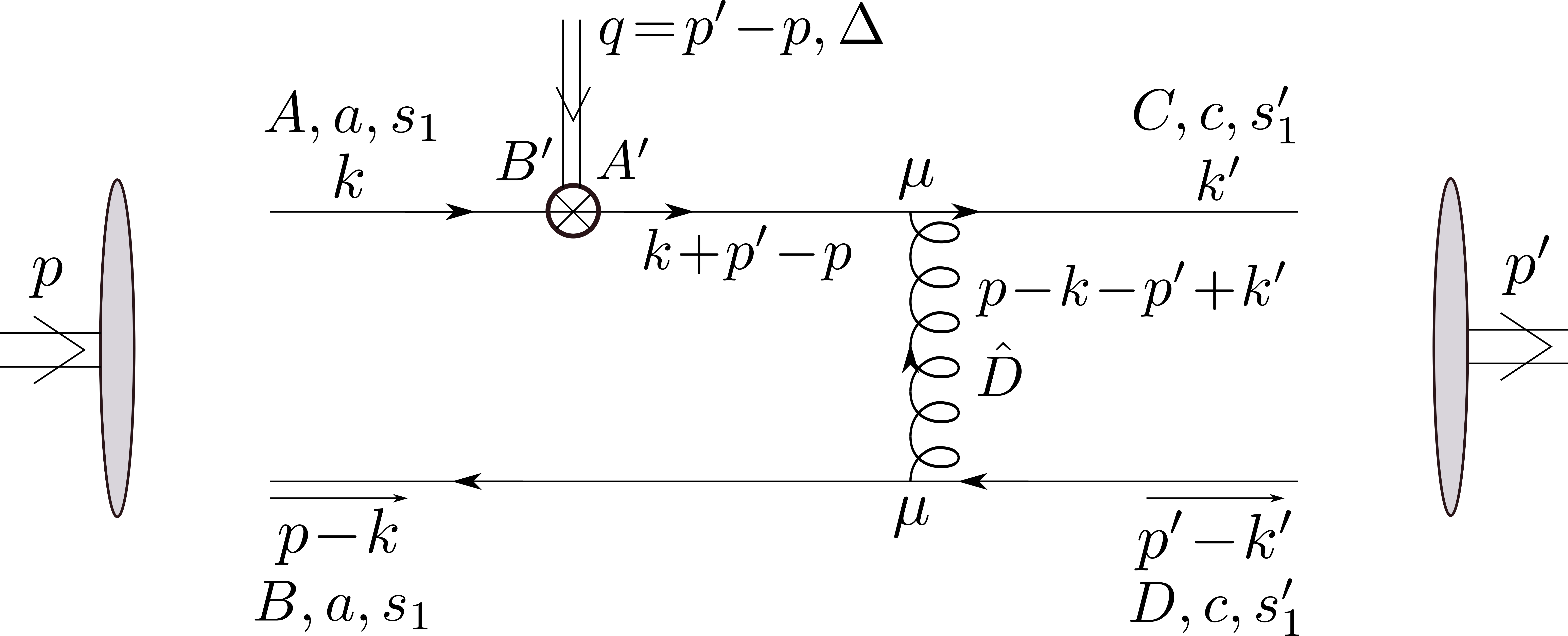}
\caption{}
\end{subfigure}
\vspace{0.25cm}
\hrule
\vspace{0.5cm}
\begin{subfigure}{\textwidth}
\centering
\includegraphics[width=0.55\textwidth]{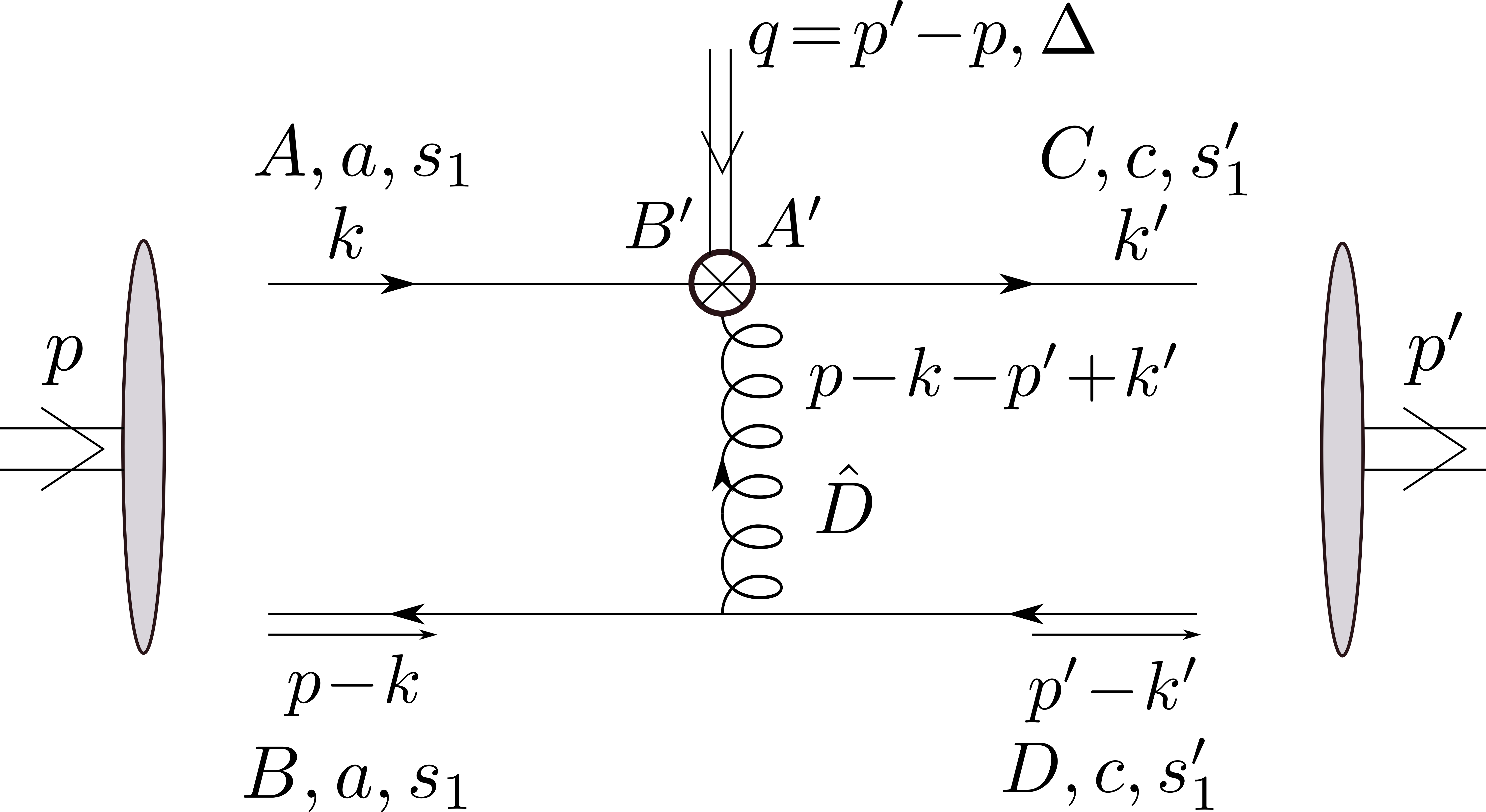}
\caption{}
\end{subfigure}
\caption{Standard Feynman diagrams with on-shell external particles contributing to the asymptotic pion form factor in the parton model. The grey blobs represent a linear combination of free fermions as in \eqref{QCDpion}. The indices $A$,$B$,$C$,$D$,$A'$,$B'$ indicate flavor, $a,c$ color, $s_1,s'_1$ polarization, and $\hat{D}$ are in the adjoint representation of the gauge group.}
\label{FFdiags}
\end{figure}

The computation of asymptotic form factors is well known since the pioneering work of Brodsky and Lepage \cite{osti_1447331} on exclusive QCD processes. In this section we give a simple presentation of the subject that allows to compute the case of general $\ell$ at lowest order. Physically, for large momentum, the form factor is very small since it gives the amplitude for a rare process where a (virtual) particle (\eg\ photon) strikes a pion and the pion absorbs the particle without being destroyed. It can be computed in the parton model by considering that the particle strikes a parton that has to interchange gluons with the other partons so that all partons move in the new direction. The largest contribution comes from the part of the wave function with smallest number of partons, in this case 2. The results depend on the parton distribution function and are also valid at large energies $\sqrt{s}\gg \Lambda_{QCD}$. Since this is the only available result from QCD we use it with some caveats as discussed in section \ref{sec:nummeth}. In this section we describe the computation of the generic $\ell$ case thinking that it might help to extend the method to higher energies and angular momenta. We start by considering a generic current contracted with a vector $\Delta$ as 
\beq\label{ABcurrent}
j^\ell_{B'A'} = \bar{\psi}_{B'} \slashed\Delta D_\Delta^{\ell-1} \psi_{A'}   
\eeq
where $A'$ and $B'$ are flavor indices. We then take a
simple ``pion state" in QCD as a superposition of two free quarks (we follow the normalization in \cite{Peskin:1995ev})
\beq\label{QCDpion}
\ket{\pi^a(p)} = \int_0^1\!\!\! dx \int\!\! d^2k_\perp\, X_{s_1s_2} \psi(x,k_\perp) \half \tau^a_{AB} \sqrt{2\omega_k}\sqrt{2\omega_{p-k}}a^\dagger_{As_1,k} b^\dagger_{Bs_2,p-k}  \ket{0} 
\eeq    
where $k= x p + k_\perp$, $k'= x' p' + k'_\perp$, the indices $s_{1,2}$ denote the polarizations, and color indices of quark and anti-quark are contracted. When not contracted color indices of quark and gluons are denoted with lower-case letters (not to be confused with pion isospin indices as in $\pi^a(p)$).
In all calculations we take the large energy limit where $s=q^2=(p-p')^2\simeq -2pp' \rightarrow\infty$ and we drop all terms involving $k_\perp$, $k'_\perp$. This is explained in more detail in appendix \ref{appendix:FF} where we compute one Feynman diagram for pedagogical reasons.  
The wave function $\psi(x,k_\perp)$ is a low energy quantity. We sum over spin assuming a choice $v_p=-\gamma_5 u_p$ such that $u_p^s\bar{v}^s_p = (\slashed p+m_q)\gamma_5$ as required to match the one piece of information we have from the axial current form factor
\beq
\bra{0} j_\mu^{a,5}(0)\ket{\pi^{b}(p)} = -i \delta^{ab}\, p_\mu\, f_\pi 
\eeq  
It also requires the normalization
\beq
\int_0^1 dx \int d^2k_\perp\, \sqrt{x(1-x)} \psi(x,k_\perp) = -\frac{i}{2N_c}\, f_\pi 
\eeq
From \cite{osti_1447331} we take, for very high energy, in our normalization
\beq
\sqrt{x(1-x)} \psi(x,k_\perp) = -\frac{i}{2N_c}\, f_\pi\ \phi(x)\ \tilde{\psi}(k_\perp),\ \ \ \ \ \ \ \ \int d^2 k_\perp \tilde{\psi}(k_\perp) = 1 
\eeq
where $\phi(x)$ is usually taken to be $\phi(x)=6 x(1-x)$ \cite{osti_1447331}. 
The calculation for the insertion of a current \eqref{ABcurrent} follows, at lowest order in perturbation theory from the Feynman diagrams (a) and (b) in fig. \ref{FFdiags} using the vertices of fig. \ref{Vertices}. For the contribution of all four diagrams of type $(a)$,\ie\  inserting the current in any of the fermion lines, we get (see appendix for one of the diagrams)
\beq
\mathbb{T}^{(a)}_{ABCDA'B'}= \frac{16 i g^2}{s} C_2(N_c)  (-ip\Delta)^\ell \frac{x^{\ell-1}}{(1-x)(1-x')} 
\left[\delta_{AB'}\delta_{A'C}\delta_{BD}+(-)^\ell \delta_{A'B}\delta_{AC}\delta_{B'D}\right]
\eeq
with $C_2(N_c)=\frac{N_c^2-1}{2}$. Similarly we find (two diagrams of type $(b)$ inserting the current on the top or bottom lines) 
\beq
\mathbb{T}^{(b)}_{ABCDA'B'}= \frac{8 i g^2}{s} C_2(N_c)  (-ip\Delta)^\ell \frac{x'{}^{\ell-1}-x^{\ell-1}}{(1-x)(1-x')(x'-x)} \left[\delta_{AB'}\delta_{A'C}\delta_{BD}+(-)^\ell \delta_{A'B}\delta_{AC}\delta_{B'D}\right]
\eeq
resulting in the form factor
\beq\label{phiphi}
\bra{\pi^{b}_{p'}} j^\ell_{B'A'}\ket{\pi^a_p}=\frac{f_\pi^2}{4N_c^2}\int_0^1\int_0^1 dx\,dx'\, \phi(x)\,\phi^*(x')\, \frac{1}{4}\tau^{a}_{BA}\tau^{b}_{CD}\,\mathbb{T}_{ABCDA'B'}
\eeq
where $\phi(x)=6x(1-x)$ and $\mathbb{T}=\mathbb{T}^{(a)}+\mathbb{T}^{(b)}$. We find ($\alpha_s=g^2/(4\pi)$.)
\beqa
\lefteqn{\bra{\pi^{b}_{p'}} j^\ell_{B'A'}\ket{\pi^a_p} =}&& \\
&&\frac{72i\pi\alpha_s f_\pi^2}{t} \frac{C_2(N_c)}{N_c^2}(-ip\Delta)^\ell \left[\tau^{b}\tau^{a}+(-)^\ell \tau^a\tau^{b}\right]_{A'B'} 
\left\{\frac{1}{\ell+1}+\frac{2(H_\ell-1)}{\ell+2}\right\} \nonumber
\eeqa
where we introduced the harmonic numbers $H_\ell=\sum_{p=1}^\ell \frac{1}{p}$. Two cases have to be considered now. 

\noindent {\bf Isospin 1:}\\
We take the current to be
\beq
j^{a}_\ell = \half \tau^a_{B'A'}\, j^\ell_{B'A'}
\eeq
We define the form factor $F_\ell^I(s)$ from the equation:
\beq
\bra{\pi^{a}_{p'}} j^{b}_{\ell} \ket{\pi^{c}_{p}} = -2 (-ip\Delta)^{\ell}\, \epsilon^{abc} F_\ell^1(t=(p-p')^2)
\eeq 
Using \eqref{phiphi} we obtain (for $\ell$ odd, it vanishes for $\ell $ even):
\beq
F^1_\ell(t) \simeq_{t\rightarrow\infty} -\frac{72\pi\alpha_sf_\pi^2}{t} \frac{C_2(N_c)}{N_c^2} 
\left\{\frac{1}{\ell+1}+\frac{2(H_\ell-1)}{\ell+2}\right\}
\eeq
with $H_\ell=\sum_{p=1}^\ell \frac{1}{p}$. Setting $\ell=1$ and $N_c=3$ in the previous formula $\left( C_2(N_c)=\frac{N_c^2-1}{2}\right)$ we find
\beq
F^1_1(t) \simeq_{s\rightarrow\infty} -\frac{16\pi\alpha_s f_\pi^2}{t} 
\eeq
in agreement with Brodsky-Lepage \cite{osti_1447331,Pire:1996bc}. 
The higher values of $\ell$ are not used in this paper but they could be useful for later reference.

\noindent {\bf Isospin 0:}\\
We take the current to be
\beq
j_{\ell} = \delta_{B'A'}\, j^\ell_{B'A'}
\eeq
We define the form factor $F_\ell^I(s)$ from the equation:
\beq\label{F0elldef}
\bra{\pi^{a}_{p'}} j_{\ell} \ket{\pi^{b}_{p}} = -2 (-ip\Delta)^{\ell}\, \delta^{ab} F_\ell^0(t=(p-p')^2)
\eeq 
Using \eqref{phiphi} we obtain (for $\ell$ even, it vanishes for $\ell $ odd):
\beq
F^0_\ell(t) \simeq_{t\rightarrow\infty} -\frac{144i\pi\alpha_sf_\pi^2}{t} \frac{C_2(N_c)}{N_c^2}   \left\{\frac{1}{\ell+1}+\frac{2(H_\ell-1)}{\ell+2}\right\}
\eeq
with $H_\ell=\sum_{p=1}^\ell \frac{1}{p}$. The case $\ell=2$, is relevant for the gravitational form factor:
\beq
F^0_{2q}(t) \simeq_{t\rightarrow\infty} -\frac{84i\pi\alpha_sf_\pi^2}{t} \frac{C_2(N_c)}{N_c^2}  
\eeq
where we added the subindex $q$ to indicate it comes from the quark part. 

\subsubsection{The gravitational form factor}
\begin{figure}[H]
\centering
\includegraphics[width=0.6\textwidth]{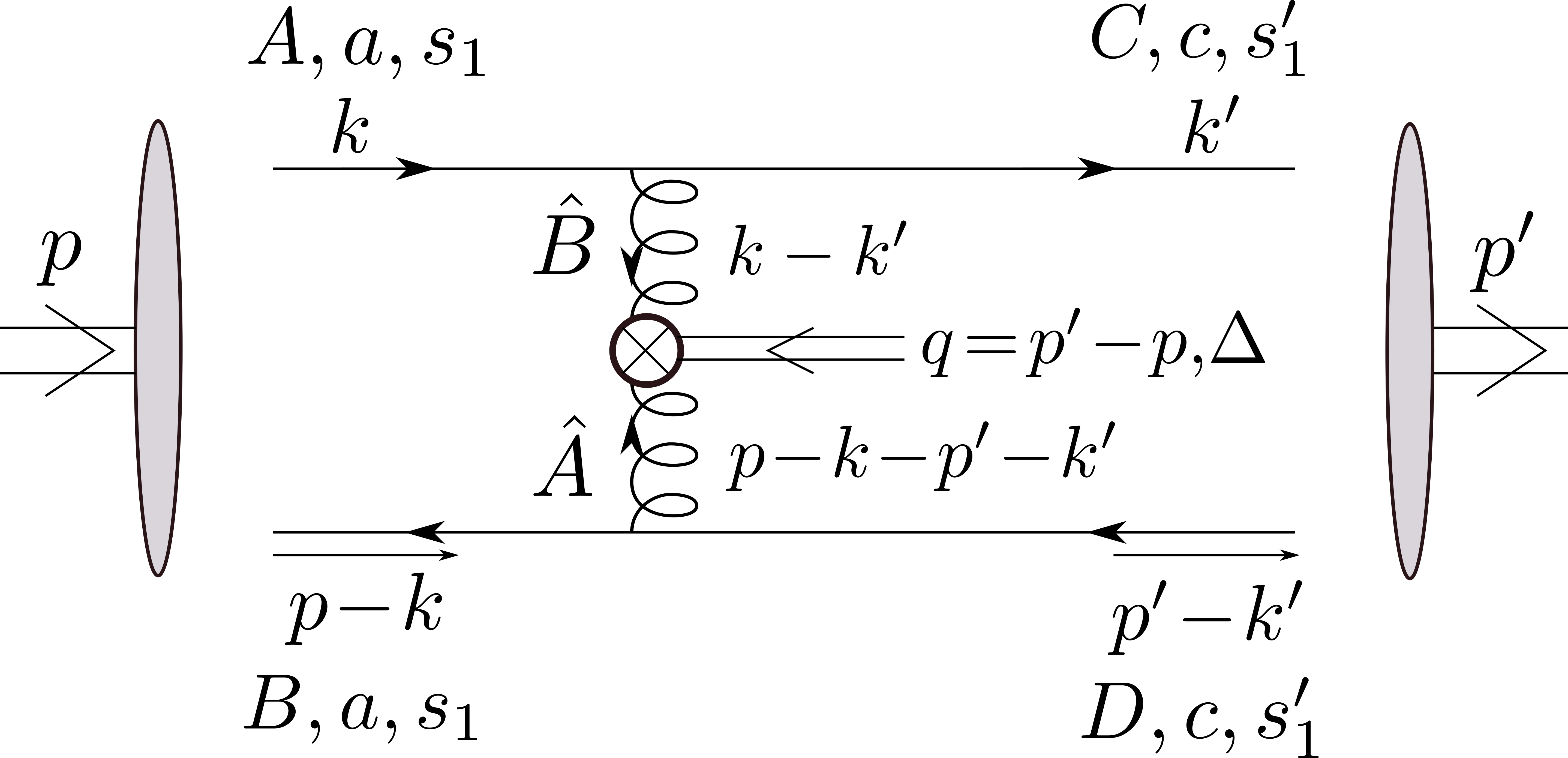}
\caption{Feynman diagrams contributing energy momentum asymptotic pion form factor (gluon contribution).}
\label{FFdiags3}
\end{figure}

The gravitational form factor is the form factor associated with the energy momentum tensor\footnote{In this section lowercase indices are in the adjoint of $SU(N_c)$}
\beq
T^{\mu\nu}\Delta_\mu \Delta_\nu = -F^{a\mu\alpha}F^{a\nu}{}_{\alpha} \Delta_\mu\Delta_\nu + \bar{\psi}_A \slashed\Delta iD_\Delta \psi_A
\eeq
The kernels were computed in \cite{Tong_2021}. Since in that paper the final integral with the parton distribution was not performed, to avoid problems with different normalization we recompute the result here. In fact, up to a factor $i$ the quark part was already computed in the previous subsection. The gluonic part can be computed from the diagram in fig.\ref{FFdiags3} giving
\beq
\mathbb{T}_{ABCD} = -\frac{8g^2}{t}C_2(N_c) (-ip\Delta)^2\left\{\frac{1}{x(1-x)}+\frac{1}{x'(1-x')}\right\} \delta_{AC}\delta_{BD}
\eeq
and thus 
\beq
\bra{\pi^{I_3}_{p'}} -F^{a\mu\alpha}F^{a\nu}{}_{\alpha} \Delta_\mu\Delta_\nu  \ket{\pi^{I'_3}_{p}} \simeq_{t\rightarrow\infty} -\frac{48\pi\alpha_sf_\pi^2}{t} \frac{C_2(N_c)}{N_c^2}  (-ip\Delta)^2 \delta^{I_3I'_3}
\eeq 
and, from \eqref{F0elldef}:
\beq
F^0_{2g}(t) \simeq_{t\rightarrow\infty} \frac{24\pi\alpha_sf_\pi^2}{t} \frac{C_2(N_c)}{N_c^2}  
\eeq
Putting both together we find
\beq
F^0_{2}(t)= F^0_{2g}(t)+i F^0_{2q}(t) \simeq_{t\rightarrow\infty} \frac{108\pi\alpha_sf_\pi^2}{t} \frac{C_2(N_c)}{N_c^2} 
\eeq
For the case $N_c=3$ of interest here we find
\beq
F^0_{2}(t) \simeq_{t\rightarrow\infty} \frac{48\pi\alpha_sf_\pi^2}{t} 
\eeq
As explained more in section \ref{sec:nummeth}, it turns out that around $s\sim 2\,\GeV$ this estimate is somewhat low due to the $f_2$ resonance so only the order of magnitude is relevant.  

\subsubsection{Scalar form factor}
We were not able to find much information about the asymptotic form of the scalar form factor in the literature. A calculation similar to the one for the vector form factor vanishes in the chiral limit since one gets the trace of a product of an odd number of Dirac matrices. The subleading term contains a power of the quark mass $m_q$. Namely, the scalar current changes the chirality of the fermion and we need a mass insertion to restore it. The resulting computation has a divergence at the ends of the interval $x\rightarrow 0,1$. The fermion propagator regulates it at an energy $p^2=p'{}^2=m^2_\pi$. These qualitative arguments suggest a behavior
\beq
F^0_{0}(t) \sim_{t\rightarrow\infty} 4\pi\alpha_sf_\pi^2 \frac{m_q^2}{t} \frac{C_2(N_c)}{N_c^2}  \ln\left(\frac{s}{m_\pi^2}\right)
\eeq     
up to a numerical factor. The other power of $m_q$ is from the definition of the current. This is too small for the precision of our numerics where we use another estimate.   

\section{Two point functions}\label{sec:jj}

In this section we consider the two point functions of various currents discussed in the previous section. Since the results needed for the numerical computations of the paper are available in the literature, this section can be omitted in a first read of the paper. It contains calculations that can be useful in future implementaions of the bootstrap with more partial waves. The leading term in the expansion of the two point functions comes from the free theory, namely the UV fixed point of the theory but the results are already not trivial at this order since we are considering composite operators and renormalization is required. 

\subsection{Definitions}\label{2ptdef}
Given an arbitrary local operator or current $j(x)$ we define an integrated operator:
\beq
\cO(P) = \int\frac{d^4 x}{(2\pi)^4} e^{-iPx} \ j(x) 
\eeq
The two currents correlator, or vacuum polarization is defined as
\beqa
\Pi(s) = i\int \frac{d^4 x}{(2\pi)^4} e^{iP(x-y)} \bra{0} \hat{T}\{j(x)^\dagger j(y)\} \ket{0} 
\eeqa
and the spectral density as
\beq
\rho(s) = 2\,\Im \Pi(s+i\epsilon) = \int \frac{d^4 x}{(2\pi)^4} e^{iPx} \bra{0} j(x)^\dagger j(0)\ket{0} 
\eeq
where $s=P^2$. Notice that
\beq
\bra{0} (\cO(P'))^\dagger\ \cO(P)\ket{0} = \delta^{(4)}(P'-P) \rho(s) 
\eeq
which is the identity required by the form factor bootstrap.

\subsection{Asymptotic form}

As in the case of the form factors, the calculation requires the asymptotic form of the the two current correlator for $|s|\rightarrow\infty$. Here $|s|=s_0=2\,\GeV$ and we make the assumption that the asymptotic form given by the free theory (Feynman diagram in fig.\ref{jjdiags1}) is already dominant.

\subsubsection{Two quark operators of arbitrary $\ell$}
\begin{figure}[H]
\centering
\includegraphics[width=0.35\textwidth]{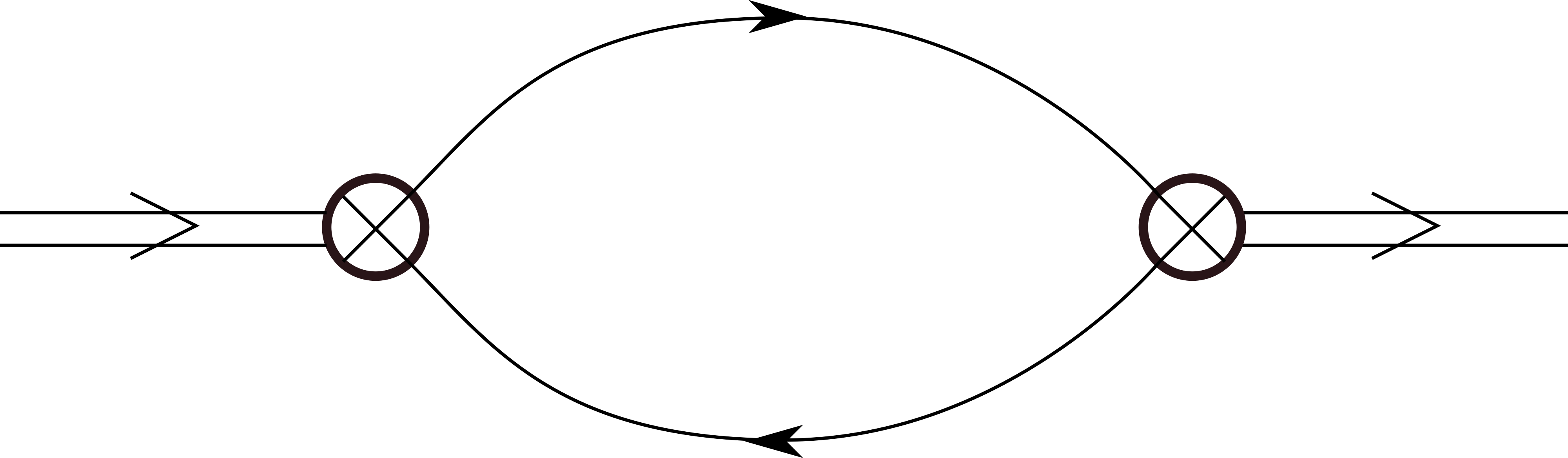}
\caption{Standard Feynman diagrams contributing to the two point function of the current at lowest order. }
\label{jjdiags1}
\end{figure}

Consider the current and operator
\beqa
j_{\ell,AB}(x) &=& \bar{\psi}_{A}(x)\, \slashed\Delta\, D_\Delta^{\ell-1}\psi_{B}(x) \\
\cO_{q,AB}^\Delta &=& \int\frac{d^4 x}{(2\pi)^4} e^{-iqx} \ j_{\ell,AB}(x) 
\eeqa
where $A,B$ are flavor indices (color indices are contracted) and $\Delta=\frac{1}{\sqrt{2}}(0,1,i,0)$. This implies $\Delta^2=0=\bar{\Delta}^2$, $\Delta\bar{\Delta}=-1$ and also, in the center of mass where $q$ is time-like, $q\Delta=0=q\bar{\Delta}$. A standard one loop diagram (fig.\ref{jjdiags1}) using the vertex $(b)$ of fig.\ref{Vertices} gives, in the free theory:
\beqa
\Pi(s)_{\ell,ABCD} &=& i\int \frac{d^d x}{(2\pi)^d} e^{iP(x-y)} \bra{0} \hat{T}\{(j_{\ell,AB}(x))^\dagger j_{\ell,CD}(y)\} \ket{0} \nonumber\\
&=& -\frac{N_c}{(2\pi)^4}\, \frac{\delta_{AC}\delta_{BD}}{2 \pi^2}\, \frac{\ell+1}{2^\ell\, \ell} B(\ell+1,\ell+1) \, s^\ell \ln(-\frac{s}{\mu^2})
\eeqa
Also $B(a,b)=\frac{\Gamma(a)\Gamma(b)}{\Gamma(a+b)}$. From here we find 
\beq
\rho(s)_{\ell,ABCD} = 2\Im \Pi(s+i\epsilon)_{\ell,ABCD} = 
\frac{N_c}{(2\pi)^4}\, \frac{\delta_{AC}\delta_{BD}}{\pi}\, \frac{\ell+1}{2^\ell\, \ell} B(\ell+1,\ell+1) \, s^\ell 
\eeq

and the generic sum rule
\beq
\frac{1}{s_0^{n+\ell+1}} \int_4^{s_0} \rho(x)_{\ell,ABCD}\, x^n\, dx = \frac{N_c}{(2\pi)^4}\, \frac{\delta_{AC}\delta_{BD}}{\pi}\, \frac{\ell+1}{2^\ell\, \ell} \frac{B(\ell+1,\ell+1)}{n+\ell+1} , \ \ n\ge 0
\eeq
It might be interesting to note that for $\ell\gg 1$ a standard Stirling approximation shows that the right hand side is very small
\beq
\frac{1}{s_0^{n+\ell+1}} \int_4^{s_0} \rho(x)_{\ell,ABCD}\, x^n\, dx \simeq \frac{N_c}{(2\pi)^4}\, \frac{\delta_{AC}\delta_{BD}}{\pi}\, \sqrt{\frac{\pi}{l}}\frac{1}{2^{3\ell+1}}  \frac{1}{n+\ell+1} , \ \ n\ge 0, \ell\gg 1 
\eeq
This is a reasonable approximation even if $\ell$ is small.

\subsubsection{Energy momentum tensor}\label{sec:FFTmunu}

The energy momentum tensor is one of the most studied operators in Quantum Field Theory. For recent results in QCD see \cite{narison_2004} and also the very recent paper \cite{caronhuot2023twopoint} that uses energy momentum tensor correlators to study glueballs. Now we are interested in the vacuum polarization
\beq
\Pi_2(s) = i\int \frac{d^4x}{(2\pi)^4}\,e^{iqx}\,  \bra{0}\hat{T}\{ T^{--}(x) T^{++}(0)\}\ket{0}
\eeq
From \cite{Zoller_2012} we find
\beqa
i\int d^4x\,e^{iqx}\, \bra{0}\hat{T}\{ T^{\mu\nu}(x) T^{\rho\sigma}(0)\}\ket{0} &\simeq& C_0^T(s)\left\{ P^{\mu\rho}P^{\nu\sigma} +P^{\mu\sigma}P^{\nu\rho} -\frac{2}{3}P^{\mu\nu}P^{\rho\sigma}\right\} \non \\
&& + C_0^{S}(s)\, P^{\mu\nu}P^{\rho\sigma}
\eeqa 
with
\beqa
P^{\mu\nu} &=& q^2 \eta^{\mu\nu} -q^\mu q^\nu \\
C_0^T(s) &=& \frac{1}{16\pi^2}\left[-\frac{n_g}{10}-\frac{1}{20}n_f d_R + \frac{\alpha_s}{\pi} \left\{\frac{1}{18}C_A n_g-\frac{7}{144}n_fT_F n_g\right\}\right] \, \ln\left(-\frac{s}{\mu^2}\right) \nonumber \\ 
&=& \frac{1}{16\pi^2}\left(-\frac{11}{10}+\frac{17}{18}\frac{\alpha_s}{\pi}\right)\, \ln\left(-\frac{s}{\mu^2}\right)
\eeqa
From here, and working in the frame where $q=(\sqrt{s},0,0,0)$ we obtain
\beq
\Pi_2(s) = \frac{1}{(2\pi)^4}\ 2 s^2C_0^T(s)
\eeq 
Notice that the power $s^2$ is determined by the symmetry and conservation law, therefore also appears as $s\rightarrow 0$ implying that $\frac{1}{s^2}\Pi_2(s)$ is analytic.
For $SU(N_c)$ gauge theory with $n_f$ flavors in the fundamental we have $d_R=N_c$, $n_g=N_c^2-1$, $C_A=2T_FN_c$, $C_F=\frac{T_Fn_g}{N_c}$, $T_F=\half$. If we set $n_f=0$ these results agree with those of \cite{caronhuot2023twopoint}. In any case, for $N_c=3$ and $n_f=2$, the results of \cite{Zoller_2012} reduce to  
\beq
\Pi_2(s) =\frac{1}{(2\pi)^4}\  \frac{1}{8\pi^2}\left(-\frac{11}{10}+\frac{17}{18}\frac{\alpha_s}{\pi}\right)\,s^2\, \ln\left(-\frac{s}{\mu^2}\right) + \ldots 
\eeq
namely
\beq
\rho_2(s) =_{s\rightarrow\infty} \frac{1}{(2\pi)^4}\  \frac{1}{4\pi}\left(\frac{11}{10}-\frac{17}{18}\frac{\alpha_s}{\pi}\right)\,s^2\, \ln\left(-\frac{s}{\mu^2}\right)
\eeq
giving the FESR
\beqa
\frac{1}{s_0^{n+3}} \int_4^{s_0} \rho_2(x) x^n dx&=&\frac{1}{(2\pi)^4}\frac{1}{4\pi}\frac{1}{n+3}\left(\frac{11}{10}-\frac{17}{18}\frac{\alpha_s}{\pi}\right), \ \ \ n\ge -2
\eeqa
that we impose for $n=-2,-1$. 

\subsubsection{Four quark scalar current}

\begin{figure}[H]
\centering
\begin{subfigure}{0.35\textwidth}
\centering
\includegraphics[width=\textwidth]{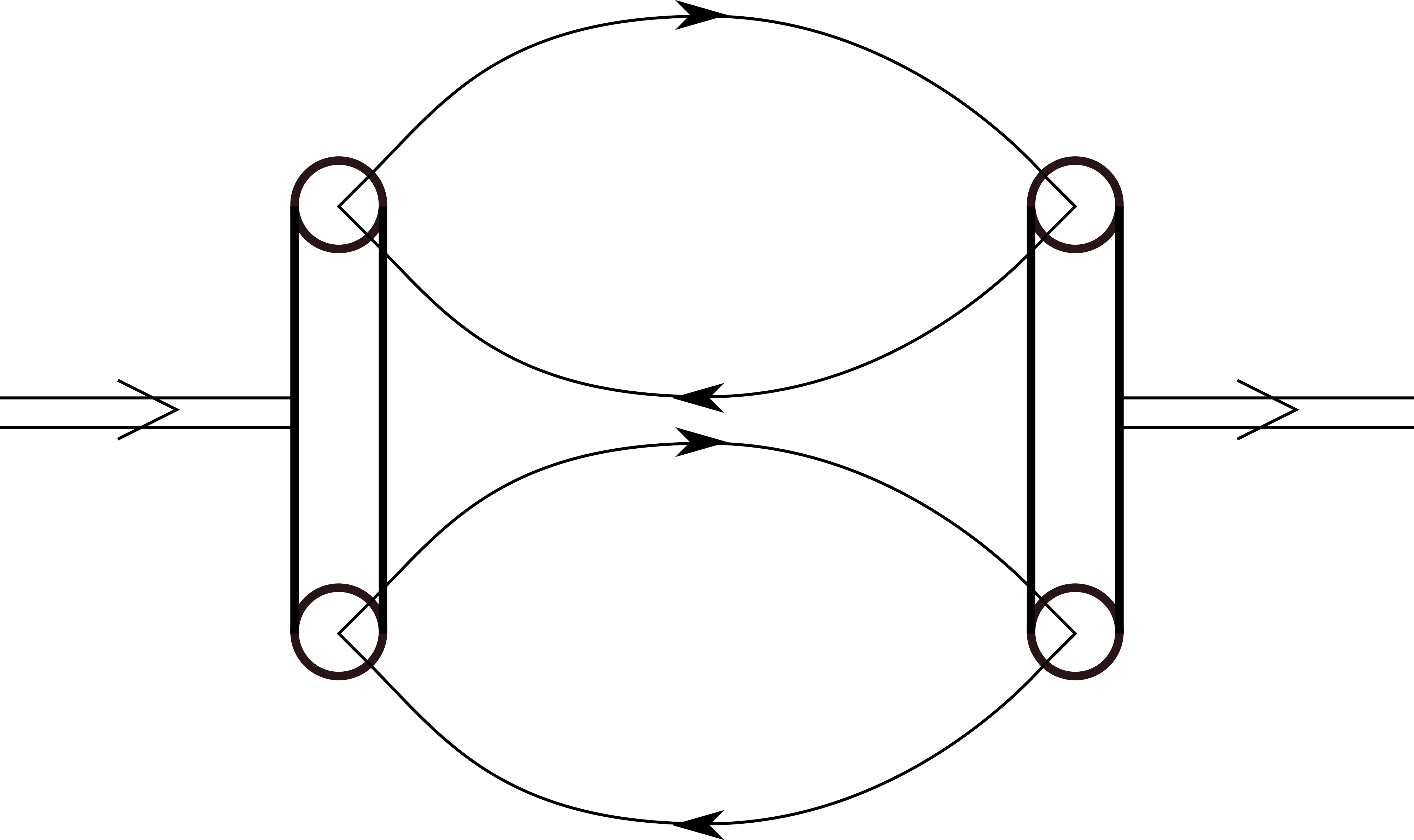}
\caption{}
\end{subfigure}
\hspace{0.5cm}
\begin{subfigure}{0.35\textwidth}
\centering
\includegraphics[width=\textwidth]{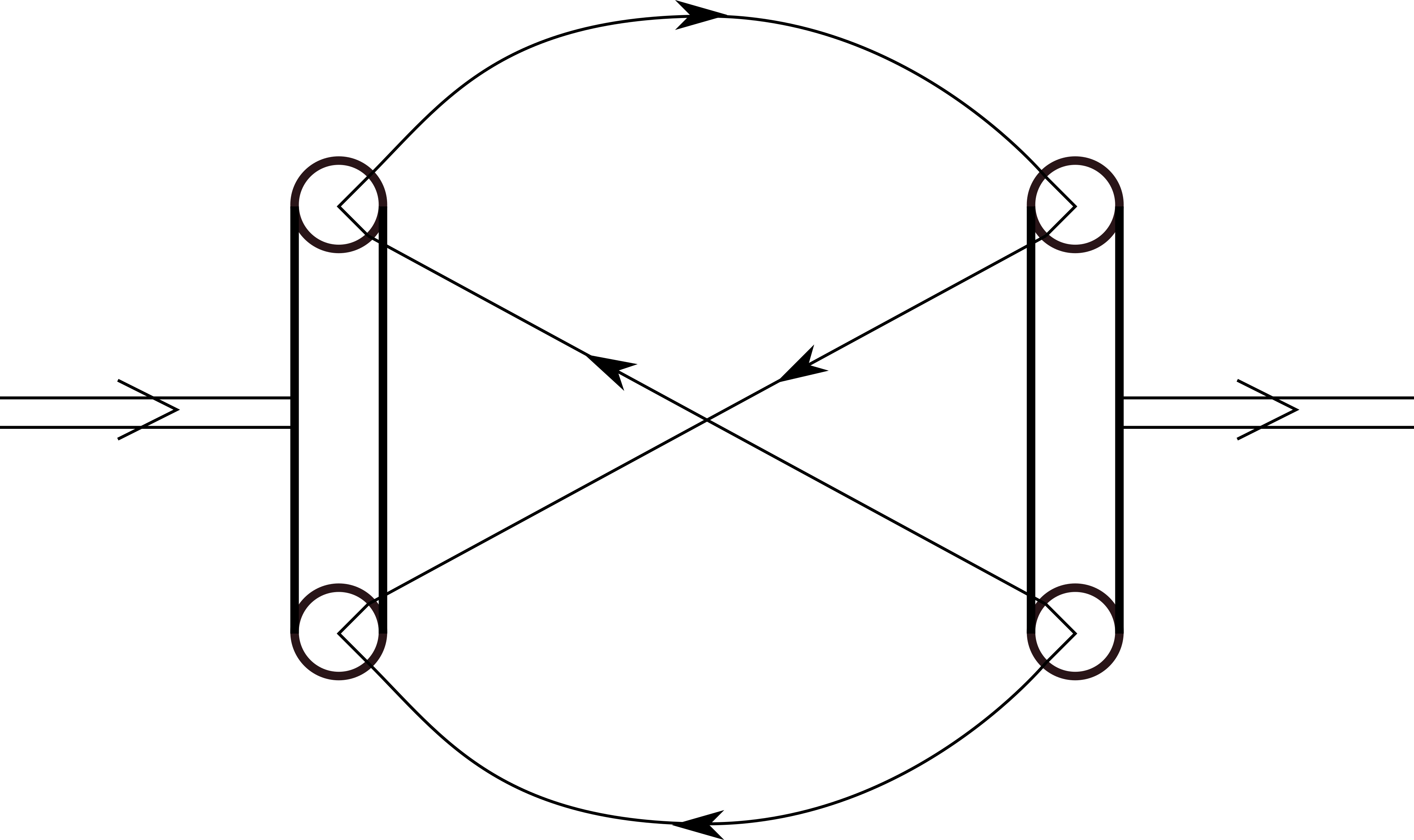}
\caption{}
\end{subfigure}
\caption{Standard Feynman diagrams contributing to the two point function of the four fermion current at lowest order. }
\label{jjdiags2}
\end{figure}

Consider the scalar current relevant for the isospin $I=2$ case
\beq
j_{ABCD} = (\bar{\psi}_A\psi_B)(\bar{\psi}_C\psi_D)
\eeq
where the fields in parenthesis are color singlets. The indices $A,B,C,D=1\ldots N_f$ are flavor indices. We are interested in the correlator 
\beq
\Pi(s) = i \int \frac{d^dx}{(2\pi)^d}\, e^{-iPx}\, \bra{0}\hat{T}\left\{(j_{A'B'C'D'}(x))^\dagger j_{ABCD}(0) \right\} 
\eeq
where $s=P^2$. Using Wick's theorem, in the free theory, namely the Feynman diagrams of fig.\ref{jjdiags2} where the blobs represent current insertions, we find
\beq
\begin{aligned}
&\Pi(s) =  i \int \frac{d^dx}{(2\pi)^d}\, e^{-iPx}\,\times  \\
& \left\{N_c^2 \left(\delta_{AA'}\delta_{BB'}\delta_{CC'}\delta_{DD'}+\delta_{AC'}\delta_{CA'}\delta_{BD'}\delta_{DB'}\right) \tr[\Delta_F(-x)\Delta_F(x)] \tr[\Delta_F(-x)\Delta_F(x)] \right.\\
&\left. -N_c \left(\delta_{AC'}\delta_{CA'}\delta_{BB'}\delta_{DD'}+\delta_{AA'}\delta_{CC'}\delta_{BD'}\delta_{DB'}\right) \tr[\Delta_F(-x)\Delta_F(x)\Delta_F(-x)\Delta_F(x)]\right\}
\end{aligned}
\eeq
where $\Delta_F(x)$ is the fermion propagator 
\beq
\Delta_F(x) = \int\frac{d^dk}{(2\pi)^d} \frac{i(\slashed{k}+m)}{k^2-m^2+i\epsilon}\, e^{-ikx} = (i \slashed{\partial}_x +m) \Delta(x) 
\eeq
where $\Delta(x)$ is the scalar propagator. This results in 
\beqa
\Pi(s) &=&  i \int \frac{d^dx}{(2\pi)^d}\, e^{-iPx}\,[\partial_\mu \Delta(x)\partial^\mu\Delta(x)+m^2\Delta(x)^2]^2\times  \non \\
&& \left\{16 N_c^2 \left(\delta_{AA'}\delta_{BB'}\delta_{CC'}\delta_{DD'}+\delta_{AC'}\delta_{CA'}\delta_{BD'}\delta_{DB'}\right) \right. \\
&& \left. -4 N_c \left(\delta_{AC'}\delta_{CA'}\delta_{BB'}\delta_{DD'}+\delta_{AA'}\delta_{CC'}\delta_{BD'}\delta_{DB'}\right) \right\} \non 
\eeqa
Since we consider the case of light quarks, we ignore the quark mass and use the massless propagator and the general formula
\beq
\int\frac{d^dk}{(2\pi)^d} \frac{e^{-ikx}}{(k^2+i\epsilon)^a} = i\frac{\Gamma(\frac{d}{2}-a)}{\Gamma(a)}  \frac{(-1)^a}{4^a\pi^{\frac{d}{2}}(-x^2)^{\frac{d}{2}-a}}
\eeq 
we obtain
\beq
\int \frac{d^dx}{(2\pi)^d}\, e^{-iPx}\,[\partial_\mu \Delta(x)\partial^\mu\Delta(x)]^2 = -i \frac{\Gamma(\frac{d}{2})^4}{16\pi^{2d}}
\frac{\Gamma(2-\frac{3}{2}d)}{\Gamma(2d-2)}\frac{1}{4^{2d-2}\pi^{\frac{d}{2}}(-P^2-i\epsilon)^{2-\frac{3}{2}d}}
\eeq
Replacing $d=4-\epsilon$, expanding in $\epsilon$ and renormalizing we get
\beqa
\Pi(s) &=& i \int \frac{d^4x}{(2\pi)^4}\, e^{-iPx}\, \bra{0}\hat{T}\left\{(j_{A'B'C'D'}(x))^\dagger j_{ABCD}(0) \right\} \non\\
&=& -\left\{16 N_c^2 \left(\delta_{AA'}\delta_{BB'}\delta_{CC'}\delta_{DD'}+\delta_{AC'}\delta_{CA'}\delta_{BD'}\delta_{DB'}\right) \right. \non\\
&& \left. -4 N_c \left(\delta_{AC'}\delta_{CA'}\delta_{BB'}\delta_{DD'}+\delta_{AA'}\delta_{CC'}\delta_{BD'}\delta_{DB'}\right) \right\} \non\\
&& \frac{1}{(2\pi)^4} \frac{1}{2^{18}\,3^2\, 5\,\pi^6}\, s^4\ln\left(-\frac{s}{\mu^2}\right)
\eeqa
One check is that for the currents $j_\pm=\frac{1}{\sqrt{2}} \sum_{\Gamma=1,\gamma_5} (\bar{s}\Gamma s)\,(\bar{u}\Gamma u\pm\bar{d}\Gamma d)$ we get
\beq
\Pi(s) = - 2\frac{16 N_c^2}{(2\pi)^4 }\frac{1}{2^{18}\,3^2\, 5\,\pi^6}\, s^4\ln\left(-\frac{s}{\mu^2}\right) = - \frac{1}{(2\pi)^4} \frac{1}{40960\, \pi^6}\, s^4\ln\left(-\frac{s}{\mu^2}\right)
\eeq
in agreement with \cite{narison_2004,Latorre:1985uy} 

In our case we are interested in
\beqa
j(x) &=& (\bar{u}(x)d(x))\,(\bar{u}(x)d(x)) \\
\cO(P) &=& \int\frac{d^4 x}{(2\pi)^4} e^{-iPx} \ j(x) 
\eeqa
which gives
\beqa
\Pi(s) &=& i \int\frac{d^4 x}{(2\pi)^4} e^{-iqx} \bra{0} \hat{T}\left\{j(x)j(0)\right\}\ket{0} \non\\ 
&=& -8N_c(4N_c-1)\frac{1}{(2\pi)^4}\, \frac{1}{2^{18}\, 3^2\, 5}\, s^4\ln(-\frac{s}{\mu^2}) \non\\
&=& -\frac{1}{(2\pi)^4}\,\frac{11}{491520\,\pi^6}\,  s^4\ln(-\frac{s}{\mu^2}) 
\eeqa
and from here
\beq\label{rhoaa}
\rho(s) = 2\Im \Pi(s+i\epsilon) = \frac{1}{(2\pi)^4} \frac{11}{245760\,\pi^5}\ s^4
\eeq
One may suspect that the result may have some ambiguity from renormalization, but that is not the case since $\rho(s)$ should be finite. To check that, one can alternatively compute
\beq
\cO_q\ket{0} =\! \int\!\prod_{j=1}^4 \frac{d^3 k_j}{(2\pi)^3\sqrt{2\omega_j}}\ \delta^{(4)}({\sum_{j=1}^4} k_j-q)\!\!\! \sum_{s_1,s_2,s_3,s_4}\!\!\!\! 
(\bar{u}^{s_1}_{k_1} v^{s_2}_{k_2}) (\bar{u}^{s_3}_{k_3} v^{s_4}_{k_4}) 
a^\dagger_{k_1s_1a}b^\dagger_{k_2s_2a}a^\dagger_{k_3s_3b}b^\dagger_{k_4s_4b}\ket{0}
\eeq
where $a^\dagger_{k_1s_1a}$ creates a $u$-quark with momentum $k_1$ polarization $s_1$ and color index $a$, and $b^\dagger_{k_1s_1a}$ does the same for an anti $d$-quark.
Notice that the time part of the delta function (total energy fixed) implies that $|\vec{k}_j|< q^0$ and therefore the integrals are over finite size balls giving a finite value to $\rho(s)$ in 
\beq
\bra{0} \cO^\dagger(P') \cO(P) \ket{0} = \rho(P^2)\, \delta^{(4)}(P'-P)
\eeq   
Using the ``multi-ball integrals" in the appendix one can evaluate $\rho(s)$ from this last equation giving 
\beqa
\rho(s) &=& \frac{8N_c(4N_c-1)}{(2\pi)^{12}} \int \prod_{j=1}^4\frac{d^3 k_j}{(2\pi)^3\,2\omega_j}\ \delta^{(4)}(\sum_{j=1}^4 k_j-q)  (k_1k_2)(k_3k_4) \non\\
&=&  \frac{1}{(2\pi)^4} \frac{11}{245760\,\pi^5}\ s^4
\eeqa
in agreement with eq.(\ref{rhoaa}). Now, using the value of $\Pi(s)$ we can write the sum rule
\beq
\frac{1}{s_0^{n+5}} \int_4^{s_0} \rho_{S2}(x)\, x^n\, dx = \frac{1}{(2\pi)^4} \frac{11}{245760\,\pi^5}\frac{1}{n+5}, \ \ n\ge 0  
\eeq

\section{Numerical method} \label{sec:nummeth}

In section \ref{sec:gtbsummary}, we have explained the main ideas of the gauge theory bootstrap and now we are going to describe the numerical implementation. This section is perhaps better read in conjunction with the mathematica and matlab programs provided with this paper.  Let us briefly recall the steps involved in this framework:
\begin{itemize}
\item We find an (approximate) parameterization of the space of all S-matrices that satisfy the constraints of analyticity, crossing and global symmetries in terms of a large but finite number of real parameters. Within that space we identify a finite volume convex set of all S-matrices that satisfy the unitarity constraints.
\item We require that the partial waves, in the Mandelstam region, approximate the Weinberg model ($\chi$PT at tree level) within a certain tolerance and the spectral density has the correct kinematics at low energy. This reduces the allowed space of amplitudes to a much smaller one.
\item  We introduce information of the UV gauge theory (QCD) through the sum rules and asymptotic form factors. The allowed space of amplitudes is further reduced, with the amplitudes on the boundary displaying behaviors (phase shifts, form factors) agreeing with experiments.
\end{itemize}
To understand the resulting convex space we maximize linear functionals using tools of convex optimization, and examine the extremal amplitudes/form factors living on the boundary of the allowed space. The idea is that given enough physical constraints from low energy and high energy, the allowed space of amplitudes consistent with the bootstrap constraints should become a very small set consisting of amplitudes with similar behavior, and eventually allowing to identify a unique theory. In principle, the maximization functional should be irrelevant in achieving this. In practice, we focus on functionals given by linear combinations of partial waves evaluated in the unphysical region (Mandelstam triangle) to carry out this projection, and use the low energy physical information $f_{\pi}=92$ MeV to identify the theory that we wish to compute. It would however be interesting for future work to check different functionals for this purpose.

In this section, we describe in detail the numerical method to carry out the gauge theory bootstrap computation. We have also provided a Mathematica notebook {\tt GTB\_numerics.nb} that implements the numerical computation described in this section, as well as a Matlab notebook {\tt GTB\_numerics.m} that contains the optimization program for the bootstrap. To carry out the optimization, we use CVX \cite{cvx,gb08,boyd2004convex}, a package for specifying and solving convex programs \cite{cvx,gb08}. Interested readers can download the notebooks, install the CVX package\footnote{We use the MOSEK solver \cite{mosek} for fast numerical optimizations.}, and run the programs to reproduce the results we present in this paper. The programs can also be found in the github repository for this paper: \href{https://github.com/hyfysics/gauge-theory-bootstrap/tree/main/arxiv-2403.10772}{\color{blue}arxiv-2403.10772}.

\subsection{Kernels for partial wave computations}\label{kernel section}

To impose the unitarity condition \eqref{uni}, evaluate the maximization functionals \eqref{fnals}, and imposing the low energy matching \eqref{chiralratio}, we need to evaluate the partial waves $f^I_{\ell}(s)$ from the bootstrap variables $\{T_0,\sigma_{1,2},\rho_{1,2}\}$. In this section, we compute such expression:
\begin{equation}\label{fx}
\begin{aligned}
f^I_\ell(s)=&\hat{\Phi}^I_\ell(s)T_0+\int_4^{\infty}\!\! dx\Phi^{I}_{\ell,1}(s,x)\sigma_1(x)+\int_4^{\infty}\!\! dx\Phi^{I}_{\ell,2}(s,x)\sigma_2(x)\\	
+&\int_4^{\infty}\!\! dxdy\tilde{\Phi}^{I}_{\ell,1}(s,x,y)\rho_1(x,y)+\int_4^{\infty}\!\! dxdy\tilde{\Phi}^{I}_{\ell,2}(s,x,y)\rho_2(x,y)
\end{aligned}
\end{equation}
where $f^I_{\ell}(s)$ is an analytic function of $s$ with cut $s>4,s<0$.

To arrive at the expression \eqref{fx}, the idea is to compute the partial waves projection of different parts of the amplitudes (single and double dispersion analytic functions of the variables $s,t,u$) and combine them into the appropriate combinations with fixed isospins.

We start by rewriting the Mandelstam representation of the amplitude \eqref{Adef} as
\begin{equation}\label{Arewrite}
A(s,t,u) = T_0 + H_1(s) + H_2(t)+H_2(u)+ \tilde{H}_1(s,t)+\tilde{H}_1(s,u)+\tilde{H}_2(t,u)
\end{equation}
with
\begin{equation}\label{HHtdef}
\begin{aligned}
H_i(\nu)&=\frac{1}{\pi}\int_4^{\infty} \!\! dx \left[\frac{1}{x-\nu}-\frac{1}{x-\nu_0}\right]\sigma_i(x)\\
\tilde{H}_i(\nu_1,\nu_2)&=\frac{1}{\pi^2}\int_4^{\infty} \!\! dx dy\left[\bigg(\frac{1}{x-\nu_1}-\frac{1}{x-\nu_0}\bigg)\bigg(\frac{1}{y-\nu_2}-\frac{1}{y-\nu_0}\bigg)\right]\rho_i(x,y)
\end{aligned}
\end{equation}
where we simply subtracted each integration kernel at a point $s,t,u=\nu_0$ and redefined the bootstrap variables $\{T_0,\sigma_1,\sigma_2,\rho_1,\rho_2\}$. Keep in mind the symmetry $\rho_2(x,y)=\rho_2(y,x)$. Neglecting the indices $i$, let us consider the partial wave projection for each part of the amplitude \eqref{Arewrite} and write down the following Froissart-Gribov type formula:
\begin{subequations}\label{fsingle}
\begin{eqnarray}
f^{(s)}_{\ell}(s)&\equiv& \frac{1}{4}\int_{-1}^{1}d\mu P_{\ell}(\mu)H(s)dx=\frac{\delta_{l0}}{2}\mathcal{K}(s,x)\sigma(x)\\
f^{(t)}_{\ell}(s)&\equiv& \frac{1}{4}\int_{-1}^{1}d\mu P_{\ell}(\mu)H(t)dx=\int_4^{\infty}\!\! dx\mathcal{A}(s,x)\sigma(x)-\frac{\delta_{l0}}{2}\int \!\! dx\mathcal{W}_0(x)\sigma(x)\\
f^{(u)}_{\ell}(s)&\equiv& \frac{1}{4}\int_{-1}^{1}d\mu P_{\ell}(\mu)H(u)dx=(-1)^l\int_4^{\infty}\!\! dx\mathcal{A}(s,x)\sigma(x)-\frac{\delta_{l0}}{2}\int \!\! dx\mathcal{W}_0(x)\sigma(x)\nonumber\\
\end{eqnarray}
\end{subequations}
and
\begin{subequations}\label{fdouble}
\begin{eqnarray}
f^{(s,t)}_{\ell}(s)&\equiv&\frac{1}{4}\int_{-1}^{1}d\mu P_{\ell}(\mu)\tilde{H}(s,t)dx\nonumber\\
&=&\int_4^{\infty}\!\! dxdy\mathcal{A}(s,y)\mathcal{K}(s,x)\rho(x,y)-\frac{\delta_{l,0}}{2}\mathcal{K}(s,x)\mathcal{W}_0(y)\rho(x,y)\\
f^{(s,u)}_{\ell}(s)&\equiv&\frac{1}{4}\int_{-1}^{1}d\mu P_{\ell}(\mu)\tilde{H}(s,u)dx\nonumber\\
&=&(-1)^l\int_4^{\infty}\!\! dxdy\mathcal{A}(s,y)\mathcal{K}(s,x)\rho(x,y)-\frac{\delta_{l,0}}{2}\mathcal{K}(s,x)\mathcal{W}_0(y)\rho(x,y)\\
f^{(t,u)}_{\ell}(s)&\equiv&\frac{1}{4}\int_{-1}^{1}d\mu P_{\ell}(\mu)\tilde{H}(t,u)dx\nonumber\\
&=&\int_4^{\infty}\!\! dxdy \mathcal{A}(s,y)\mathcal{W}(s,y,x)\rho(y,x)+(-1)^l\int_4^{\infty}\!\! dxdy \mathcal{A}(s,y)\mathcal{W}(s,y,x)\rho(x,y)\nonumber\\&&+\frac{\delta_{l,0}}{2}\mathcal{W}_0(x,y)\rho(x,y)
\end{eqnarray}
\end{subequations}
where on the LHS we use the superscripts to indicate the complex variables of the analytic functions $H,\tilde{H}$ being projected.
In the expressions \eqref{fdouble} and \eqref{fsingle}, we have defined the kernels:
\begin{subequations}\label{kernels}
\begin{eqnarray}
\mathcal{A}(s,x)&=&\frac{1}{\pi(s-4)}Q_{\ell}\Big(1+\frac{2x}{s-4}\Big)\\
\mathcal{W}(s,y,x)&=&\frac{1}{\pi}\left[ \frac{1}{x+y+s-4}-\frac{1}{x-\nu_0}\right]\\
\mathcal{W}_{00}(x,y)&=&\frac{1}{\pi^2}\frac{1}{(x-\nu_0)(y-\nu_0)}\\
\mathcal{W}_0(x)&=&\frac{1}{\pi}\frac{1}{x-\nu_0}\\
\mathcal{K}(s,x)&=&\frac{1}{\pi}\Big(\frac{1}{x-s}-\frac{1}{x-\nu_0}\Big)\label{Kkernal}
\end{eqnarray}
\end{subequations}
with $x,y\in(4,\infty)$, $s$ complex and $Q_\ell$ is the second Legendre function satisfying
\begin{equation}
Q_{\ell}(z)=\frac{1}{2}\int_{-1}^1\!\! d\mu\frac{P_{\ell}(\mu)}{z-\mu}
\end{equation}
Notice that the kernel $\mathcal{K}$ in \eqref{Kkernal} has an imaginary part for $s>4$ and therefore will give rise to an imaginary part of the partial waves in the physical region.

The expressions \eqref{fsingle} and \eqref{fdouble} allow us to define the kernels
\begin{equation}
\begin{aligned}
\Phi^{(s)}_{\ell}(s,x)&=\frac{\delta_{l0}}{2}\mathcal{K}(s,x)\\
\Phi^{(t)}_{\ell}(s,x)&=\mathcal{A}(s,x)-\frac{\delta_{l0}}{2}\mathcal{W}_0(x)\\
\Phi^{(u)}_{\ell}(s,x)&=(-1)^l\mathcal{A}(s,x)-\frac{\delta_{l0}}{2}\mathcal{W}_0(x)\label{Phikernels}
\end{aligned}
\end{equation}
\begin{equation}
\begin{aligned}
\tilde{\Phi}^{(s,t)}_\ell(s,x,y)&=\mathcal{A}(s,y)\mathcal{K}(s,x)-\frac{\delta_{l,0}}{2}\mathcal{K}(s,x)\mathcal{W}_0(y)\\
\tilde{\Phi}^{(s,u)}_\ell(s,x,y)&=(-1)^l\mathcal{A}(s,y)\mathcal{K}(s,x)-\frac{\delta_{l,0}}{2}\mathcal{K}(s,x)\mathcal{W}_0(y)\\
\tilde{\Phi}^{(t,u)}_\ell(s,x,y)&=\mathcal{A}(s,y)\mathcal{W}(s,y,x)\mathcal{T}_{x,y}+(-1)^l\mathcal{A}(s,y)\mathcal{W}(s,y,x)+\frac{\delta_{l,0}}{2}\mathcal{W}_0(x,y)\label{Phikernels2}
\end{aligned}
\end{equation}
where $\mathcal{T}_{x,y}$ interchanges the variables of a double spectral density $\rho(y,x)=\mathcal{T}_{x,y}\rho(x,y)$.
Namely, integrating the spectral densities $\sigma,\rho$ with these kernels \eqref{Phikernels}, \eqref{Phikernels2} give the partial waves \eqref{fsingle} and \eqref{fdouble} by projecting out different parts of the amplitudes to a definite angular momentum $\ell$.

With these basic kernels, we can now make combinations which gives rise to kernels to be integrated with the spectral densities $\sigma_{1,2}$ and $\rho_{1,2}$ respectively. We have first the kernels for the single spectral densities:
\begin{subequations}\label{K1}
\begin{eqnarray}
\Phi^{(s)}_{\ell,1}(s,x)&=&\Phi^{(s)}_\ell(s,x)\\ \Phi^{(t)}_{\ell,1}(s,x)&=&\Phi^{(t)}_\ell(s,x)\\
\Phi^{(u)}_{\ell,1}(s,x)&=&\Phi^{(u)}_\ell(s,x)\\
\Phi^{(s)}_{\ell,2}(s,x)&=&\Phi^{(t)}_\ell(s,x)+\Phi^{(u)}_\ell(s,x),\\ \Phi^{(t)}_{\ell,2}(s,x)&=&\Phi^{(s)}_\ell(s,x)+\Phi^{(u)}_\ell(s,x),\\
\Phi^{(u)}_{\ell,2}(s,x)&=&\Phi^{(s)}_\ell(s,x)+\Phi^{(t)}_\ell(s,x)
\end{eqnarray}
\end{subequations}
where the subscript $1,2$ indicate the corresponding single spectral density $\sigma_{1,2}$. Similarly, we have
\begin{subequations}\label{K2}
\begin{eqnarray}
\tilde{\Phi}^{(s,t)}_{\ell,1}(s,x,y)&=&\tilde{\Phi}^{(s,t)}_{\ell}(s,x,y)+\tilde{\Phi}^{(s,u)}_{\ell}(s,x,y)\\
\tilde{\Phi}^{(t,s)}_{\ell,1}(s,x,y)&=&\tilde{\Phi}^{(s,t)}_{\ell}(s,x,y)\mathcal{T}_{x,y}+\tilde{\Phi}^{(t,u)}_{\ell}(s,x,y)\\
\tilde{\Phi}^{(u,t)}_{\ell,1}(s,x,y)&=&\tilde{\Phi}^{(t,u)}_{\ell}(s,x,y)\mathcal{T}_{x,y}+\tilde{\Phi}^{(s,u)}_{\ell}(s,x,y)\mathcal{T}_{x,y}\\
\tilde{\Phi}^{(s,t)}_{\ell,2}(s,x,y)&=&\tilde{\Phi}^{(t,u)}_{\ell}(s,x,y)\\
\tilde{\Phi}^{(t,s)}_{\ell,2}(s,x,y)&=&\tilde{\Phi}^{(s,u)}_{\ell}(s,x,y)\\
\tilde{\Phi}^{(u,t)}_{\ell,2}(s,x,y)&=&\tilde{\Phi}^{(s,t)}_{\ell}(s,x,y)\mathcal{T}_{x,y}
\end{eqnarray}
\end{subequations}
where again, the subscript $1,2$ indicate the corresponding double spectral density $\rho_{1,2}$ that it should be multiplied with. The superscript indicate the complex variables involves in the analytic function $\tilde{H}$ before projecting out to a partial wave. For example, the superscript $(t,s)$ arises from the integration kernel after projecting out a fixed angular momentum for analytic function $\tilde{H}(t,s)$. This switch of complex variables in the analytic functions $\tilde{H}$ is due to the interchange of $s,t,u$ variables in \eqref{a10} for defining amplitudes of a fixed isospin.

The final kernels for computing the partial waves of fixed isospin from the spectral densities $\sigma_{1,2},\rho_{1,2}$ is given by (we omit the variables $s,x,y$ here)
\begin{equation}
\tilde{\Phi}^{I}_{\ell,i}=\begin{bmatrix}
\tilde{\Phi}^{0}_{\ell,i}\\
\tilde{\Phi}^{1}_{\ell,i}\\
\tilde{\Phi}^{2}_{\ell,i}
\end{bmatrix}=\begin{bmatrix}
3 \tilde{\Phi}^{(s,t)}_{\ell,i}+\tilde{\Phi}^{(t,s)}_{\ell,i}+\tilde{\Phi}^{(u,t)}_{\ell,i}\\
\tilde{\Phi}^{(t,s)}_{\ell,i}+\tilde{\Phi}^{(u,t)}_{\ell,i}\\
\tilde{\Phi}^{(t,s)}_{\ell,i}-\tilde{\Phi}^{(u,t)}_{\ell,i}
\end{bmatrix},\;\;i=1,2
\end{equation}
\begin{equation}
\Phi^{I}_{\ell,i}=\begin{bmatrix}
\Phi^{0}_{\ell,i}\\
\Phi^{1}_{\ell,i}\\
\Phi^{2}_{\ell,i}
\end{bmatrix}=\begin{bmatrix}
3 \Phi^{(s)}_{\ell,i}+\Phi^{(t)}_{\ell,i}+\Phi^{(u)}_{\ell,i}\\
\Phi^{(t)}_{\ell,i}+\Phi^{(u)}_{\ell,i}\\
\Phi^{(t)}_{\ell,i}-\Phi^{(u)}_{\ell,i}
\end{bmatrix},\;\;i=1,2
\end{equation}
as well as the coefficient to multiply the constant $T_0$
\begin{equation}
\hat{\Phi}^I_\ell=\begin{bmatrix}
\hat{\Phi}^{0}_{\ell,i}\\
\hat{\Phi}^{1}_{\ell,i}\\
\hat{\Phi}^{2}_{\ell,i}
\end{bmatrix}=\frac{\delta_{l0}}{2}\begin{bmatrix}
(N+2)\\0\\2
\end{bmatrix}
\end{equation}
The combinations here can be understood directly from the combinations in \eqref{a10} for defining amplitudes with fixed isopins.
Integrating the constant $T_0$, spectral densities $\sigma_{1,2}$ and $\rho_{1,2}$ according to \eqref{fx} give the analytic partial waves with definite isospins.

\subsection{Mapping to the circle}

For numerically computing the above integrals, we map the cut plane of $\nu$ (standing for $s,t,u$) onto a unit disk
\begin{equation}\label{znu}
z(\nu)=\frac{\sqrt{4-\nu_0}-\sqrt{4-\nu}}{\sqrt{4-\nu_0}+\sqrt{4-\nu}}
\end{equation}
where the point $\nu=\nu_0$ is mapped to the center of the unit $z$ disk.\footnote{this is the same map used in the $\rho$ variable parametrization of the S-matrix bootstrap \cite{Paulos:2017fhb}} The region $x\in(4,\infty)$ is mapped to the upper half circle of $z=e^{i\phi}, \phi\in\left[0,\pi\right]$:
\begin{equation}\label{xphi}
x(\phi)=\nu_0+\frac{8-2\nu_0}{1+\cos(\phi)},\;\; x\in(4,\infty)
\end{equation}
Under this map, the integrals in \eqref{fx} simply become
\begin{equation}
\int_4^{\infty}\!\! dx\to\int_0^{\pi}\! x'(\phi)d\phi
\end{equation}
where
\begin{equation}\label{xp}
x'(\phi)=\frac{2(4-\nu_0)\sin\phi}{(1+\cos\phi)^2}
\end{equation}
The variables $x,y$ in the kernels \eqref{kernels} can be simply replaced by $x(\phi_1),y(\phi_2)$. We will rename the spectral densities $\sigma_{1,2}(\phi_1),\rho_{1,2}(\phi_1,\phi_2)$ accordingly.

Under the map \eqref{znu}, the region above the cut is mapped to the upper half circle of $z$ and below the cut mapped to the lower half circle of $z$. Let us now look at the kernel $\mathcal{K}$ which has a jump across the cut $s>4$. Under this map, we can then replace
\begin{equation}
\int_4^{\infty}\! dx\mathcal{K}(\nu,x)\to\int_0^{\pi}\! d\phi K(z,\phi)
\end{equation}
with
\begin{equation}\label{kernel2}
K(z,\phi)=\frac{2z\sin\phi}{\pi+\pi z^2-2\pi z\cos\phi}
\end{equation}
where we absorbed \eqref{xp} into the definition of $K$. The kernel $K$ allows to compute analytic functions on the unit disk of $z$ (and vanishes at $z=0$) using their imaginary part on the upper half circle:
\begin{equation}
g(z)=\int_0^{\infty}d\phi K(z,\phi)\text{Im}g(\phi)
\end{equation}
Evaluating the real part $\text{Re}\,g(\phi)$ on the upper half circle requires computing a principal part integral
\beq\label{disp}
\begin{aligned}
\text{Re}g(\phi) = \fint_0^\pi d\phi_1\hat{K}(\phi,\phi_1)\text{Im}g(\phi_1)=
\frac{1}{\pi} \fint_0^\pi d\phi_1 \frac{\sin\phi_1}{\cos \phi-\cos\phi_1} \text{Im}g(\phi_1)
\end{aligned}
\eeq 
To perform this integral we extend $\text{Im}g(\phi_1)$ to the range $-\pi\le \phi_1\le \pi$ by assuming that it is anti-symmetric $\text{Im}g(-\phi_1)=- \text{Im}g(\phi_1)$ so that the integrand is symmetric. For the spectral densities $\sigma_{1,2}(\phi_1)$, $\rho_{1,2}(\phi_1,\phi_2)$ this agrees with them being the imaginary part of an amplitude that changes sign across the cut.

\medskip

The expression \eqref{fx} for computing the physical partial waves now becomes ($I=0,1,2$)
\begin{equation}\label{pwintegral}
\begin{aligned}
f^I_\ell(s)=&\hat{\Phi}^I_\ell(s)T_0+\int_0^{\pi}\!\! d\phi_1\Phi^{I}_{\ell,1}(s,\phi_1)\sigma_1(\phi_1)+\int_0^{\pi}\!\! d\phi_1\Phi^{I}_{\ell,2}(s,\phi_1)\sigma_2(\phi_1)\\	
+&\int_0^{\pi}\!\! d\phi_1d\phi_2\tilde{\Phi}^{I}_{\ell,1}(s,\phi_1,\phi_2)\rho_1(\phi_1,\phi_2)+\int_0^{\pi}\!\! d\phi_1d\phi_2\tilde{\Phi}^{I}_{\ell,2}(s,\phi_1,\phi_2)\rho_2(\phi_1,\phi_2)
\end{aligned}
\end{equation}
where we just renamed the kernels as $\Phi,\tilde{\Phi}$ when using variables $\phi_1,\phi_2$ and absorb the Jacobian \eqref{xp} into the definitions of $\Phi$ and $\tilde{\Phi}$.

\bigskip

Consider now the analytic form factor \eqref{Fdisp} which can be parameterized by its imaginary part $\text{Im}F(x)$ (neglecting the indices for now) above the cut $s>4$. The same map \eqref{znu} and \eqref{xphi} applies and we can write its real part $\text{Re}F(s)$ in the physical region $s>4$ (denoted as $\text{Re}F(\phi)$) using the above dispersion relation \eqref{disp}
\begin{equation}\label{FFana}
\text{Re}F(\phi)=1+\fint_0^\pi d\phi_1\hat{K}(\phi,\phi_1)\text{Im}F(\phi_1)+\int_{0}^{\pi} K_0(\phi_1)\text{Im}F(\phi_1)
\end{equation}
with
\begin{equation}
K_0(\phi)=\frac{\nu_0\sin\phi}{\pi(8-\nu_0+\nu_0\cos\phi)}
\end{equation}
computing the difference between subtracting at $s=0$ and $s=\nu_0$. \footnote{recall the normalization $F(0)=1$ and that the point $s=\nu_0$ is mapped to the center of the disk where the kernel \eqref{kernel2} vanishes} Similarly, the spectral density $\rho^I_{\ell}$ can also be parametrized by the angle $\phi_1$ as
\begin{equation}
\rho^I_{\ell}(x)\to\rho^I_{\ell}(\phi)\;.
\end{equation}

\medskip

To summarize, after mapping to the circle, all our bootstrap variables \eqref{ffApara} are parameterized by the angles $\phi_1,\phi_2\in (0,\pi)$: (where $\rho_2(\phi_1,\phi_2)=\rho_2(\phi_2,\phi_1)$)
\begin{equation}\label{ffAparaphi}
\{T_0,\;\; \sigma_{\a=1,2}(\phi_1),\;\; \rho_{\a=1,2}(\phi_1,\phi_2),\;\; \text{Im}F^I_{\ell}(\phi_1), \;\; \rho^I_{\ell}(\phi_1)\}
\end{equation}
which we will discretize in the next subsection.

\subsection{Numerical discretization}

We consider a hybrid numerical discretization through interpolation points for the bootstrap variables $\{T_0,\sigma_1,\sigma_2,\rho_1,\rho_2,\text{Im}F^I_{\ell}\}$ and BSpline basis functions for the spectral density $\{\rho^I_{\ell}\}$. The latter provides a good numerical control of the low energy kinematic of the spectral density given in \eqref{rholowenergy} which we will describe in subsection \ref{BSplinesection} below.

\subsubsection{Interpolation points for amplitudes and form factors}
To discretize the bootstrap variables $\{T_0,\sigma_1,\sigma_2,\rho_1,\rho_2,\text{Im}F^I_{\ell}\}$, we choose equally spaced $M$ points in the range $\phi\in(0,\pi)$:
\begin{equation}\label{phij}
\phi_j=\Big(j-\frac{1}{2}\Big)\Delta_\phi,\;\; j=1,2,\ldots,M,\;\;\;\Delta_\phi=\frac{\pi}{M}
\end{equation}
The index $j$ is shifted to avoid taking the point exactly at threshold\footnote{Note that according to the map \eqref{xphi}, the discrete set of angles \eqref{phij} correspond to different distributions of $x_j\in(0,4)$ depending on the center $\nu_0$ for the map. We have tested the convergence of the numerics for a few values of $\nu_0$ and the results in this paper are obtained using $\nu_0=-20$.}. This corresponds, through \eqref{xphi}, to a set of points in the interval $(4,\infty)$
\begin{equation}\label{xj}
s_j=x(\phi_j),\;\;j=1,\ldots,M
\end{equation}
which is also the set of points where we will evaluate the partial waves in the physical region.
We thus have the discrete set of bootstrap variables (where $\rho_{2,j_1j_2}=\rho_{2,j_2j_1}$)
\begin{equation}
\{T_0,\sigma_{1,j_1},\sigma_{2,j_1},\rho_{1,j_1j_2},\rho_{2,j_1j_2},\text{Im}F^I_{\ell,j_1}\},\;\; j_1,j_2=1,2,\ldots,M
\end{equation}


For the bootstrap procedure, we also have to evaluate the partial waves at an unphysical point $s_*$ (can be complex in general) $f^I_{\ell}(s_*)$. This will be used for both defining functionals of the type \eqref{fnals}, as well as for the requirement of matching with the tree level partial waves \eqref{chiralconstraints} in the low energy unphysical region $s\in(0,4)$.

The partial waves evaluation at a generic point $s_*$ (away from the cut $s>4$) can be done directly by taking an unphysical point $s_*$ and replacing the integrals in \eqref{pwintegral} with sums:
\begin{equation}\label{fsstar}
f^I_\ell(s_*)=fT_0+f_{1,j_1}\sigma_{1,j_1}+f_{2,j_1}\sigma_{2,j_1}+f_{1,j_1j_2}\rho_{1,j_1j_2}+f_{2,j_1j_2}\rho_{2,j_1j_2}
\end{equation}
where
\begin{equation}\label{Manum1}
\begin{aligned}
&f=\hat{\Phi}^I_\ell(s_*),\;
f_{1,j_1}=\Delta_{\phi}\Phi^{I}_{\ell,1}(s_*,\phi_{j_1}),\;
f_{2,j_1}=\Delta_{\phi}\Phi^{I}_{\ell,2}(s_*,\phi_{j_1}),\\
&f_{1,j_1j_2}=\Delta^2_{\phi}\tilde{\Phi}^{I}_{\ell,1}(s_*,\phi_{j_1},\phi_{j_2}),\;\;
f_{2,j_1j_2}=\Delta^2_{\phi}\tilde{\Phi}^{I}_{\ell,2}(s_*,\phi_{j_1},\phi_{j_2})
\end{aligned}
\end{equation}

The evaluation of the partial waves in the physical region $s>4$ above the cut is slightly trickier since the kernel $\mathcal{K}$ (or equivalently $K$ in \eqref{kernel2}) has real and imaginary parts in the physical region. In this case, we can write the discretized version of the kernel $K$ \eqref{kernel2} (evaluated on the upper half circle) as
\beq\label{K1a}
K(\phi_j,\phi_{j_1}) = \hat{K}_{jj_1} + \frac{i}{2\pi} \delta_{jj_1}
\eeq
where the principal integral \eqref{disp} becomes the sum
\beq 
\text{Re}g(\phi_j) = \sum_{j_1=-M}^M \hat{K}_{jj_1} \text{Im}g(\phi_{j_1})
\eeq 
and
\beq\label{K1b}
\hat{K}_{j_1,j_2} = -\frac{1}{2M} (1-(-)^{j_1-j_2}) \cot\left(\frac{\pi}{2M}(j_1-j_2)\right)
\eeq
namely the discretized kernel that relates the real and imaginary part of an analytic function at the boundary of the unit disk. 

With this, we can now evaluate the partial waves in the physical region. In practice, for imposing the unitarity constraints, we evaluate the rescaled partial waves
\begin{equation}\label{hdef}
h^I_{\ell}(s)=\pi\sqrt{\frac{s-4}{s}}f^I_{\ell}(s)
\end{equation}
at discrete points $s_{j}=x(\phi_j)\in(4,\infty)$. This is given by
\begin{equation}\label{pwboot}
h^I_{\ell,j}=h^I_{\ell,j}T_0+h^{I}_{\ell,1;j,j_1}\sigma_{1,j_1}+h^{I}_{\ell,2;j,j_1}\sigma_{2,j_1}+h^{I}_{\ell,1;j,j_1,j_2}\rho_{1,j_1,j_2}+h^{I}_{\ell,2;j,j_1,j_2}\rho_{2,j_1,j_2}
\end{equation}
where
\begin{equation}\label{fphyskernels}
\begin{aligned}
&h^I_{\ell,j}=\omega_j\hat{\Phi}^I_\ell(s_j),\;
h^{I}_{\ell,1;j,j_1}=\omega_j\Delta_{\phi}\Phi^{I}_{\ell,1}(s_j,\phi_{j_1}),\;
h^{I}_{\ell,2;j,j_1}=\omega_j\Delta_{\phi}\Phi^{I}_{\ell,2}(s_j,\phi_{j_1}),\\
&h^{I}_{\ell,1;j,j_1,j_2}=\omega_j\Delta^2_{\phi}\tilde{\Phi}^{I}_{\ell,1}(s_j,\phi_{j_1},\phi_{j_2}),\;\;
h^{I}_{\ell,2;j,j_1,j_2}=\omega_j\Delta^2_{\phi}\tilde{\Phi}^{I}_{\ell,2}(s_j,\phi_{j_1},\phi_{j_2})
\end{aligned}
\end{equation}
with
\begin{equation}
\omega_j=\pi\sqrt{\frac{s_j-4}{s_j}}
\end{equation}
from the factor in \ref{hdef}.
For evaluating the sum \eqref{pwboot}, one simply uses their original definition in section \ref{kernel section}, and take the values of $s_j$ at discrete points \eqref{xj} (keep in mind that the Jacobian has been absorbed into the redefinitions of $\Phi,\tilde{\Phi}$). In particular, for the kernels that involve $\mathcal{K}$, one has to use the expressions \eqref{K1a}, \eqref{K1b} that perform the numerical principal integral which gives rise to the real and imaginary part of the partial waves in the physical region $s>4$.

As an additional numerical comment, due to the symmetry of the spectral density $\rho_{2,j_1j_2}=\rho_{2,j_2j_1}$, in practice we symmetrize the numerical matrices that multiply $\rho_2$, namely $f_{2,j_1j_2}$ and $\tilde{\Phi}^{I}_{\ell,2;j,j_1,j_2}$ with respect to the $j_1,j_2$ indices thus reducing the number of variables.

\medskip

The numerical computation of the form factors is completely analogous. The form factor is parametrized by its imaginary part on the physical region at discretized points $\text{Im}F(s_j)\to\text{Im}F(\phi_j)$. According to 
\eqref{FFana}, the real part can be computed with
\begin{equation}\label{Manum3}
\text{Re}F_j=1+\big(\hat{K}_{j,j_1}+K_{0,j_1}\big)\text{Im}F_{j_1}
\end{equation}
where
\begin{equation}
K_{0,j_1}=K_0(\phi_{j_1})
\end{equation}

\subsubsection{BSpline basis functions for spectral density}\label{BSplinesection}

The spectral densities $\rho^I_{\ell}(s)$ enter the positive semidefinite matrix \eqref{pos} and are subject to the SVZ sum rules up to the energy scale $s_0$. As we will describe in the next section, in the numerics, we compute the positive matrix for $s\in(4,s_0)$ so we will consider the discretization of the functions $\rho^I_{\ell}(s)$ for $s\in(4,s_0)$.\footnote{According to the finite energy sum rules, the values of $\rho$ above $s_0$ are assumed to take the perturbative QCD form (\ie\ ignoring the condensates). See section \ref{SVZasympt}.} This is done by expanding the $\rho^I_{\ell}(s)$ with the BSpline basis functions which allows a minimal control the low energy kinematic behavior \eqref{rholowenergy}. See \cite{He:2025gws} for an alternative, more powerful implementation of this behavior.

To do this, we first rewrite the bootstrap variables $\rho^I_{\ell}(s)$ as
\begin{equation}\label{rhohat}
\rho^I_{\ell}(s)=\kappa_{\ell}(s)\hat{\rho}^I_{\ell}(s)
\end{equation}
where the $\kappa_{\ell}(s)$ are the kinematic factors in \eqref{rholowenergy}:
\begin{subequations}\label{kappadef}
\begin{eqnarray}
\kappa_{0}(s) &=& \frac{1}{(2\pi)^4}\, \frac{3}{16\pi}\,\left(1-\frac{4}{s}\right)^{\frac{1}{2}} \\ 
\kappa_{1}(s) &=& \frac{1}{(2\pi)^4}\,\frac{s}{24\pi}\, \left(1-\frac{4}{s}\right)^{\frac{3}{2}} \\ 
\kappa_{2}(s) &=&  \frac{1}{(2\pi)^4}\,\frac{s^2}{160\pi}\left(1-\frac{4}{s}\right)^{\frac{5}{2}}
\end{eqnarray}
\end{subequations}
that depend only on $s$ and $\ell$. These factors are the square of the rescaling kinematic factor between $F_{\ell}(s)$ and $\mathcal{F}_{\ell}(s)$ (see \eqref{FFS0scale}, \eqref{FFP1scale} and \eqref{FFD0scale}) so that the inequality $\rho\ge|\mathcal{F}|^2$ from the positive matrix \eqref{pos} is equivalent to
\begin{equation}\label{straightFrhohat}
\hat{\rho}(s)\ge |F(s)|^2\;.
\end{equation}
 Below the four pion threshold ($4\le s\le 16$) we have exact saturation $\hat{\rho}(s)=|F(s)|^2$. We can use this to extend the function $\hat{\rho}(s)$ continuously to the region $0<s<4$ by defining it simply as $\hat{\rho}(s)=F^2(s)$, $0<s<4$. Since $F(0)=1$, this extended function obeys the boundary condition $\hat{\rho}(s=0)=1$ that we can impose in the bootstrap. It implies that, near threshold $\hat{\rho}(s)\simeq 1$ and therefore the desired behavior \eqref{rholowenergy}.

The (extended) function $\hat{\rho}(s)$ can be expanded using the BSpline basis with degree $p=3$:
\begin{equation}\label{b}
\hat{\rho}^I_{\ell}(s)=\sum_{i=1}^{N_B}\rho^I_{\ell,i}B_{i}(s),\;\; s\in[0,s_{n_0}]
\end{equation}
where $B_i(s)$ denotes the BSpline basis functions with degree $p=3$. $s_{n_0}$ here denotes the last discrete point in \eqref{xj} below $s_0$, namely the last numerical evaluation point for the positive matrix computation. See \eqref{numpos} below. The BSpline basis $B_i(s)$ can be defined using a series of knots which we choose to be equally spaced on $[0,s_{n_0}]$. So for $N_B$ basis functions we have the knots: 
\begin{equation}
\{0,0,0,0,\ldots,s_{n_0},s_{n_0},s_{n_0},s_{n_0}\}
\end{equation}
where $\ldots$ indicates the equally spaced inner knots. The BSpline basis functions thus defined are non-negative for the whole range $s\in[0,s_{n_0}]$ and we can impose the condition $\rho^I_{\ell,i}\ge 0,i=1,\ldots,N_B$ to ensure the positivity of the spectral density $\hat{\rho}$.

With this parametrization, the bootstrap numerical variables are now the coefficients $\rho^I_{\ell,i}$ in \eqref{b} and the boundary condition $\hat{\rho}(0)=1$ amounts to setting
\begin{equation}\label{b11}
\rho^I_{\ell,1}=1
\end{equation}
The $\hat{\rho}$ can then be evaluated on the set of points \eqref{xj} up to $s_{n_0}$:
\begin{equation}\label{BSplineM}
\hat{\rho}^I_{\ell,j}\equiv\hat{\rho}^I_\ell(\phi_j)=B_{j,i}\rho^I_{\ell,i},\;\; B_{j,i}=B_i(s_j),\;\;j=1,\ldots,n_0
\end{equation}
and then enters the positive semidefinite matrices and the sum rules. For implementing the sum rules, we compute the discrete sum:
\begin{equation}\label{SRdiscrete}
\int_0^{\phi_0} x'(\phi) d\phi \rho^I_\ell(\phi) x(\phi)^n \to \sum_{j\le n_{0}}\Delta_{\phi} \left|\frac{\partial x(\phi)}{\partial \phi}\right|_{\phi=\phi_j}\kappa_\ell(s_j) \hat{\rho}^I_{\ell,j}s_j^n
\end{equation}

\medskip

The numerical matrices involved in the above computations (\eqref{Manum1}, \eqref{fphyskernels}, \eqref{Manum3}, \eqref{BSplineM}, \eqref{SRdiscrete}) are all included in the Mathematica notebook  {\tt GTB\_numerics.nb} we provide with this paper.

\subsection{Numerical gauge theory bootstrap}\label{NGTB}

With the above numerical setup, let us now describe the concrete implementation of the gauge theory bootstrap. Together with this submission we provide a matlab code {\tt GTB\_numerics.m} that implements the procedure. Interested readers can download the file, install the \href{https://cvxr.com/cvx/}{CVX} package \cite{cvx,gb08} together with the MOSEK solver \cite{mosek} to run the program and reproduce the results in this paper. In that case, this section can be used as a guide to the numerical code.

\paragraph{S-matrix/form factor bootstrap} In the numerical procedure, we have a discrete set of variables ($\rho_{2,j_1j_2}$ is symmetric in $j_1\leftrightarrow j_2$)
\begin{equation}\label{va}
\begin{aligned}
v_a=&\{T_0,\sigma_{1,j_1},\sigma_{2,j_1},\rho_{1,j_1j_2},\rho_{2,j_1j_2},\text{Im}F^I_{\ell,j_1},\rho^I_{\ell,i}\},\\
&j_1,j_2=1,\ldots,M,\;i=1,\ldots,N_B
\end{aligned}
\end{equation}
where $j_1,j_2$ label a discrete set of interpolating points and $i$ indexes a discrete set of basis functions. The index $a$ runs over all variables. 
As described above, $\{T_0,\sigma_{1,2},\rho_{1,2}\}$ parameterize the amplitude and therefore also the rescaled partial waves \eqref{pwboot} 
in the physical region that we evaluate at the discrete set of points $s_j$ \eqref{xj} and up to a maximal spin $\ell=L$ ($L=19$ in our paper). 
If we introduce an index $n=(I,\ell,j)$ to label the partial waves $h_n=h^I_\ell(s_j)$ and take into account that the parameterization is linear, we can compute two matrices $A$ and $B$ such that 
\begin{equation}\label{hvinitial}
\text{Re}h_n=A_{n,a}v_a,\;\; \text{Im}h_n=B_{n,a}v_a
\end{equation}
Since the unitarity condition is valid for each partial wave and each value of $s$, the discretized version is simply
\begin{equation}\label{uninum}
\text{Re}h_n^2+\text{Im}h_n^2\le 2\text{Im}h_n,\;\;\; \forall n
\end{equation}

As for the form factors and spectral densities, in this paper, we consider only the quantum numbers
\begin{equation}
\{\ell,I\}\in\{S0,P1,D0\}
\end{equation}
The form factors are parametrized by the imaginary part $\text{Im}F(s_j)$. Using \eqref{Manum3}, we can compute the real part $\text{Re}F(s_j)$ on the physical line. 
To impose \eqref{pos}, we take $s_0 = 2\,\GeV$, and require for all $s_j\in(4,s_0)$ the positive semidefinite conditions:
\begin{equation}\label{numpos}
\begin{aligned}
\begin{pmatrix}
1 &\ S^I_{\ell}(s_j)\ &\ \ \cF^I_{\ell}(s_j) \\
S^{I*}_{\ell}(s_j) &\ 1\ & \ \ \cF^{I*}_{\ell}(s_j) \\
\cF^{I*}_{\ell}(s_j) &\ \cF^I_{\ell}(s_j)\ &\ \ \rho^I_{\ell}(s_j)
\end{pmatrix}\succeq 0\\
\\
\{\ell,I\}\in\{S0,P1,D0\},\;\; 4<s_j< 2\,\GeV
\end{aligned}
\end{equation}
In summary, we have a total number of variables
\begin{equation}
\mathbf{M}=1+2M+M^2+\frac{M(M+1)}{2}+3M+3N_B
\end{equation}
subject to $\frac{3 M L}{2}$ unitarity constraints \eqref{uninum} and positivity constraints \eqref{numpos}. The numerical results given in the main body of the paper, \ie\ in section \ref{sec:phaseshifts}. are obtained for $M=50,L=19,N_B=40$. In appendix \ref{numtest}, we consider other values to check consistency.

Let us remark that in the numerical computations, we use the regularization procedure we introduced in \cite{He:2021eqn} in the pure S-matrix bootstrap context by bounding the norm of the double spectral densities with a large regulator $M_{reg}$
\begin{equation}
||\left[\rho_{1,j_1,j_2},\rho_{2,j_1,j_2}\right]||_4\le M_{reg}
\end{equation}
The detail motivation of the regularization is explained in \cite{He:2021eqn}. To summarize briefly, the regularization avoids unphysical highly oscillating double spectral densities $\rho_{1,2}$ in the primal problem and renders the dual problem feasible. The value of the $M_{reg}$ is tested with the pure S-matrix bootstrap to ensure convergence. In producing the results in this paper, we take $M_{reg}=10^2$ and consider other values in appendix \ref{numtest} to check consistency.

\paragraph{Rescaling the positive matrix}
In actual numerical computations, we make improvements by rescaling and redefining the positive matrix. Recall that the positive semidefinite matrix \eqref{pos} arise from the overlapping of three states:
\begin{equation}
|\psi_1\rangle=|\text{out}\rangle,\;\; |\psi_2\rangle=|\text{in}\rangle,\;\; |\psi_3\rangle=\mathcal{O}|0\rangle
\end{equation}
\beq\label{Ba}
\begin{array}{c c} &
\begin{array}{c c c} \!\!\!\!\!\!\!\!\!\!\!\!\!\!|\psi_1\rangle \ \ \  & \ \  |\psi_2\rangle  \ \ &  \ \ \ \ |\psi_3\rangle \\
\end{array}
\\
\begin{array}{c c c}
\langle \psi_1|  \\
\langle \psi_2| \\
\langle \psi_3|
\end{array}
&
\left(
\begin{array}{c c c}
\ \ 1 \ \ &\ \ \ S^I_\ell(s)\ &\ \ \cF^I_\ell(s) \\
S^{I*}_\ell(s) &\ \ 1\ &\ \ \cF^{I*}_\ell(s)\ \\
\cF^{I*}_\ell(s) &\ \cF^I_\ell(s) &\ \ \rho_\cO(s)
\end{array}
\right) \succeq 0
\end{array} 
\eeq
We can perform a change of basis
\begin{equation}
|\psi_2\rangle\to -i|\psi_2\rangle+i|\psi_1\rangle
\end{equation}
and rescale the states $|\psi_2\rangle,|\psi_3\rangle$ by some $\ell,s$-dependent factors to define three new states: 
\begin{equation}\label{newstates}
\begin{aligned}
&|\tilde{\psi}_1\rangle=|\psi_1\rangle,\\
&|\tilde{\psi}_2\rangle=\Lambda_\ell^{-1}(s)\big(-i|\psi_2\rangle+i|\psi_1\rangle\big),\\
&|\tilde{\psi}_3\rangle=\kappa_{\ell}^{-1/2}(s)|\psi_3\rangle
\end{aligned}
\end{equation}
The overlaps of \eqref{newstates} define a new positive semidefinite matrix:
\beq\label{Baa}
\begin{array}{c c} &
\begin{array}{c c c} \!\!\!\!\!\!\!\!\!\!\!\!\!\!|\tilde{\psi}_1\rangle \ \ \ \  & \ \  |\tilde{\psi}_2\rangle  \ \ &  \ \ \ \ \ |\tilde{\psi}_3\rangle \\
\end{array}
\\
\begin{array}{c c c}
\langle\tilde{\psi}_1|  \\
\langle\tilde{\psi}_2| \\
\langle\tilde{\psi}_3|
\end{array}
&
\left(
\begin{array}{c c c}
\ \ \  \ \ 1 \ \ &\ \ \ \ \tilde{h}^ I_\ell(s)\ &\ \ F^ I_\ell(s) \\
\tilde{h}^ {I*}_\ell(s) &\ \ \  2\Im \hat{h}^ I_\ell(s)\ &\ \ 2\Im \hat{F}^ I_\ell(s) \\
F^{ I*}_\ell(s) &\ \ 2\Im \hat{F}^ I_\ell(s) &\ \ \hat{\rho}^ I_\ell(s)
\end{array}
\right) \succeq 0
\end{array} 
\eeq
Here we have taken $\kappa_{\ell}(s)$ to be \eqref{kappadef} so that after this rescaling, the form factor $F^I_{\ell}(s)$ and the redefined spectral density $\hat{\rho}^I_{\ell}(s)$ from \eqref{rhohat} enter the positive matrix. These are order 1 numbers so for machine precision semi-definite programming, the new numerical positive matrices \eqref{Baa} turn out to be better scaled compared to the original matrices \eqref{pos} that involve $\mathcal{F},\rho$. The other scale factor $\Lambda_{\ell}(s)$ in \eqref{newstates} is taken to be
\begin{equation}
\Lambda_{\ell}(s)=\left(\frac{\sqrt{s}-2}{\sqrt{s}+2}\right)^{\frac{\ell}{2}}
\end{equation}
that was previously found to be useful in the pure S-matrix bootstrap \cite{He:2021eqn}. The rest of the entries in \eqref{Baa} are given by
\begin{equation}
\begin{aligned}
&\tilde{h}^I_{\ell}(s)=\Lambda^{-1}_{\ell}(s)h^I_{\ell}(s),\;\;\Im \hat{h}^I_{\ell}(s)=\Lambda^{-2}_{\ell}(s)h^I_{\ell}(s),\\
&\Im \hat{F}^ I_\ell(s)=\Lambda^{-1}_{\ell}(s)\Im F^ I_\ell(s)\;.
\end{aligned}
\end{equation}
The partial waves in \eqref{hvinitial} and the unitarity constraints in \eqref{uninum} are similarly rescaled resulting in:
\begin{equation}\label{rescal}
\begin{aligned}
\text{Re}\tilde{h}_n=\tilde{A}_{n,a}v_a,\;\; &\text{Im}\tilde{h}_n=\tilde{B}_{n,a}v_a,\;\; \text{Im}\hat{h}_n=\hat{B}_{n,a}v_a,\\
\tilde{A}_{n,a}=\Lambda_n^{-1} A_{n,a},\;\; &\tilde{B}_{n,a}=\Lambda_n^{-1}B_{n,a},\;\; \hat{B}_{n,a}=\Lambda^{-2}_n B_{n,a},\\
&\text{Re}\tilde{h}_n^2+\text{Im}\tilde{h}_n^2\le 2\text{Im}\hat{h}_n,\;\;\; \forall n
\end{aligned}
\end{equation}
In the numerical computations (in {\tt GTB\_numerics.nb}), we compute the linear transformation matrices $\tilde{A}_{n,a},\tilde{B}_{n,a},\hat{B}_{n,a}$ with high precision before importing them into the optimization program and multiplying them with the machine precision variables $v_a$. 

\medskip

\paragraph{Chiral symmetry breaking} The implementation of the chiral symmetry breaking condition is straightforward. Using \eqref{pwboot}, we can evaluate the first six partial waves
\begin{equation}
f^0_0,\;\;  f^0_2, \;\; f^2_0,\;\;  f^2_2, \;\;f^1_1,\;\;  f^1_3
\end{equation} 
at four unphysical points
\begin{equation}\label{unphys}
s_{*,i}=\{1/2,1,3/2,2\}
\end{equation}
The tree level partial waves \eqref{chiralratio} define the following ratios of the $S0, P1, S2$ waves:
\begin{equation}
R^{\text{tree}}_{01}(s)=3\,\frac{2s-1}{s-4},\;\; R^{\text{tree}}_{21}(s)=3\,\frac{2-s}{s-4}
\end{equation}
while all the higher tree level  partial waves vanish.
Let us define
\begin{equation}
\begin{aligned}
{\bf f}^{\chi}_{01}=&f^0_0(s_{*,i})-R^{\text{tree}}_{01}(s_{*,i})f^1_1(s_{*,i})\\
{\bf f}^{\chi}_{21}=&f^2_0(s_{*,i})-R^{\text{tree}}_{21}(s_{*,i})f^1_1(s_{*,i})\\
{\bf f}^{0}_{2}=&f^0_2(s_{*,i}),\; {\bf f}^{2}_{2}=f^2_2(s_{*,i}),\; {\bf f}^{1}_{1}=f^1_1(s_{*,i})
\end{aligned}
\end{equation}
where the bold face letter should be interpreted as a vector of four components corresponding to the evaluations at the points \eqref{unphys}. 
We then require
\begin{equation}\label{chicon}
||[{\bf f}^{\chi}_{01},{\bf f}^{\chi}_{21}]||\le \epsilon^{\chi},\;\;\; 
||[{\bf f}^0_2,{\bf f}^2_2,{\bf f}^1_1]||\le \epsilon^{\chi},\\
\end{equation}
Namely we are matching these quantities with the tree level partial waves at very low energy (the four points \eqref{unphys}) up to a tolerance $\epsilon^{\chi}$.

In producing the results of this paper, we take the tolerance to be $\epsilon^{\chi}=2\times 10^{-3}$. This choice follows from the explorations done in our previous paper \cite{GTBPRD}. To briefly summarize the idea, we have previously tested imposing the constraints \eqref{chicon} with a series of tolerances $\epsilon^{\chi}$. Such tolerance for the matching should not be too small or too large: if it is too small, the theory we are studying with the physical $f_{\pi}=92$ MeV would be excluded; if it is too large, there would be a large deviation from the tree level behavior at very low energy, manifested as the disappearance of the chiral zero in $S0$ partial wave. Through such exploration, we numerically fix a tolerance $\epsilon^{\chi}=2\times 10^{-3}$, namely the minimal tolerance before the tree-level point (black point in fig. \ref{shape}) is excluded. See section 4.2 in our previous paper \cite{GTBPRD} for discussion. Other ways to impose these constraints are possible but we are describing a very specific way in case the reader want to reproduce the results of this paper.  

Finally, as mentioned already in section \ref{BSplinesection}, the kinematic behavior \eqref{rholowenergy} of the spectral density $\rho$ can be implemented simply by the conditions \eqref{b11} on the BSpline basis expansion coefficients $\rho^I_{\ell,1}$.

\bigskip

\paragraph{SVZ sum rules} To include the QCD information at high energy, we first impose the finite energy SVZ sum rules for $S0,P1,D0$ by constraining the moments of the spectral densities $\rho^I_{\ell}$ with the perturbative QCD calculations as summarized in \eqref{S0SR}, \eqref{P1SR} and \eqref{D0SR}. Denoting $n_{SR}$ as the number of moments for each spectral density, in this paper, we take $n_{SR}=3$ which corresponds to the following moments for the $S0,P1,D0$ spectral densities:
\begin{equation}
S0:\;n=0,1,2,\;\; P1:\; n=-1,0,1,\;\; D0:\; n=-2,-1,0
\end{equation}
The values from the perturbative QCD calculations (RHS of \eqref{S0SR}, \eqref{P1SR} and \eqref{D0SR}) are:
\begin{eqnarray}
S0: \!\!\!\!\!\!&&\int_4^{s_0}\!\!\!\frac{\rho^0_0(x)dx}{s_0^2} \simeq3.1\!\times\!10^{-7}, \int_4^{s_0}\!\!\! \frac{\rho^0_0(x)xdx}{s_0^{3}} \simeq2.1\!\times\!10^{-7},
\int_4^{s_0}\!\!\! \frac{\rho^0_0(x) x^2dx}{s_0^{4}}\simeq1.6\!\times\!10^{-7} \nonumber\\
P1:\!\!\!\!\!\!&&\int_4^{s_0}\!\!\!\frac{\rho^1_1(x)dx}{xs_0^2} \simeq5.6\!\times\!10^{-5},\int_4^{s_0}\!\!\! \frac{\rho^1_1(x)dx}{s_0^{3}} \simeq2.8\!\times\!10^{-5},
\int_4^{s_0}\!\!\! \frac{\rho^1_1(x)xdx}{s_0^{4}}\simeq1.8\!\times\!10^{-5} \nonumber\\
D0:\!\!\!\!\!\!&&\int_4^{s_0}\!\!\!\frac{\rho^0_2(x)dx}{x^2s_0^3} \simeq5.1\!\times\!10^{-5}, \int_4^{s_0}\!\!\! \frac{\rho^0_2(x)dx}{xs_0^{4}} \simeq2.6\!\times\!10^{-5},
\int_4^{s_0}\!\!\! \frac{\rho^0_2(x)dx}{s_0^{5}} \simeq 1.7\!\times\!10^{-5}\nonumber\\ \label{SRvalues}
\end{eqnarray}
These values come from the identity operator contribution in the SVZ expansion up to order $\alpha_s$ (see section \ref{sec:jj} for the perturbative QCD computations). Since there are contributions from higher order $\alpha_s$ corrections, we impose the sum rules up to corrections estimated from the order $\alpha_s$ terms. 

In the concrete numerical implementations, we replace the integrals with discrete sums as given above in \eqref{SRdiscrete} and perform the numerical sum up to the largest\footnote{In this paper (see table \ref{numpara}), we take the last point to be $s_{39}\sim 1.83$ GeV.} $s_j<s_0$. Defining difference between the sum rule and the desired value as 
\begin{equation}
w^I_{\ell,n}=\frac{1}{s_{n_0}^{n+n_{\ell}}}\sum_{j\le n_0}\left|\frac{\partial x(\phi)}{\partial \phi}\right|_{\phi=\phi_j}\kappa_\ell(s_j) \hat{\rho}^I_{\ell,j} s_j^n\, \Delta_{\phi} -\text{RHS of \eqref{SRvalues}},\;\; n_\ell=\begin{cases}
2,&S0,P1\\
3,&D0
\end{cases}
\end{equation}
we have the following sum rule constraints
\begin{equation}\label{SRerr}
\begin{aligned}
||(w^0_{0,n=0},w^0_{0,n=1},w^0_{0,n=2})||&\lesssim \epsilon^{0}_{0}=1\times 10^{-7}\\
||(w^1_{1,n=-1},w^1_{1,n=0},w^1_{1,n=1})||&\lesssim \epsilon^{1}_{1}=5\times 10^{-6}\\
||(w^0_{2,n=-2},w^0_{2,n=-1},w^0_{2,n=0})||&\lesssim \epsilon^{0}_{2}=6\times 10^{-6}
\end{aligned}
\end{equation}
where the sum rule errors $\epsilon^I_{\ell}$ are estimated as described above.

In appendix \ref{SRnum}, we will examine the bootstrap results for varying number of moments $n_{SR}$ and sum rule errors $\epsilon^I_{\ell}$.

\bigskip

\paragraph{Form factor asymptotics} As was already noted in \cite{GTBPRL,GTBPRD}, it is crucial to require that the form factors become small at large momentum transfer as follows from generic QCD arguments \cite{osti_1447331,Pire:1996bc}. The computation of such asymptotic forms is explained in section \ref{FFasymform} and we have summarized the leading order expressions for $P1,D0$ in \eqref{P1asymp} and \eqref{D0asymp} and made a rough estimate for the $S0$ asymptotic form factors in \eqref{S0asymp}. Such results are supposed to be valid at very high energy where the pion appears as constituted by just two partons (quark and anti-quark). Since this is not the case at $2\,\GeV$ we use the known asymptotic behavior $1/s,\ (s\rightarrow\infty)$ for $F^1_1(s)$ $(P1)$ and $F^0_2(s)$ $(D0)$ but allow a factor of order one to multiply the coefficient. Such factor is necessary since otherwise the problem becomes unfeasible and no solution is found. Somewhat surprisingly, for $S0$ we have to require a constant behavior at infinity instead of $1/s$.     
To be concrete, we consider the discrete points $s_j>s_0$ and require
\begin{equation}\label{rdef}
\begin{aligned}
|F^0_{0}(s_j)|&\le r^0_0,\\
|s_jF^1_{1}(s_j)|&\le r^1_1 \big(16\pi\alpha_s  f_\pi^2\big),\\
|s_jF^0_{2}(s_j)|&\le  r^0_2\big(48\pi\alpha_s  f_\pi^2\big),\\
\end{aligned}
\end{equation}
where the parameters $r^I_\ell$ are chosen so that the problem becomes feasible and a solution is found. In this paper we take
\begin{equation}\label{rvalues}
r^0_0=5\times 10^{-2},\; r^1_1=2,\;\; r^0_2=6
\end{equation}
The factors $r_1^1$ and $r^0_2$ are values of order one that make the problem feasible (it is not possible to take $r^1_1=r^0_2=1$). In the case of $r^0_0$ we do not have any useful way to estimate it so we put a value close to the smallest one that makes the problem feasible. In appendix \ref{FFnum}, we show how the results vary as we change $r^I_{\ell}$ around the values \eqref{rvalues}. A more precise theoretical estimate for these asymptotic values should be desirable, especially in the case of $S0$.  

\bigskip

\paragraph{Maximization and Watsonian unitarization}

The procedure continues by choosing the two linear functionals defined in \eqref{fnals} to study its space of allowed values. This is required since we will find concrete S-matrices and form factors only at the boundary of the space. In that way we find the allowed region depicted in fig. \ref{shape}. The black dot in that figure indicates the values corresponding to the non-linear sigma model that we are trying to extend into a full consistent S-matrix. For that reason we examine a range of points (blue, green, red) close to the black dot. In practice, the plot in fig. \ref{shape} is done by scanning one functional $\mathfrak{F}_0$ and maximize/minimize the other one $\mathfrak{F}_1$. Evaluating these functionals in the unphysical region can be done as described in \eqref{fsstar}.  

As described in section \ref{Wunit}, we next perform the Watsonian unitarization that takes us closer to the boundary of the allowed space and improves the saturation of the $3\times 3$ matrices \eqref{pos}. If we use the saturation conditions \eqref{W}, or better, the saturation of the rescaled matrix \eqref{Baa}, we find 
\begin{equation}\label{hFsat}
\tilde{h}=\frac{\text{Im}\hat{F}}{F^*}
\end{equation}
which is the Watson relation \eqref{W} rewritten in terms of the rescaled matrix entries.  
Consider again the rescaled unitarity cone in \eqref{rescal}:
\begin{equation}
\text{Re}\tilde{h}_n^2+\text{Im}\tilde{h}_n^2\le 2\text{Im}\hat{h}_n,\;\;\; \forall n\nonumber
\end{equation}
Now, suppose that after an initial maximization of the functionals \eqref{fnals}, we have obtained some extremal form factors $\mathsf{F}_n,\text{Im}\hat{\mathsf{F}}_n$ which contains the UV information from the sum rules and the asymptotics behavior. 
We can use them to define some new linear functionals acting on the partial waves in the physical region
\begin{equation}\label{unif}
\mathcal{F}\left[\{\text{Re}\tilde{h}_n,\text{Im}\tilde{h}_n,\text{Im}\hat{h}_n\}\right]=\text{Re}\tilde{\mathsf{h}}_n \text{Re}\tilde{h}_n+\text{Im}\tilde{\mathsf{h}}_n \text{Im}\tilde{h}_n-\sum_n\text{Im}\hat{h}_n
\end{equation}
with
\begin{equation}\label{Ffunctional}
\text{Re}\tilde{\mathsf{h}}_n=\frac{\text{Re}\mathsf{F}_n\text{Im}\hat{\mathsf{F}}_n}{|\mathsf{F}_n|^2},\;\;\text{Im}\tilde{\mathsf{h}}_n=\frac{\text{Im}\mathsf{F}_n\text{Im}\hat{\mathsf{F}}_n}{|\mathsf{F}_n|^2}
\end{equation}
as follows from \eqref{hFsat}.
Maximizing the linear functional \eqref{unif} leads to partial waves satisfying the Watson relation with the form factor $\mathsf{F}$ from the initial maximization and thus saturating unitarity. In the appendix \ref{numW}, we show the effect of the Watsonian unitarization for the $D0$ wave.

In principle we should apply this procedure for all the channels. In this paper, since we only compute the form factors for $S0, P1, D0$, we define the new functionals with \eqref{Ffunctional} only for these three channels. As for $S2, D2, F1$ channels, we also perform the unitarization, but with $\text{Re}\tilde{\mathsf{h}}_n,\text{Im}\tilde{\mathsf{h}}_n$ defined simply with the partial waves from the initial maximization. 

\paragraph{Summary of numerical parameters}

Although our input are the values $\mpi$, $\fpi$ and the gauge theory parameters, the numerical procedure requires, by force to choose other parameters such as the number of discretization points, tolerances, factors in the asymptotic form factors, etc. We summarize those choices in table \ref{numpara} and in appendix \ref{testN} we explore how some of these choices affect the results to understand the precision of our calculation. 
\begin{table}[!h]
\begin{center}
\renewcommand{\arraystretch}{1.4}
\begin{tabular}{|c|c|c|}
\hline
&  Expressions  &Parameters \\
\hline
\hline
Discretization & & $M=50,L=19,N_B=40$\\ 
Regularization & & $M_{reg}=10^{2}$\\  
\hline
Chiral symmetry breaking & \eqref{chiralratio} & $\epsilon^{\chi}=2\times 10^{-3}$ \\
\hline
Form factor asymptotics	& \eqref{S0asymp},\eqref{P1asymp},\eqref{D0asymp} &$r^I_{\ell}$ \eqref{rvalues}\\
\hline
SVZ sum rules & \eqref{S0SR},\eqref{P1SR},\eqref{D0SR} & $n_{SR}=3$, $\epsilon^I_{\ell}$ \eqref{SRerr}\\
\hline
UV scale & & $s_0=2\;\text{GeV}$\\
\hline
\end{tabular}
\end{center}
\caption{Summary of numerical parameters.}\label{numpara}
\end{table}
One parameter that we do not change is the value of the UV matching $s_0=2\,\GeV$. The purpose of this paper was to extend the upper energy value of the Gauge Theory Bootstrap from $s_0=1.2\,\GeV$ in our previous paper \cite{GTBPRL,GTBPRD} to $s_0=2\,\GeV$. This is non-trivial and required implementing new techniques such as the matching of spectral densities near threshold and Watsonian unitarization. A further small change in $s_0$ will not affect the results. A large change such as going to $s_0=3\,\GeV$ or more will, we believe, again require new improved techniques that we have not found yet. The other parameter that we keep fixed is $\epsilon^\chi=0.002$, same as used in the previous work \cite{GTBPRL,GTBPRD}. Reducing the parameter would exclude the non-linear sigma model values and a larger value does not effectively impose the low energy constraints, in particular for the $S0$ wave. The $P1$ and $D0$ wave seem to get most of the input from the UV so the low energy matching is not crucial. In the long term we would like to improve the control of the low energy region further to better determine, or possibly eliminate this parameter.       

\section{Conclusions}\label{sec:conclusions}

In a previous paper \cite{GTBPRL,GTBPRD} we proposed the Gauge Theory Bootstrap, a method to study the strongly coupled QCD dynamics of pions at energies where the low energy effective field theory is no longer applicable. The idea was to find the most general S-matrix consistent with bootstrap constraints that matches the low and high energy theories. In particular we matched the gauge theory at an energy $\sqrt{s_0}=1.2\,\GeV$. The results of \cite{GTBPRL,GTBPRD} were encouraging: we obtained phase shifts in agreement with experiment and, in particular, the $\rho$ vector meson.  However one obvious question was if this agreement would persist if we increased $s_0$. In fact one might have suspected that the mass of the $\rho$ and other features of the phase shifts were due to the scale $s_0$. In part motivated by this we apply here the same method but using a matching energy $\sqrt{s_0}=2\,\GeV$ roughly doubling the energy range.  We also compute six partial waves instead of just three as before. The results are even better. The phase  shifts agree with the ones found before, showing that $s_0$ is not introducing a scale. Moreover, in the new range of energies we find one strong resonance, the $f_2(1270)$ in the $D0$ channel and a weak new resonance in the $P1$ channel that we identified with the $\rho(1450)$. Overall, the new results are a strong indication that the method works and successfully locks into the correct amplitudes. Although this is very encouraging a lot of work remains to be done. In QCD we should include axial and pseudoscalar currents to introduce more constraints. Improvements need to be done for bettering identifying the point corresponding to the low energy QCD S-matrix perhaps even computing the value of $f_\pi$. Even  more interesting could be to start exploring the space of gauge theories and see what happens if we change the number of colors or flavors, or if we take the pion lighter or more massive. Finally the numerics should be simplified and streamlined to make these explorations easier. 

In summary, if we assume chiral symmetry breaking and confinement we can argue that the strongly coupled physics of gauge theories is dominated by pions and nucleons. However, effective field theory methods based on symmetry considerations alone do not allow us to solve the theory since even at moderate energies we need a large number of couplings (or Wilson coefficients) that we do not know how to compute from the gauge theory. We believe the Gauge Theory Bootstrap we proposed is a way to build a bridge between the low and high energy theories and allows to solve the strongly coupled QCD dynamics. 

\section*{Acknowledgements}\label{sec:ack}

We want to thank all the colleagues that provided us with many comments, suggestions and encouragement after we published the previous paper and that we tried to reflect in this new publication. In particular we are grateful to B. Bellazini, F. Boudjema, L. C\'ordova, C. Delaunay, V. Gorbenko, K. Häring, J. Henriksson, A. Herderschee, C. Herzog, D. Karateev, Z. Komargodski, L. Lellouch, H. Osborn, M. Paulos, J. Pel\'aez, J. Penedones, M. Procura, J. Qiao, F. Riva, S. Rychkov, A. Shapere, N. Su, P. Tourkine, A. Vainshtein, A. Vichi, S. Zhiboedov, and J.B. Zuber.
This work was supported in part by DOE through grant DE-SC0007884 and under the QuantiSED Fermilab consortium.

\appendix

\numberwithin{equation}{section}

\section{Scattering lengths from chiral Lagrangian coefficients}\label{sec:param2}

In \cite{GASSER1984142} Gasser and Leutwyler computed the scattering lengths in terms of the low energy parameters $\bar{l}_j$ using one-loop chiral perturbation theory obtaining the result:
\beqa\label{EFTa}
a_0^{(0)} &=& \frac{7m_\pi^2}{32\pi f_\pi^2}\left\{1+\frac{5m_\pi^2}{84\pi^2f_\pi^2}\left(\bar{l}_1+2\bar{l}_2-\frac{3}{8}\bar{l}_3+\frac{21}{10}\bar{l}_4+\frac{21}{8}\right)\right\} \\
b_0^{(0)} &=& \frac{1}{4\pi f_\pi^2}\left\{1+\frac{m_\pi^2}{12\pi^2f_\pi^2}\left(2\bar{l}_1+3\bar{l}_2+\frac{3}{2}\bar{l}_4-\frac{13}{16}\right)\right\} \\
a_0^{(2)} &=& -\frac{m_\pi^2}{16\pi f_\pi^2}\left\{1-\frac{m_\pi^2}{12\pi^2f_\pi^2}\left(\bar{l}_1+2\bar{l}_2-\frac{3}{8}\bar{l}_3-\frac{3}{2}\bar{l}_4+\frac{3}{8}\right)\right\} \\
b_0^{(2)} &=& -\frac{1}{8\pi f_\pi^2}\left\{1-\frac{m_\pi^2}{12\pi^2f_\pi^2}\left(\bar{l}_1+3\bar{l}_2-\frac{3}{2}\bar{l}_4-\frac{5}{16}\right)\right\} \\
a_1^{(1)} &=& \frac{1}{24\pi f_\pi^2}\left\{1-\frac{m_\pi^2}{12\pi^2f_\pi^2}\left(\bar{l}_1-\bar{l}_2-\frac{3}{2}\bar{l}_4+\frac{65}{48}\right)\right\} \\
b_1^{(1)} &=& \frac{1}{288\pi^3f_\pi^4}\left(-\bar{l}_1+\bar{l}_2+\frac{97}{120}\right) \\
a_2^{(0)} &=& \frac{1}{1440\pi^3f_\pi^4}\left(\bar{l}_1+4\bar{l}_2-\frac{53}{8}\right) \\
a_2^{(2)} &=& \frac{1}{1440\pi^3f_\pi^4}\left(\bar{l}_1+ \bar{l}_2-\frac{103}{40}\right) \label{EFTab}
\eeqa
Notice that we rewrote their formulas in terms of renormalized parameters and used the physical pion mass and $f_\pi$. 
It makes sense then to recompute the scattering lengths using the low energy parameters obtained in table \ref{table:lj} (we set $\bar{l}_3=0$ since we are not able to evaluate it independently). The results are presented in table \ref{table:GTBII}. They agree reasonably well with the direct extraction as given in table \ref{table:scatlength} (denoted as GTB here). 

\begin{table}[H]
\begin{center}
\renewcommand{\arraystretch}{1.4}
\begin{tabular}{|c|c|c|c|}
\hline
&{\bf GTB}& {\bf GTBII} & PY \\
\hline
$a^{(0)}_0$&\gtbblue{0.174},   \gtbred{0.191}    &\gtbblue{0.21},    \gtbred{0.21}   & $0.230\pm 0.010$                \\ 
$a^{(2)}_0$&\gtbblue{-0.0357}, \gtbred{-0.0403}  &\gtbblue{-0.043},  \gtbred{-0.043} & $-0.0422\pm 0.0022$             \\  
$b^{(0)}_0$&\gtbblue{0.285},   \gtbred{0.282}   &\gtbblue{0.26},    \gtbred{0.26}   & $0.268\pm 0.010$                \\
$b^{(2)}_0$&\gtbblue{-0.061},  \gtbred{-0.071}  &\gtbblue{-0.080},  \gtbred{-0.079} & $-0.071\pm0.004$                \\
$a^{(1)}_1$&\gtbblue{28.7},    \gtbred{30.7}    &\gtbblue{36.3},    \gtbred{36.7}   & $38.1\pm 1.4\times 10^{-3}$     \\
$b^{(1)}_1$&\gtbblue{12.4},    \gtbred{7.28}   &\gtbblue{2.62}      \gtbred{3.43}    & $4.75\pm 0.16 \times 10^{-3}$   \\
$a^{(0)}_2$&\gtbblue{14.6},    \gtbred{14.9}    &\gtbblue{14.8},    \gtbred{15.0}   & $18.0\pm 0.2\times 10^{-4}$     \\
$a^{(2)}_2$&\gtbblue{3.41},    \gtbred{2.53}   &\gtbblue{3.4},    \gtbred{2.5}    & $2.2\pm 0.2\times 10^{-4}$      \\
\hline
\end{tabular}
\end{center}
\caption{Scattering lengths, units of $m_\pi$, consistency check. GTBII are computed using in \eqref{EFTa}--\eqref{EFTab} the values from table \ref{table:lj}. There is reasonable agreement between the three columns including phenomenological data from PY \cite{Pelaez:2004vs}.} \label{table:GTBII}
\end{table}

\section{Tests on numerical parameters}\label{testN}

As we have described in section \ref{NGTB} and summarized in table \ref{numpara}, there are several numerical parameters introduced for implementing the gauge theory bootstrap procedure. In this appendix, we show the numerical tests that explore different values for these parameters.

\subsection{Numerical discretization}\label{numtest}

\begin{figure}[H]
\centering
\includegraphics[width=0.9\textwidth]{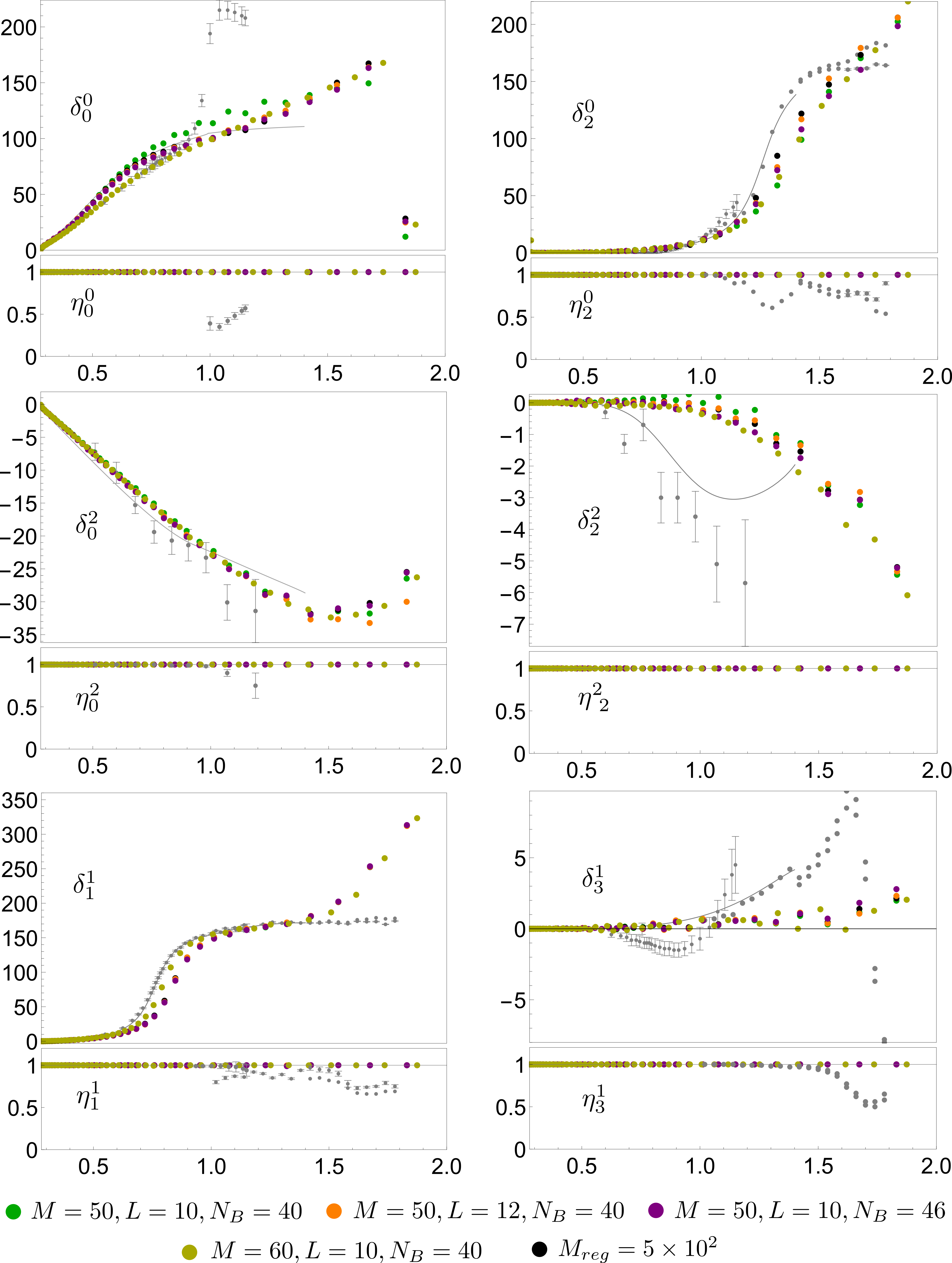}
\caption{Phase shifts $\delta^I_{\ell}$ and inelasticity $\eta^I_{\ell}$ from GTB for varying the numerical parameters $M,L,N_B,M_{reg}$.}
\label{6pwconverge}
\end{figure}

The first set of numerical parameters are for discretizing the bootstrap variables parametrizing the amplitudes, form factors and spectral densities. This includes $M,L,N_B$ as well as $M_{reg}$ for the regularization. In figure \ref{6pwconverge}, we show the changes on the phase shifts $\delta^I_{\ell}$ and inelasticities $\eta^I_{\ell}$ (as in fig. \ref{6pw}) when changing each of these parameters individually in comparison with the values used in this paper. We focus on the green point in fig. \ref{shape}. 
The green dots represent the original results (as shown in fig. \ref{6pw}) using the set of numerical parameters $M=50, L=19, N_B=40, M_{reg}=10^{2}$. We then test the cases with $L=23$ (orange), $N_B=46$ (purple), $M=60$ (yellow) and $M_{reg}=5\times10^{2}$ where in each case we keep all the other parameters unchanged. The results show a good overall stability of the numerics.

\subsection{Form factor asymptotic bound $r^I_\ell$}\label{FFnum}

\begin{figure}[h!]
\centering
\begin{subfigure}[b]{0.49\textwidth}
\centering
\includegraphics[width=\textwidth]{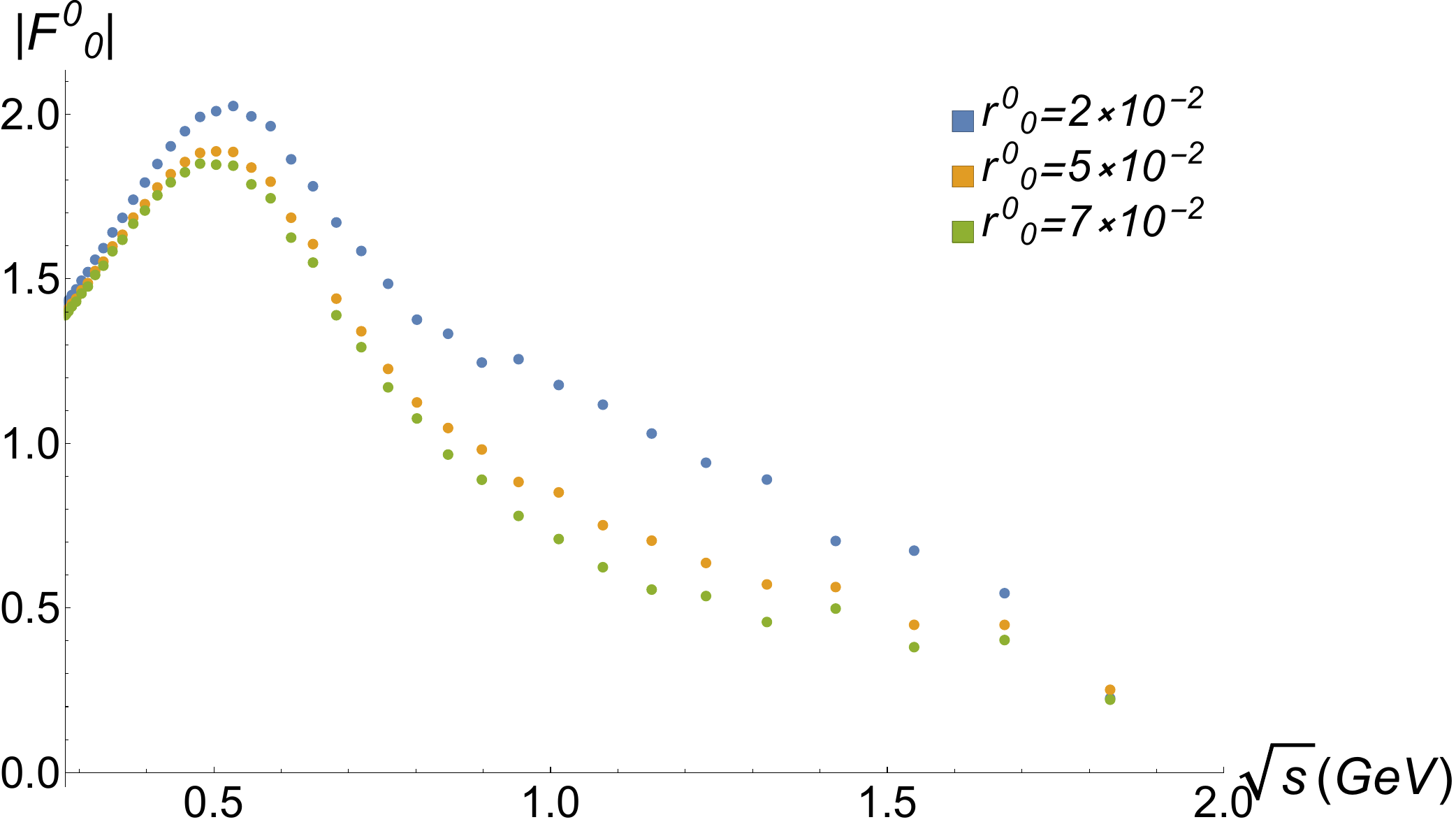}
\caption{$|F^0_{0}|$}
\end{subfigure}
\begin{subfigure}[b]{0.49\textwidth}
\centering
\includegraphics[width=\textwidth]{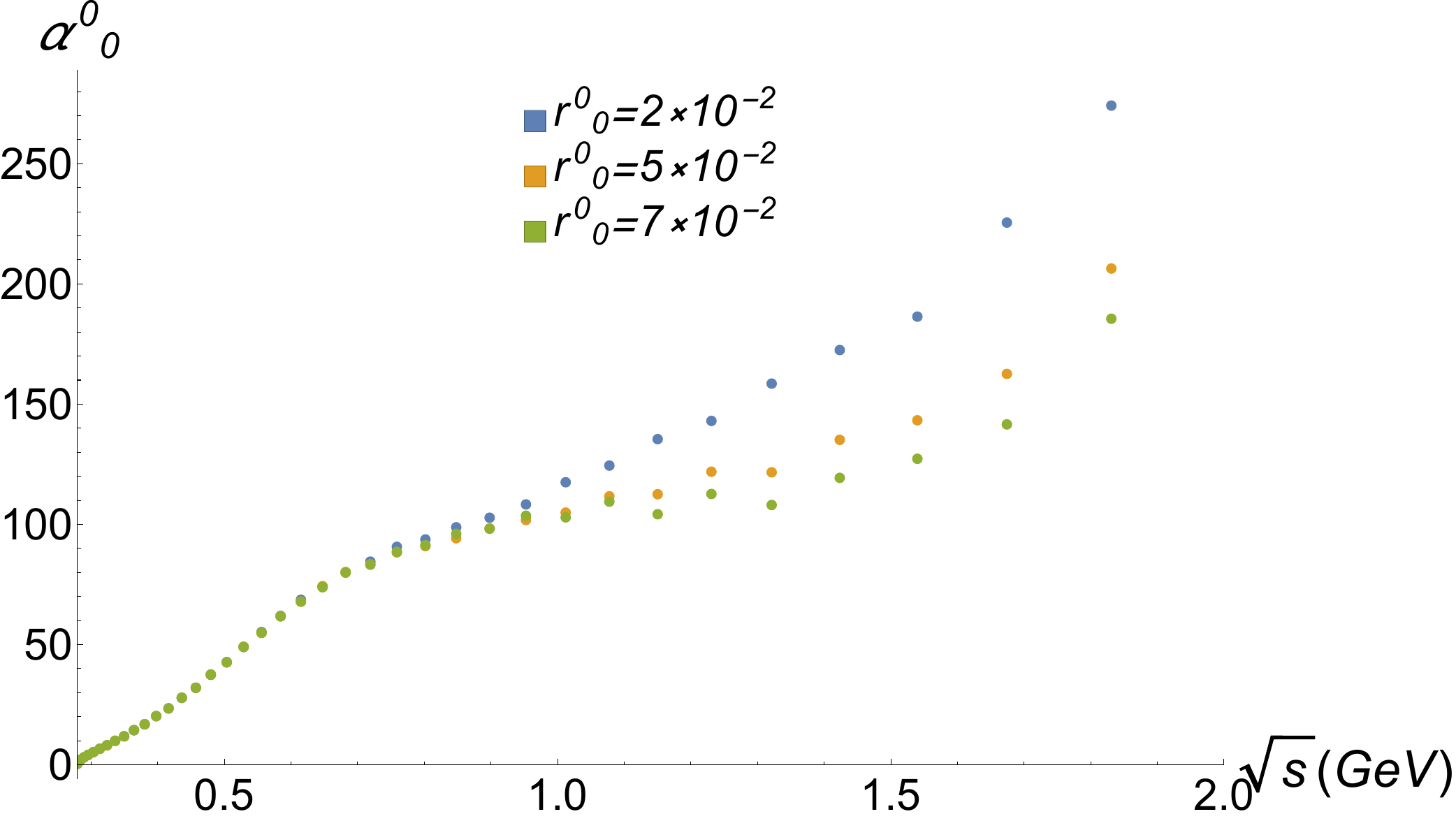}
\caption{$\alpha^0_{0}$}
\end{subfigure}
\vspace{1cm}

\begin{subfigure}[b]{0.49\textwidth}
\centering
\includegraphics[width=\textwidth]{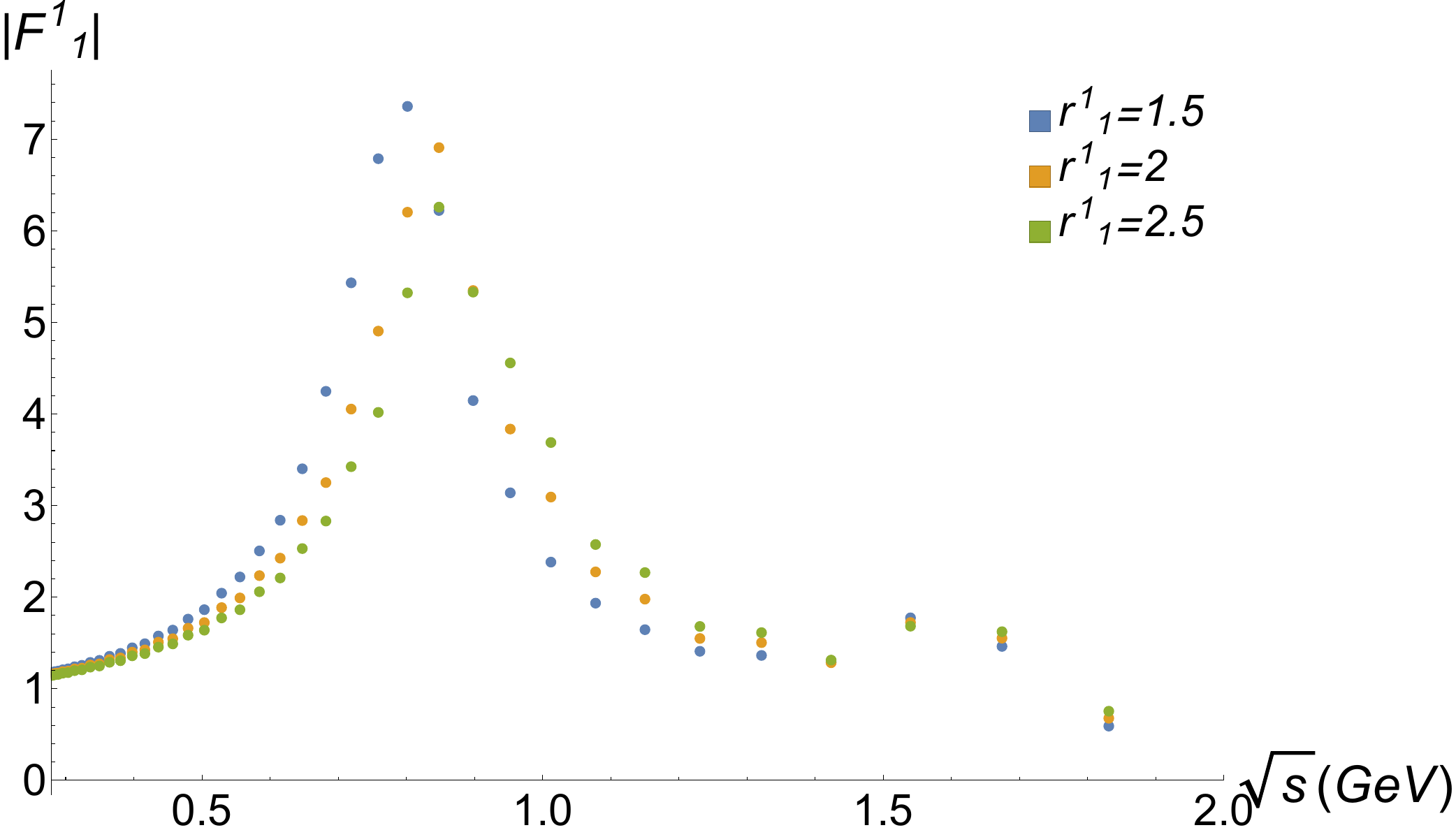}
\caption{$|F^1_{1}|$}
\end{subfigure}
\begin{subfigure}[b]{0.49\textwidth}
\centering
\includegraphics[width=\textwidth]{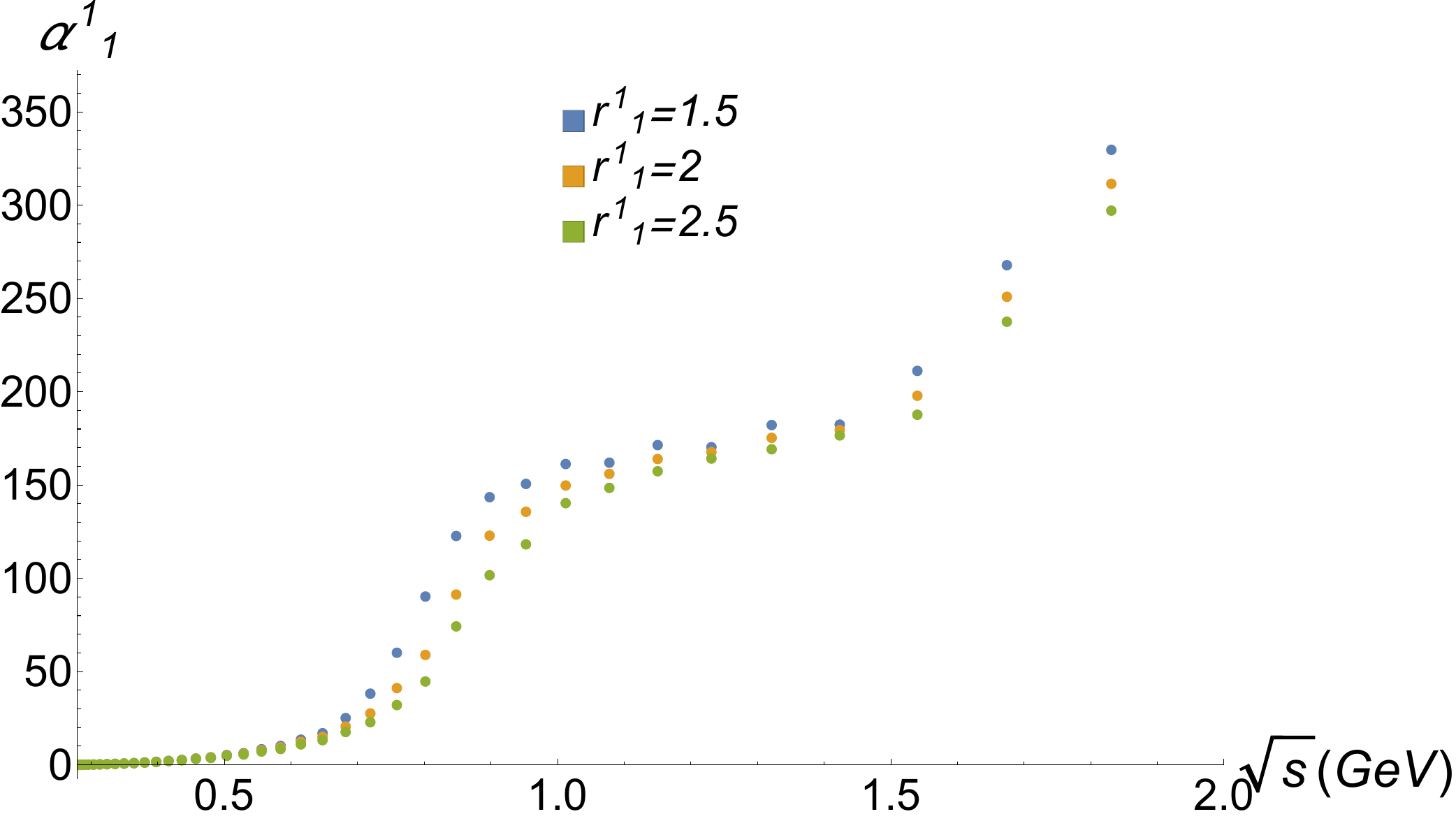}
\caption{$\alpha^1_{1}$}\label{P1alpha11}
\end{subfigure}
\vspace{1cm}

\begin{subfigure}[b]{0.49\textwidth}
\centering
\includegraphics[width=\textwidth]{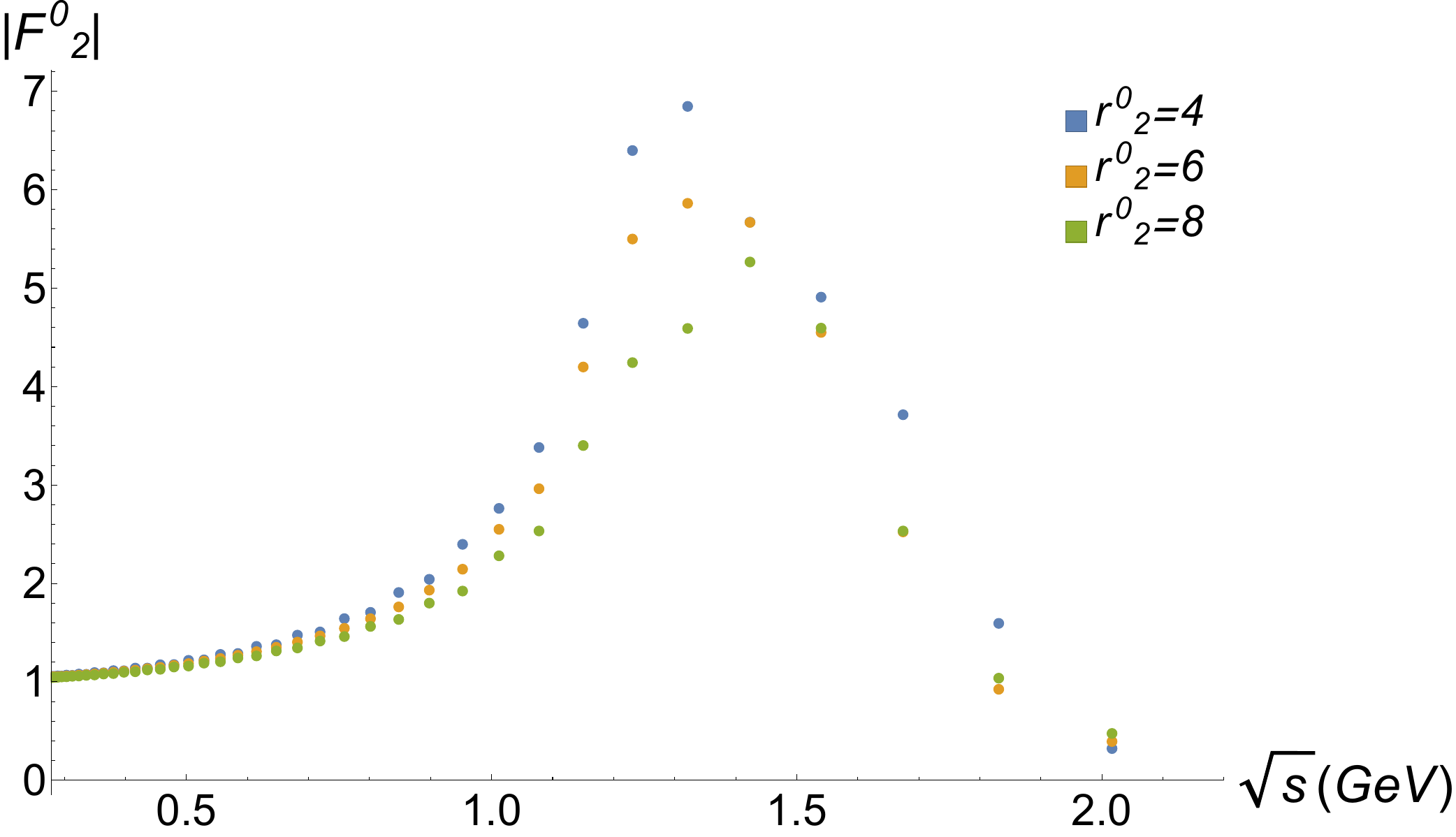}
\caption{$|F^0_{2}|$}
\end{subfigure}\
\begin{subfigure}[b]{0.49\textwidth}
\centering
\includegraphics[width=\textwidth]{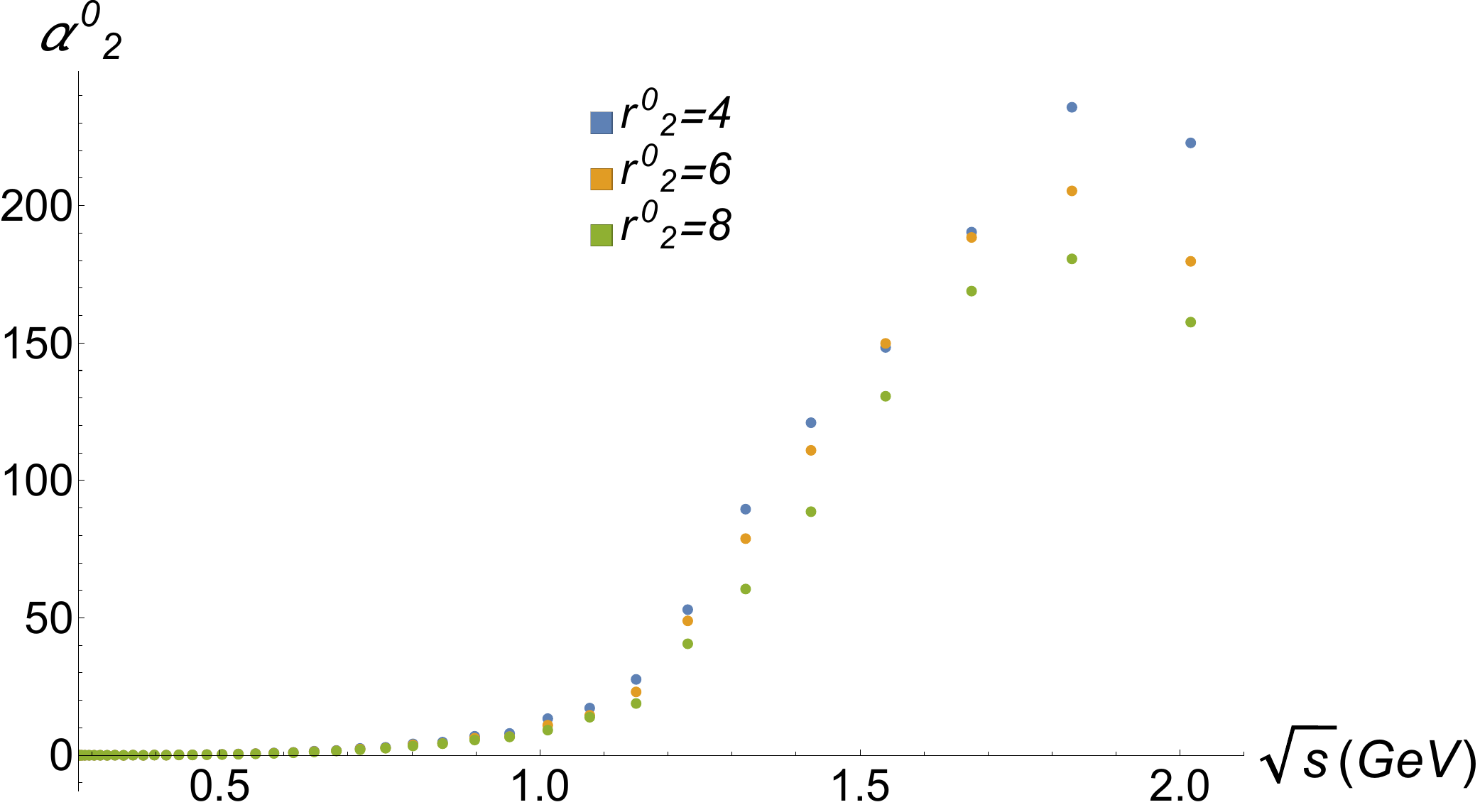}
\caption{$\alpha^0_{2}$}
\end{subfigure}

\caption{The modulus $|F^I_\ell|$ and phase $\alpha^I_\ell$ of the $S0, P1, D0$ form factors for three values of $r^I_\ell$.}
\label{Fasym02r}

\end{figure}

To impose the form factor asymptotic behavior, we bound the form factors in the high momentum transfer region. To do this, we introduced the relaxation parameter $r^I_{\ell}$ in \eqref{rdef} from the perturbative QCD calculations. In figure \ref{Fasym02r}, we show the modulus $|F^I_{\ell}|$ and the phase $\alpha^I_{\ell}$ for the $S0, P1, D0$ form factors for three different values of $r^I_{\ell}$: the first one (blue) is the minimal value where the numerical program is feasible; second one (orange) is slightly relaxed from the first value and we use in this paper, and the third one (green) is a more relaxed one. The results from the three values of $r^I_{\ell}$ have a good qualitative agreement but small variation quantitatively. For example, from fig. \ref{P1alpha11} one can estimate a $\sim 80\;\text{MeV}$ variation in the $\rho(770)$ meson mass for $r^1_1=1.5-2.5$. This is within the error of our computation (whose main source currently is the correct identification of the point corresponding to QCD in fig.\ref{shape}) but we expect to improve it in future work.

\subsection{Sum rules error $\epsilon^I_\ell$ and number of moments $n_{SR}$}\label{SRnum}

\begin{figure}[h!]
\centering
\begin{subfigure}[b]{0.49\textwidth}
\centering
\includegraphics[width=\textwidth]{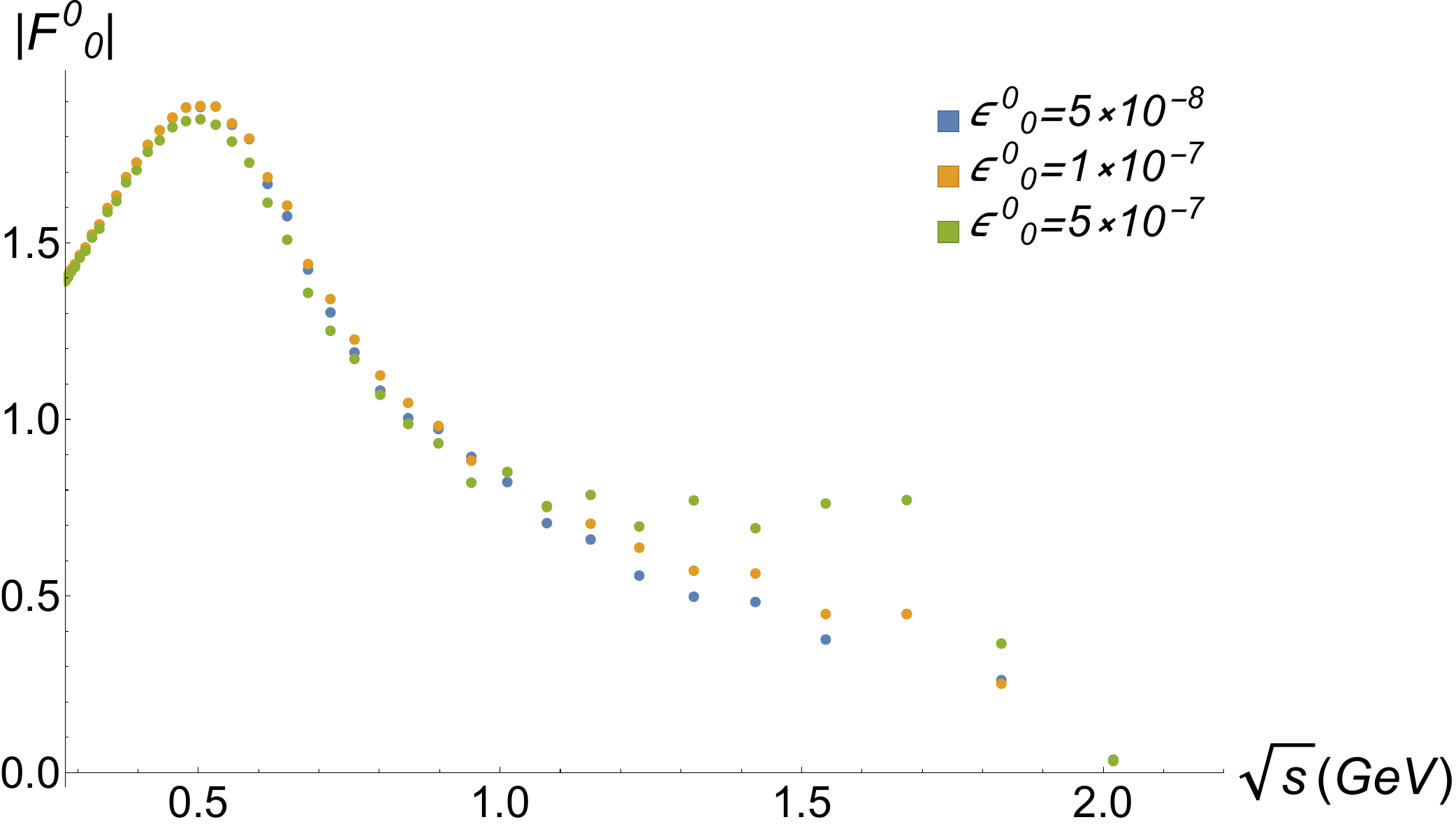}
\caption{$|F^0_{0}|$}
\end{subfigure}
\begin{subfigure}[b]{0.49\textwidth}
\centering
\includegraphics[width=\textwidth]{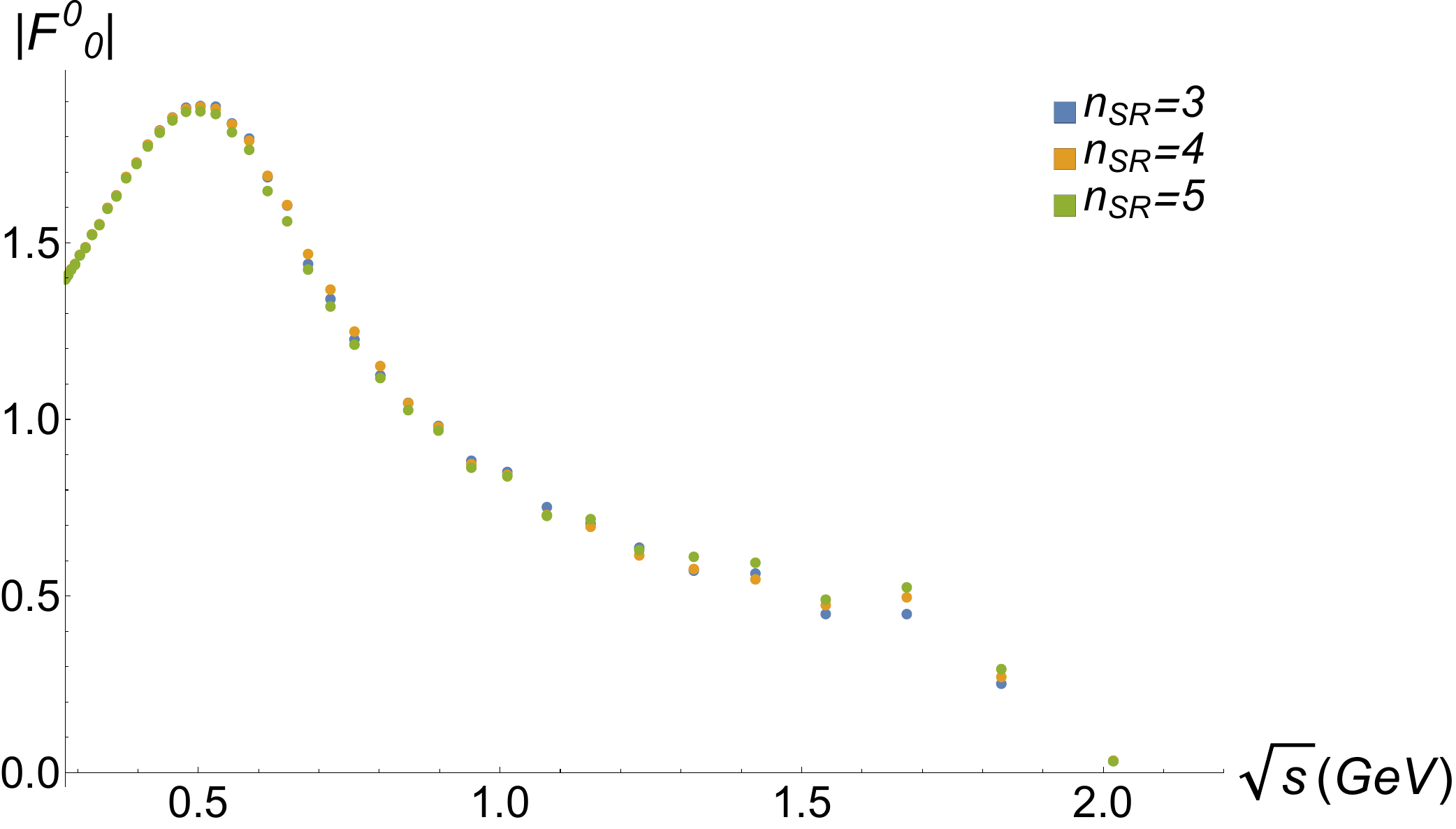}
\caption{$|F^0_{0}|$}
\end{subfigure}
\vspace{1cm}

\begin{subfigure}[b]{0.49\textwidth}
\centering
\includegraphics[width=\textwidth]{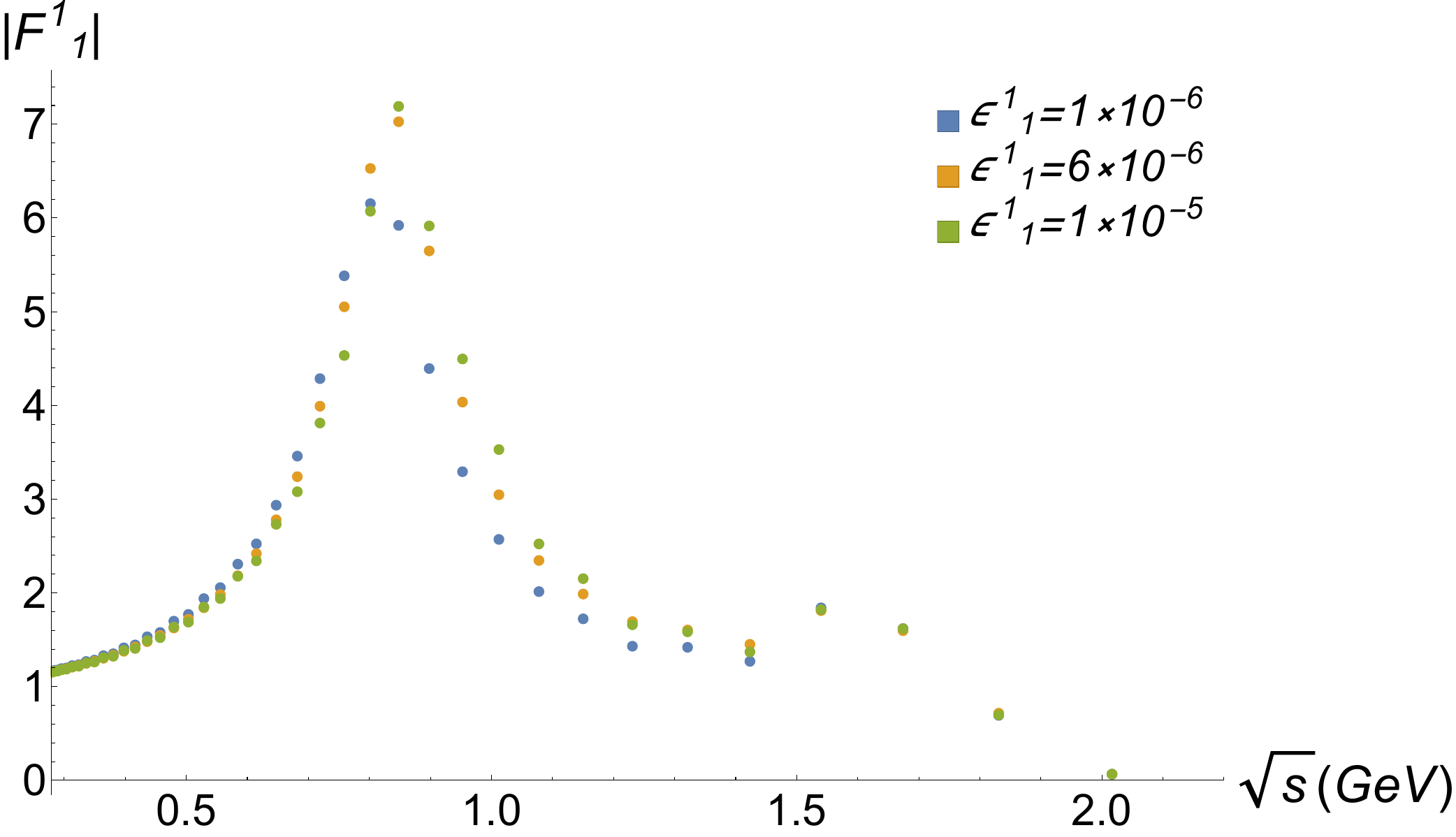}
\caption{$|F^1_{1}|$}
\end{subfigure}
\begin{subfigure}[b]{0.49\textwidth}
\centering
\includegraphics[width=\textwidth]{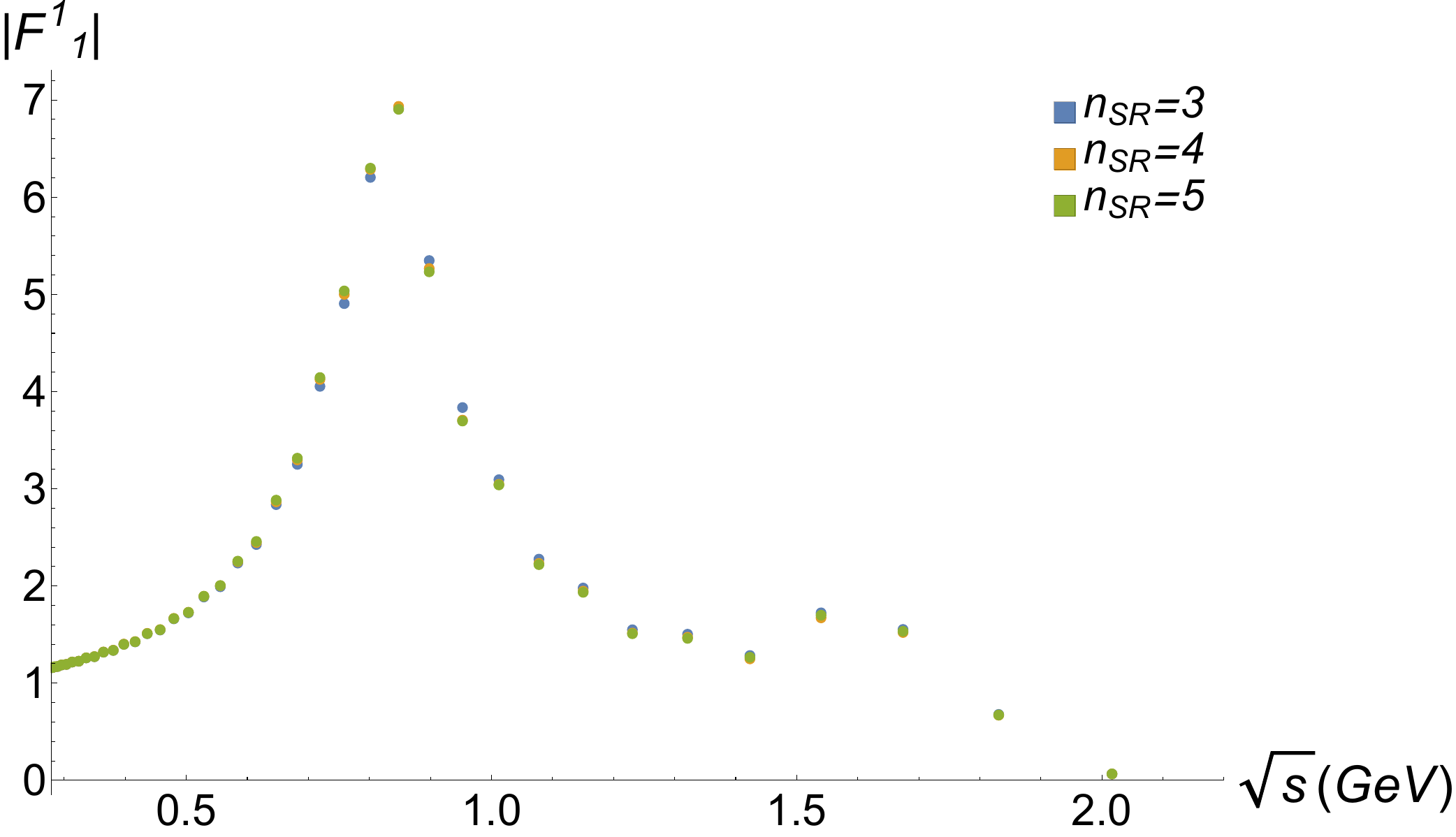}
\caption{$|F^1_{1}|$}
\end{subfigure}
\vspace{1cm}

\begin{subfigure}[b]{0.49\textwidth}
\centering
\includegraphics[width=\textwidth]{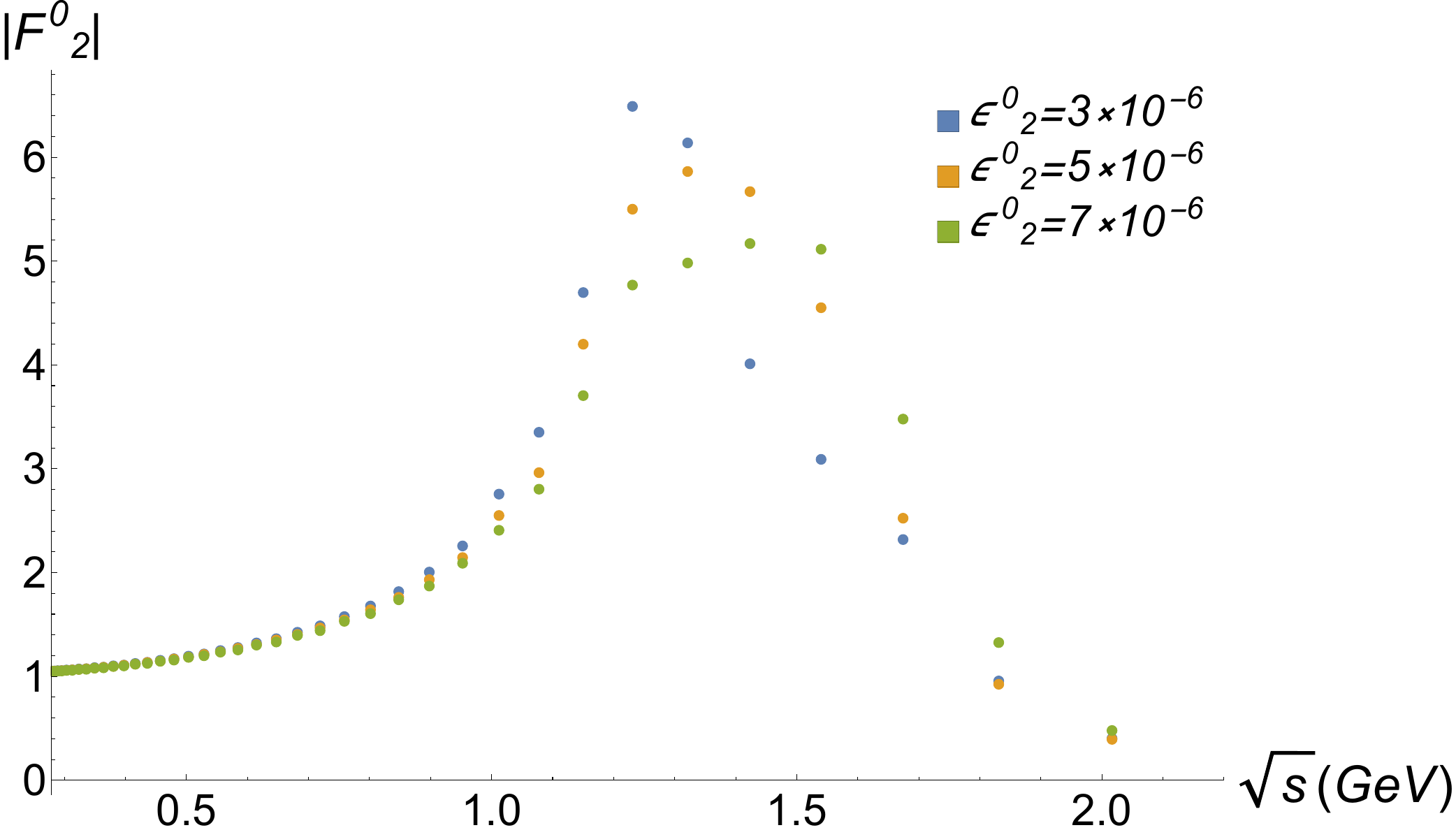}
\caption{$|F^0_{2}|$}
\end{subfigure}
\begin{subfigure}[b]{0.49\textwidth}
\centering
\includegraphics[width=\textwidth]{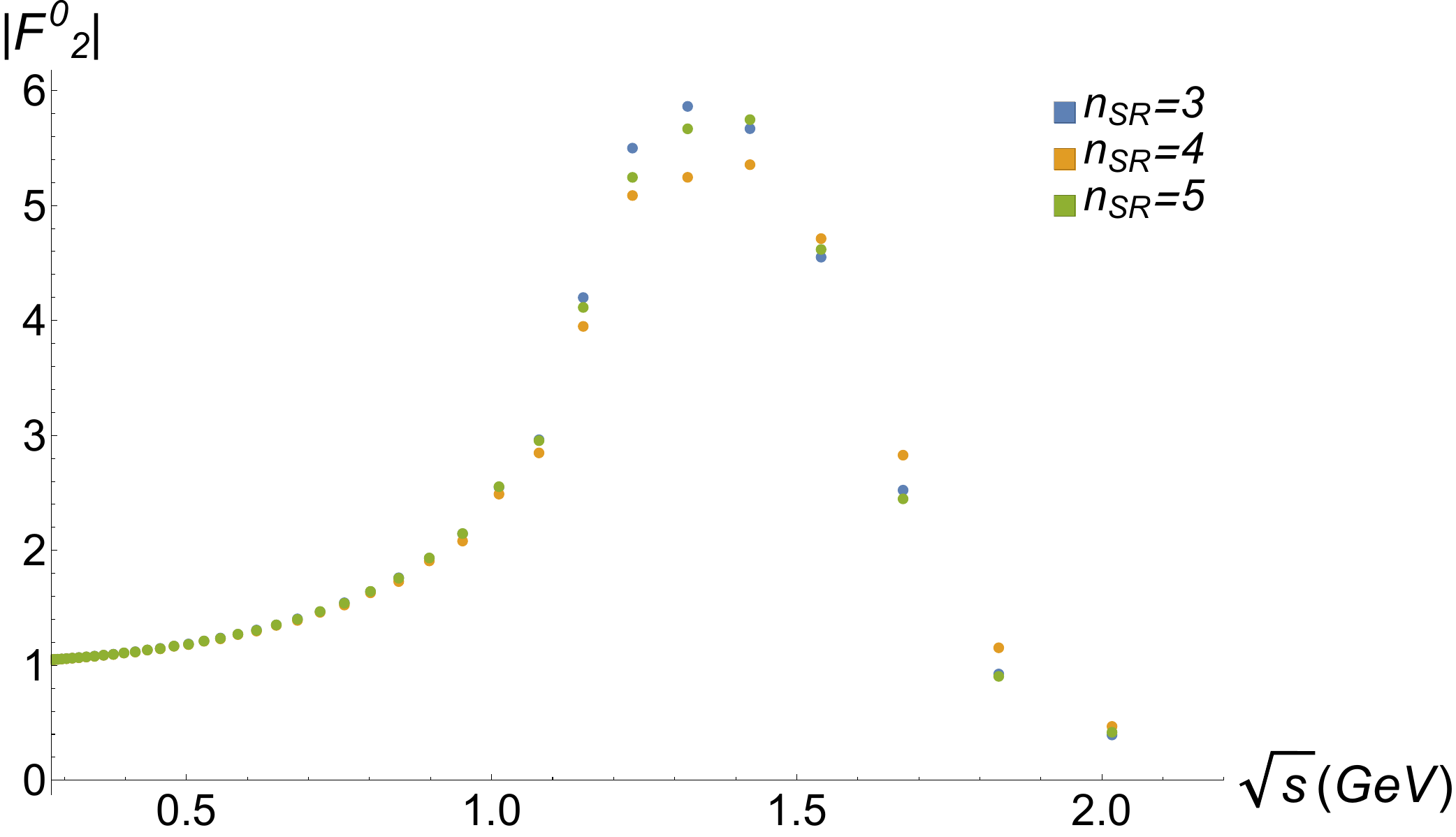}
\caption{$|F^0_{2}|$}
\end{subfigure}
\caption{The modulus $|F^I_\ell|$ of the $S0,P1,D0$ form factors for varying number of moments $n_{SR}$ and varying sum rule errors $\epsilon^I_\ell$.}
\label{SRnerr}
\end{figure}

For implementing the SVZ sum rules, we take $n_{SR}$ moment for each spectral density and characterize the difference with the leading perturbative QCD calculations by an error $\epsilon^I_{\ell}$ defined in \eqref{SRerr} which is around $10-30\%$ of the values of the moments themselves \eqref{SRvalues}. In figs. \ref{SRnerr} we show the results of $|F^I_{\ell}|$ for different number of moments $n_{SR}=3,4,5$ and varying values of $\epsilon^I_{\ell}$ around our estimate in \eqref{SRerr}. It is clear that increasing the number of moments does not change the results. Furthermore, a small variation of $\epsilon^I_{\ell}$ does not have significant influence. Of course, the error cannot be too big (e.g. bigger than the sum rules themselves). In any case, the sum rules implementation shows very good stability.

\subsection{Watsonian unitarization}\label{numW}

\begin{figure}[H]
\centering
\begin{subfigure}[b]{0.49\textwidth}
\raggedright
\includegraphics[width=\textwidth]{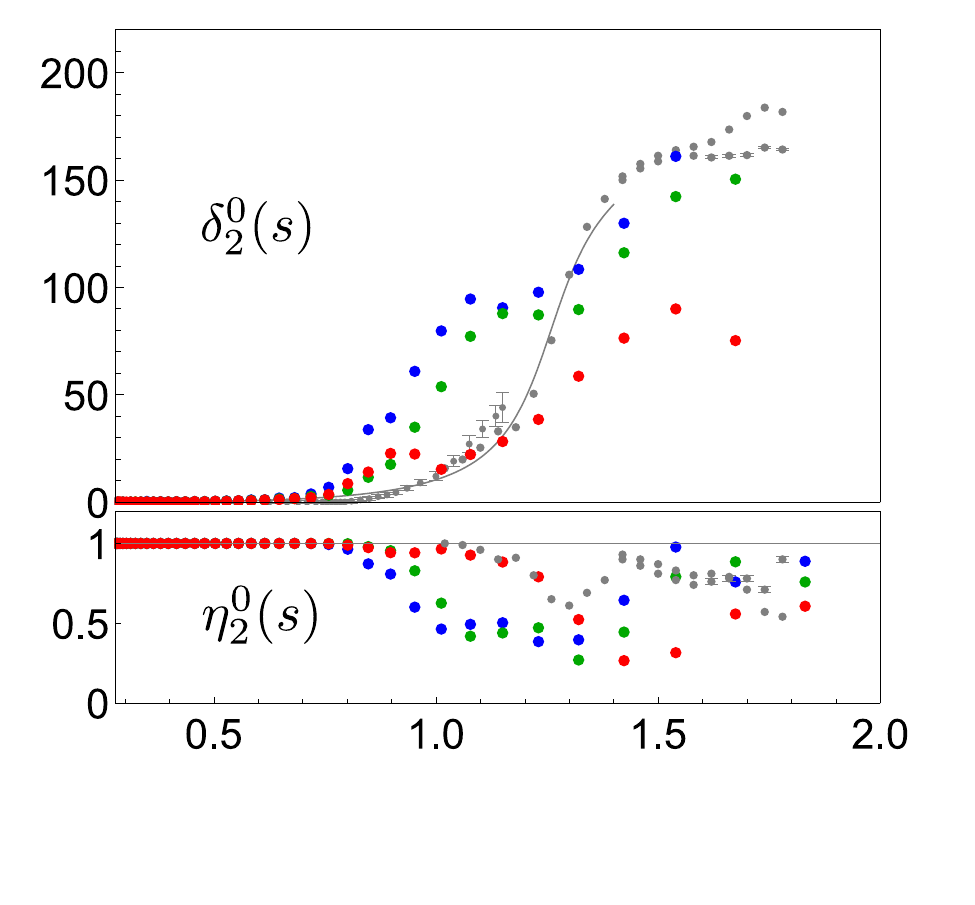}
\caption{}\label{BeforeW}
\end{subfigure}
\begin{subfigure}[b]{0.49\textwidth}
\raggedright
\includegraphics[width=\textwidth]{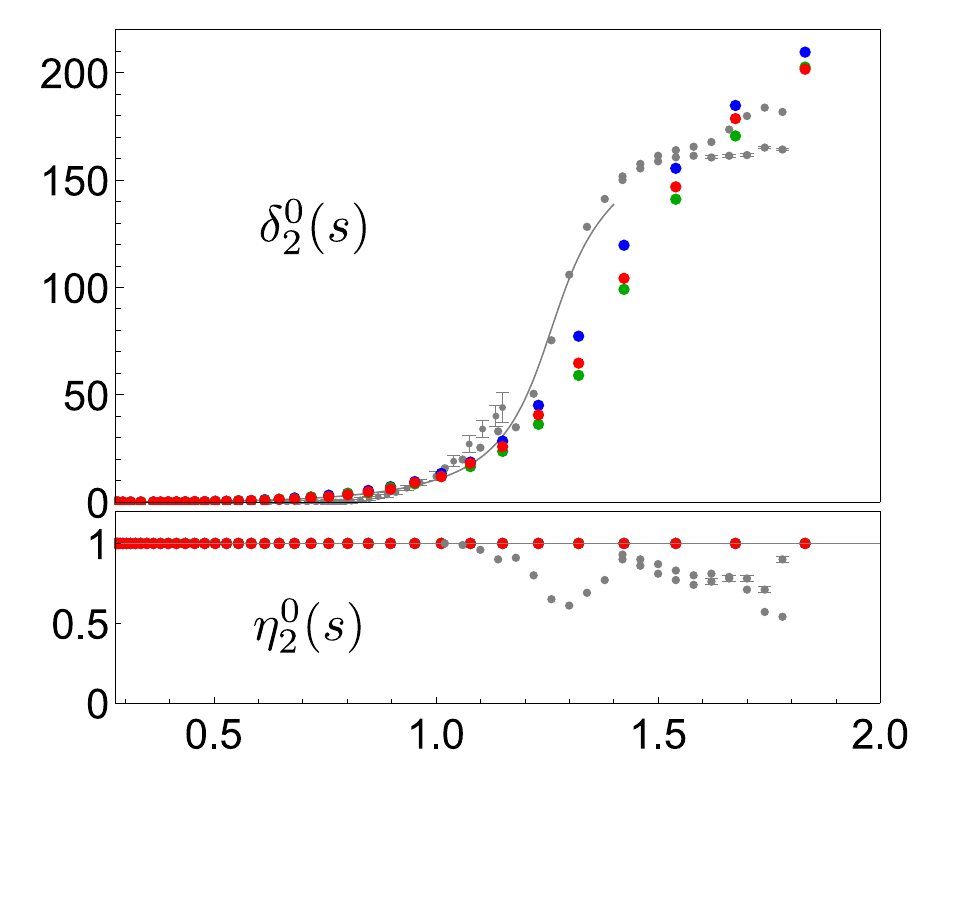}
\caption{}\label{AfterW}
\end{subfigure}
\caption{The $D0$ phase shift $\delta^0_2(s)$ and elasticity $\eta^0_2(s)$ before (\ref{BeforeW}) and after (\ref{AfterW}) the Watsonian unitarization.}
\label{WD0}
\end{figure}

As described in sections \ref{Wunit} and \ref{NGTB}, after an initial maximization, we use a Watsonian unitarization step to get closer to the boundary of the allowed space and improve saturation of the constraints. This step should help imprint UV information contained in the form factors into the phase shifts as can be seen, for example, for the case of the $D0$ wave. In fig. \ref{WD0} we show the $D0$ phase shift $\delta^0_2$ and elasticity $\eta^0_2$ before (\ref{BeforeW}) and after (\ref{AfterW}) the Watsonian unitarization for the three points (blue, green, red) in fig. \ref{shape}. It is clear that the Watsonian step greatly improves the results. 
 
\section{Form factors asymptotics}\label{appendix:FF}

The computation of asymptotic form factors is standard in QCD. However, for completeness we compute in this appendix the Feynman diagram (a) of fig.\ref{FFdiags}.  The diagram is computed as a tree level diagrams with external quark states giving what is called the kernel that has to be inserted between pion states:   
\beqa
\mathbb{T}^{(a)}_{\mathrm{1-diag}}&=&(ig)^2\left(\bar{u}_{k'}^{s'_1} \gamma^{\mu} \frac{i(\slashed k+\slashed p'-\slashed p)}{(k+p'-p)^2}\slashed\Delta u_k^{s_1}\right)\left(\bar{v}_{p-k}^{s_1}\gamma_\mu v_{p'-k'}^{s'_1}\right) \frac{(-i)}{(p-k-p'+k')^2} \times \nonumber\\
&&\ \ (-ik\Delta)^{\ell-1}\ t^{\hat{D}}_{ca} t^{\hat{D}}_{ac}\  \delta_{AB'}\delta_{A'C}\delta_{BD}\eeqa 

Writing $k=xp+k_\perp$ and $k'=x'p'+k'_\perp$ and taking the high energy limit where $s=(p-p')^2\simeq -2pp'\rightarrow \infty$ is the only term we keep, we perform the replacements
\beq
(k+p'-p)^2 \rightarrow (1-x) s, \ \ (p-k-p'+k')^2\rightarrow (1-x)(1-x')s, \ \ (k\Delta)\rightarrow x(p\Delta), 
\eeq
\beq
\bar{u}^{s'_1}_{k'}\rightarrow \sqrt{x'} \bar{u}^{s'_1}_{p'} \ \ u^{s_1}_{k}\rightarrow \sqrt{x} u^{s_1}_p\ \  \bar{v}^{s_1}_{p-k}\rightarrow \sqrt{1-x} \bar{v}^{s_1}_{p}\ \  v^{s'_1}_{p'-k'}\rightarrow \sqrt{1-x'} v_{p'}
\eeq
We can now further simplify the result with the identities (valid since we choose $v^s_p = -\gamma_5 u^s_p$):
\beq
v_{p'}^{s} \bar{u}^{s}_{p'} = -\gamma_5(\slashed p'+m_q),\ \ \ \ u_p^s\bar{v}^s_p = (\slashed p+m_q)\gamma_5
\eeq
we find, using, $t^{\hat{D}}_{ca} t^{\hat{D}}_{ac} = C_2(N_c) = \frac{N_c^2-1}{2}$, the result
\beq
\mathbb{T}^{(a)}_{\mathrm{1-diagram}}=\sqrt{x(1-x)x'(1-x')}\,  \delta_{AB'}\delta_{A'C}\delta_{BD}\ \frac{8ig^2C_2(N_c)}{s}(-ip\Delta)^{\ell-1}\frac{x^{\ell-1}}{(1-x)(1-x')}
\eeq
which is used in the main text. There we included the other three diagrams of the same form where the current is inserted in the other fermionic lines.

\section{Some integrals}\label{sec:int}
In this appendix we include some integrals that can be done by standard methods and that we find useful in the main text. 
\subsection{Feynman type integrals}
\beqa
\int\frac{d^dk}{(2\pi)^d}\, \frac{k^{2\ell}}{(k^2-\tilde{\Delta}+i\epsilon)^n} &=& i\frac{(-1)^{\ell+n}}{(4\pi)^{\frac{d}{2}}}\frac{\Gamma(\frac{d}{2}+\ell)\Gamma(n-\frac{d}{2}-\ell)}{\Gamma(n)\Gamma(\frac{d}{2})(\tilde{\Delta}-i\epsilon)^{n-\frac{d}{2}-\ell}}\\
\int\frac{d^dk}{(2\pi)^d}\, \frac{(ik\Delta)^{\ell}(ik\bar{\Delta})^{\ell}}{(k^2-\tilde{\Delta}+i\epsilon)^n} &=& i\frac{(-1)^{n}}{2^\ell(4\pi)^{\frac{d}{2}}}\frac{\ell! \, (\Delta\bar{\Delta})^\ell\,\Gamma(n-\frac{d}{2}-\ell)}{\Gamma(n)(\tilde{\Delta}-i\epsilon)^{n-\frac{d}{2}-\ell}}
\eeqa  		
where $\Delta^2=0=\bar{\Delta}^2$. Thee are useful to compute current correlators.  

\subsection{Multi-ball integrals}
In the case of four particle states it is useful to compute integrals of the form 
\beqa
\int \prod_{j=1}^4 d^3k_j\ \delta(\sum_{j=1}^4|\vec{k}_j|-1) \delta^{(3)}(\sum_{j=1}^4\vec{k}_j) &=& \frac{11\pi^3}{26880} \\
\int \prod_{j=1}^4 d^3k_j\ \delta(\sum_{j=1}^4|\vec{k}_j|-1) \delta^{(3)}(\sum_{j=1}^4\vec{k}_j)\, (1-\hat{k}_1\hat{k}_2)(1-\hat{k}_3\hat{k}_4)&=& \frac{\pi^3}{1440} 
\eeqa
that measures in some way the phase space available to four massless particles at given energy and in the center of mass. As an example, one way to compute these integrals is ($\lambda>0$ is an arbitrary constant)
\beqa
&&\int \prod_{j=1}^4 d^3k_j\ \delta(\sum_{j=1}^4|\vec{k}_j|-1) \delta^{(3)}(\sum_{j=1}^4\vec{k}_j)\nonumber\\ &=&e^{\lambda}\!\!
\int\!\!\frac{d^4 x}{(2\pi)^4} e^{-ix_0} \prod_{j=1}^4 d^3k_j\ e^{-(\lambda-ix_0)|\vec{k}_j| + i \vec{k}_j\vec{x}}  \non\\
&=& 2^{12}\pi^4\int\!\!\frac{d^4 x}{(2\pi)^4} e^{\lambda-ix_0} \frac{(\lambda-ix_0)^4}{((\lambda-ix_0)^2+x^2)^8} \non\\
&=& \frac{2^{12}\pi^4\Gamma(\frac{13}{2})}{\Gamma(8)(4\pi)^{\frac{3}{2}}}\int \frac{dx_0}{2\pi} \frac{e^{\lambda-ix_0}}{(\lambda-ix_0)^9} \non\\
&=& \frac{11 \pi^3}{26880} 
\eeqa

\bibliographystyle{utphys}
\bibliography{references}
\end{document}